\documentclass{article}
\usepackage[utf8]{inputenc}
\usepackage{amsmath, bbm}
\usepackage[margin=2.5cm]{geometry}
\usepackage{xcolor}
\usepackage[normalem]{ulem}
\usepackage{graphicx}
\usepackage{cite}
\usepackage{authblk}
\usepackage{amsmath,amsfonts,amsbsy,amssymb}
\usepackage{sansmath}
\usepackage{mathrsfs}
\usepackage{epstopdf}
\usepackage{url}
\usepackage{enumerate}

\definecolor{link}{rgb}{0.0, 0.0, 0.0 }
\usepackage[bookmarks = true,
			pdfstartview = FitH,
			colorlinks = true,
			urlcolor=link,
			citecolor=link,
			linkcolor=link,
			hyperfootnotes=false]{hyperref}

\begin{document}
\title{Dynamics of photosynthetic light harvesting systems interacting with N-photon Fock states}
\author[1,2]{Liwen Ko\thanks{liwen.jko@berkeley.edu}}
\author[1,2]{Robert L. Cook}
\author[1,2]{K. Birgitta Whaley\thanks{whaley@berkeley.edu}}
\affil[1]{Department of Chemistry, University of California, Berkeley, CA 94720, USA}
\affil[2]{Kavli Energy Nanoscience Institute at Berkeley, Berkeley, CA 94720, USA}
\maketitle

\begin{abstract}
    We develop a method to simulate the excitonic dynamics of realistic photosynthetic light harvesting systems including non-Markovian coupling to phonon degrees of freedom, under excitation by N-photon Fock state pulses. This method combines the input-output formalism and the hierarchical equations of motion (HEOM) formalism into a double hierarchy of coupled linear equations in density matrices. We show analytically that in a density matrix description, under weak field excitation relevant to natural photosynthesis conditions, an N-photon Fock state input and a corresponding coherent state input give rise to equal density matrices in the excited manifold. However, an important difference is that an N-photon Fock state input has no off-diagonal coherence between the ground and excited subspaces, in contrast with the coherences created by a coherent state input.
    We derive expressions for the probability to absorb a single Fock state photon, with or without the influence of phonons. 
    For short pulses (or equivalently, wide bandwidth pulses), we show that the absorption probability has a universal behavior that depends only upon a system-dependent effective energy spread parameter $\Delta$ and an exciton-light coupling constant $\Gamma$. This holds for a broad range of chromophore systems and for a variety of pulse shapes.  We also analyse the absorption probability in the opposite long pulses (narrow bandwidth) regime. We then derive an expression for the long time emission rate in the presence of phonons and use it to study the difference between collective versus independent emission.  Finally, we present a numerical simulation for the LHCII monomer (14-mer) system under single photon excitation that illustrates the use of the double hierarchy for calculation of Fock state excitation of a light harvesting system including chromophore coupling to a non-Markovian phonon bath. 
\end{abstract}

\section{Introduction}

Exciton dynamics in natural photosynthetic systems have been studied extensively in recent years, using a range of theoretical techniques~\cite{Ishizaki_2009, Kreisbeck_2016, Kundu_2020, Berkelbach_2012,Roden_2016}. Most of such studies assume that an initial excitation is present or created at some initial time. Behind this assumption is the implicit further assumption that light absorption and exciton transfer happen sequentially, while in reality they happen simultaneously. A related issue is that while it is well known that under weak light conditions natural photosynthetic systems have very high efficiency in utilizing absorbed photons to initiate charge transfer reactions in the reaction center~\cite{Blankenship}, the probability to absorb incoming photons in the first place is seldom discussed and is not well characterised at the microscopic level. Our group has addressed these problems previously by studying the absorption and exciton dynamics under excitation by pulses of weak coherent state light in the presence of a phonon bath \cite{Herman_2018}. Here we go beyond that study, with a theory to model the interactions and dynamics of an exciton system interacting with N-photon Fock state pulses and a phonon bath. In this work we focus on the exciton system density matrix under the influences of photon and phonon environments.   A related paper considers the dynamics of individual quantum trajectories post-selected on measuring emitted fluorescent photons~\cite{cook_trajectories_2021}.
\par
Probing a light harvesting system with N-photon Fock state pulses has the advantage that, upon observing m outgoing photons, we can deduce (ignoring experimental imperfections) that the system has exactly $N-m$ excitons, due to the excitation conserving property of the total Hamiltonian. The coherent state laser pulses commonly used in experimental studies are superpositions of different photon number (Fock) states, so they do not allow for this type of precise knowledge about the state of the photosynthetic system.
\par
A critical difference between the master equations for quantum systems interacting with Fock states and coherent states of light is that the influence of a Fock state on the system is non-Markovian \cite{Baragiola_2012}, while the influence of a coherent state of light is Markovian \cite{wiseman_milburn_2009_book} and can be treated by considering the system interacting with a classical electric field plus the quantum theory of spontaneous emission. Employing the input-output formalism \cite{Gardiner_Collett_1985, Combes_2017_review}, Baragiola et. al. \cite{Baragiola_2012} used the closely related quantum stochastic differential equation (QSDE) formalism to derive a set of Fock state master equations that propagate a physical density matrix coupled with a hierarchy of auxiliary density matrices. For completeness we present here an alternative derivation using the language of ordinary calculus to derive quantum Langevin equations in a more accessible formalism (Section \ref{sec:methods}).  A key fact that allows us to apply the input-output formalism to light harvesting systems interacting with the three-dimensional (3-d) electromagnetic field is that under the dipole approximation, the interaction with the 3-d electromagnetic field can be described as the interaction with a finite number of 1-d fields because the electric field operator is linear in the field bosonic operators. This reduction of the degrees of freedom is described in detail in Section \ref{sec:3d_field_to_1d_field}. 
\par
To model the non-Markovian effects of the phonon bath, we employ here the hierarchical equations of motion (HEOM) \cite{Tanimura_1989}. When these are combined with the Fock state hierarchy, the final master equations for the excitonic density matrix take the form of a double hierarchical structure of linearly coupled differential equations. While numerically accurate, this comes at the cost of increased computational complexity. For a system with $N$ chromophores interacting with a $N_p$-photon Fock state using the HEOM truncated at $N_c$ cutoff levels, a set of $(N_p+1)^2(N+N_c)!/(N!N_c !)$ coupled density matrix equations need to be simultaneously solved. 
Because of this cost we limit our numerical studies here to consider the 14 site LHCII monomer, a 2 site dimer and a 7-site subsystem of LHCII considered previously~\cite{Herman_2018}. 

In the interest of gaining important insights applicable also to larger systems, we additionally develop here analytical studies of the double hierarchy of equations in certain regimes.
These studies focus primarily on the case of a single Fock state photon, since the analytical solution for the reduced exciton system state is most readily obtained where there is only one photon. A key result of our analysis is the demonstration in Section \ref{sec:N_photon_Fock_vs_coh} that in the weak chromophore-light coupling limit (relevant to natural photosynthesis, since the intensity of natural sunlight is about $10^{-3}\,\text{ photons}$ per second on a single chlorophyll molecule \cite{Blankenship}),
the chromophore system dynamics under the excitation of an N-photon Fock state bears a close relationship to the dynamcis under the excitation of a single photon Fock state. 
\par
The analysis underlying this key result hinges on the natural separation of time scales between the exciton-exciton and exciton-phonon couplings, and the exciton-light dynamics.
In natural photosynthetic systems, the exciton-light coupling is about 5-6 orders of magnitude weaker than the exciton-exciton or exciton-phonon couplings. Thus the spontaneous emission occurs at a much longer (ns) time scale than the exciton-exciton and exciton-phonon dynamics, both of which occur on sub-ps time scales. Because of this separation of time scales we can ignore the effect of spontaneous emission at short times. Under this approximation, we can solve the single photon Fock state master equation exactly. The solution is most easily obtained not by solving the non-Markovian Fock state master equations or the HEOM directly, but by considering the chromophore system, the vibrations coupled to this, and the optical field together as a pure state evolving according to the Schr\"{o}dinger equation. Somewhat surprisingly, the solution to this equation is similar to the second order perturbative solution for a coherent state input. The only difference is that a Fock state input cannot induce any coherence between exciton system subspaces of different exciton number. The resulting solutions to the single photon Fock state master equations enable us to write down analytical expressions for the absorption probability and to understand its dependence on various parameters, most importantly, on pulse duration (or equivalently, inverse bandwidth). Due to the similarity between Fock state and coherent state input light, we can then further understand the coherent state absorption probability using the new Fock state absorption probability expressions.
\par
At long times, the electronic excitation decays via spontaneous emission. Note that we do not include any additional non-radiative decay pathways from the excitonic manifold in the present model. We find that due to the steady state in the excitonic manifold with respect to the phonon bath, the exciton system dynamics follows a single exponential decay to the ground state, giving us a single well-defined decay constant at long times. It is sometimes assumed that the chromophores emit independently of one another (see e.g.,~\cite{Herman_2018}), but more rigorous treatment of the light-matter interaction~\cite{dicke_coherence_1954} shows that the chromophores should emit collectively \cite{Lehmberg_1970, Gross_Haroche_1982_review, Spano_Mukamel_1989}. The collective emission rate can show enhancements ranging from $0$ to $N$, the number of chromophores.
We show that for natural photosynthetic systems the collective emission rates are usually very similar to the independent emission states, as a result of the non-uniform orientations of the dipole moments and the interaction with phonons.

\par
The remainder of the paper is organized as follows. Section \ref{sec:methods} introduces the basics of the input-output formalism and provides a detailed modeling of the absorption and energy transport problem in the language of this formalism. The near exact solutions to the Fock state master equations, with or without phonons, are derived and the input-output formalism with the HEOM presented. In Section \ref{sec:Fock_vs_coherent} we discuss the similarities and differences between excitation under a Fock state and excitation under a coherent state input optical fields, as well as the relationship of the dynamics under a single photon state and an N-photon Fock state. Section \ref{sec:analysis_on_absorption} analyzes the short time absorption probability and its dependence on various parameters such as dipole orientations, light polarization, pulse duration, and the presence or absence of exciton-phonon coupling. For short pulses, we show a universal behavior of the absorption probability by defining an effective energy spread parameter $\Delta$ that characterizes the range of system energies. For long pulses, we analyze the absorption probabilities under several different parameter regimes. Section \ref{sec:analysis_on_emission} analyzes the long time emission behavior in the presence of phonons. Numerical examples are given in these Sections using dimeric and 7-mer chromophore systems from the LHCII monomer. In Section \ref{sec:LHCII_calculation} we then describe a numerical simulation of the double hierarchy of equations describing the Fock state master equation + HEOM on the full LHCII monomer (14-mer) system. Finally in Section \ref{sec:conclusion} we provide a summary and assessment.

\section{Methods}
\label{sec:methods}
\subsection{Chromophoric System Hamiltonian}
We model the lowest two accessible electronic states of each chromophore as a two level system, corresponding to the Q\textsubscript{y} transition \cite{Blankenship}. The dipole-dipole coupling between chromophores, under the rotating wave approximation, does not change the number of electronic excitations. Therefore the chromophoric Hamiltonian will commute with the operator that counts the number of excitons in the system, and the Hamiltonian is block diagonal, with the j$^{th}$ block corresponding to the j-excitation subspace, $\mathcal{H}_j$. The excitation energy of a typical chromophore is $\sim 15,000 \,\text{cm}^{-1}$, about 75 times larger than $k_B T\approx 200 \text{ cm}^{-1}$ at room temperature. Therefore, to a very good approximation, we may regard the initial thermal state of the isolated chromophoric system as the absolute ground state of $\mathcal{H}_0$, denoted as $|g\rangle$, which is defined to have zero energy. Due to the weak system-field coupling in a natural photosynthetic system, the probability to have multiple excitations in the system is much smaller than the probability to have a single excitation, so we will only consider the $(1+N)$-dimensional subspace $\mathcal{H}_0 \bigoplus \mathcal{H}_1$. We denote the singly excited state where site $j$ is excited and all other sites are in their ground states as $|j\rangle$. The chromophoric system Hamiltonian is written as
\begin{equation}
    H_{\text{sys}} = \sum^N_{j=1} \epsilon_j |j\rangle\langle j| + \sum_{j \neq k} J_{jk}|j\rangle\langle k|,
\label{eq:system_Hamiltonian}
\end{equation}
where $\epsilon_j$ is the excited state energy of site $j$, and $J_{jk}$ is the dipole-dipole interaction between chromophores $j$ and $k$. 
The effects of the phonon environment on the site energies are discussed in Section \ref{sec:sys_vib_interaction}. The numerical values for the $\epsilon_j$ and $J_{jk}$ parameters in an LHCII monomer system are taken from \cite{Novoderezhkin_2011}. For simplicity, we shall refer to the chromophore system as the system, unless otherwise noted.

\subsection{System-light interaction as system interacting with finite number of one-dimensional electromagnetic fields}
\label{sec:3d_field_to_1d_field}
The quantized electromagnetic field in 3-dimensional space is described by harmonic oscillators indexed with a 3-dimensional wavevector and a polarization index \cite{Loudon_2000_book, Cohen-Tannoudji_photons_and_atoms}. However, it is useful to decompose the electric field into 1-dimensional fields, so that the input-output formalism can be applied \cite{Gardiner_Collett_1985,Rob_thesis}.
We will first consider the incoming N-photon Fock state in a fixed paraxial beam mode. The electric field at the center of the beam is written as
\begin{equation}
    \mathbf{E}_{\text{para}}(t) =\int_0^\infty d\omega\, \sqrt{\frac{\hbar\omega}{4\pi\epsilon_0 c}}\Tilde{\mathbf{u}}(\mathbf{0}) (ia(\omega)e^{-i\omega t} + \text{h.c.}),    
\label{eq:paraxial_Efield_main_text}
\end{equation}
(see derivation in appendix \ref{app:paraxial_quantization} or Ref.~\cite{Baragiola_2014_paraxial_quantization}), where $\Tilde{\mathbf{u}}(\mathbf{x})$ is the normalized spatial mode function such that the integral over all transverse area is unity, i.e., $\int dA_T\,|\Tilde{\mathbf{u}}|^2 = 1$. Here $a(\omega)$ is the annihilation operator for the frequency $\omega$ component of the paraxial mode. Note that the field is now indexed by the 1-dimensional parameter $\omega$.
\par
\begin{figure}[htbp]
    \centering
    \includegraphics[scale=0.6]{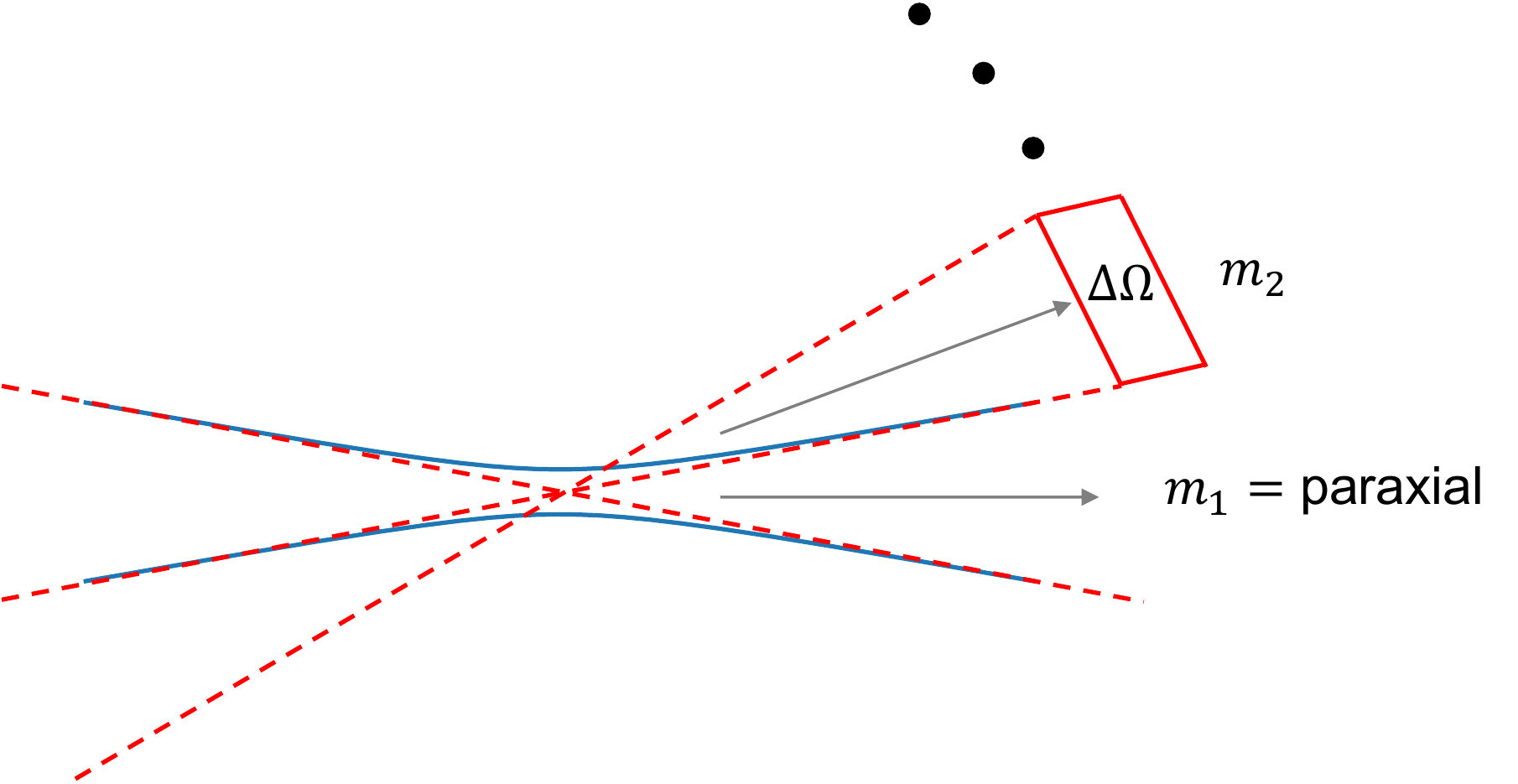}
    \caption{Small angle modal decomposition of the electric field in three dimensions defined in reference to a paraxial beam traveling from left to right.  The chromophore system will be located at the focus of this beam.   Blue solid curves show the contour of the paraxial beam mode. Red dashed lines denote the boundaries of small solid angle modes. Since the paraxial beam mode is concentrated in a small solid angle as shown, we can treat it as the small solid angle mode $m_1$. $m_2$ is another small solid angle mode propagating in the upper right direction and covering the solid angle $\Delta \Omega$.}
    \label{fig:paraxial}
\end{figure}

 To capture the spontaneous emission into other modes, we must also consider field modes other than the incoming mode. One way to do this is to partition the $4\pi$ solid angle of the 3-dimensional wavevector into finite numbers of small solid angle sections, indexed by $m$, and write the electric field at position $\mathbf{x}=\mathbf{0}$ as
\begin{subequations}
\begin{equation}
    \mathbf{E}(t) = \sum_{m,\lambda} \mathbf{E}_{m,\lambda}(t),
\end{equation}
where
\begin{equation}
    \mathbf{E}_{m,\lambda}(t) = \int_0^\infty d\omega \, \sqrt{\frac{\hbar\omega^3\Delta\Omega_m}{16\pi^3\epsilon_0 c^3}}(ia_{m,\lambda}(\omega)e^{-i\omega t}\hat{\epsilon}_{m,\lambda} + \text{h.c.})
\end{equation}
\label{eq:small_solid_angle_E_main_text}
\end{subequations}
(see appendix \ref{app:small_angle_quantization}). Here $\lambda$ indexes the two polarizations in a small solid angle section, $\Delta\Omega_m$ is the amount of solid angle (in steradian units) of section $m$, and $\hat{\epsilon}_{m, \lambda}$ is the polarization vector corresponding to $m$ and $\lambda$. One can show that in the case of the simplest transverse electromagnetic mode TEM\textsubscript{00},if we think of the boundary of the beam as the location where the beam intensity is $1/e^4\approx 2\%$ of the intensity at the center, then Eq. (\ref{eq:paraxial_Efield_main_text}) matches Eq. (\ref{eq:small_solid_angle_E_main_text}) for a particular $(m,\lambda)$. Therefore we can treat the incoming TEM\textsubscript{00} paraxial mode as one of the $(m,\lambda)$ modes in the small angle decomposition (Eq. (\ref{eq:small_solid_angle_E_main_text})), as illustrated in Figure (\ref{fig:paraxial}). 
 
\par In Eq. (\ref{eq:small_solid_angle_E_main_text}), the electric field is decomposed into a finite number of 1-D fields, which is not enough to describe all degrees of freedom in the 3-D electromagnetic field. However, appendix \ref{app:orthonormal_decomposition_free_field} shows that one can describe the 3-D electromagnetic field by a set of countably infinite 1-D fields. Then using the small solid angle decomposition (see Eq. (\ref{eq:small_solid_angle_E_main_text}) and appendix \ref{app:small_angle_quantization}), we can decompose the 3-D electromagnetic field into a finite number of small solid angle 1-D fields plus a countably infinite number of 1-D fields that are needed to describe all degrees of freedom in the 3-D field. The countably infinite number of 1-D fields are chosen to be orthogonal to each other and to the small solid angle 1-D fields in the sense described in appendix \ref{app:orthonormal_decomposition_free_field}. Under this decomposition scheme, the total system+field Hamiltonian under the dipole-electric field coupling ($-\mathbf{d}\cdot \mathbf{E}$) can be written as
\begin{equation}
    H_{\text{sys+field}} = H_{\text{sys}} -\mathbf{d}\cdot \sum_{m,\lambda}  \mathbf{E}_{m,\lambda} + \sum_{m,\lambda} \int^\infty_0 d\omega\, \hbar\omega a_{m,\lambda}^\dagger(\omega)a_{m,\lambda}(\omega) + \sum_{s}^\infty \int^\infty_0 d\omega\, \hbar\omega a_{s}^\dagger(\omega)a_{s}(\omega),
\label{eq:system+field_H_infinite_modes}
\end{equation}
where $\sum_{m, \lambda}$ sums over the finite number of small solid angle modes, and the $\sum_{s}^{\infty}$ sums over the remaining countably infinite number of 1-D field modes.
Since the last term in Eq. (\ref{eq:system+field_H_infinite_modes}) involving the infinite sum is decoupled from the rest of the Hamiltonian, we can describe the system-light interaction as system interacting with just a finite number of 1-D fields. 
 
\par An alternative way to decompose the electric field into a finite sum of 1-D fields is to write it as a sum of the x, y, and z polarization components, i.e. $\mathbf{E}(t) = E_x(t)\hat{x} + E_y(t)\hat{y} + E_z(t)\hat{z}$. $E_x$ then takes the form
\begin{equation}
    E_x(t)=\int^\infty_0 d\omega \, \sqrt{\frac{\hbar\omega^3}{6\pi^2 \epsilon_0 c^3}}(ia_x(\omega)e^{-i\omega t} + \text{h.c.})
\label{eq:polarization_mode_E_main_text},
\end{equation}
with $E_y$ and $E_z$ defined similarly (see appendix \ref{app:polarization_quantization}).

\subsection{System-light interaction in the language of input-output formalism}
\label{sec:sys-light_interaction_input_output}
To put the system-light interaction in the language of input-output formalism, we follow the procedure in \cite{Combes_2017_review}, with addition of some details specific to our modeling of light harvesting systems. Since the size of light harvesting complexes is much smaller than the wavelength of the light they interact with, we treat all chromophores as being in the same location. The dipole operator $\mathbf{d}$ of the system is written as
\begin{equation}
    \mathbf{d} = \sum_{j=1}^{N} (\Tilde{\mathbf{d}}_j \sigma^{(j)}_- + \text{h.c.}), 
\end{equation}
where $\Tilde{\mathbf{d}}_j=\langle g|\mathbf{d}|j\rangle$ is the transition dipole moment of site $j$, and $\sigma^{(j)}_-\equiv |g\rangle\langle j|$ is the lowering operator of site $j$.
The total system+field Hamiltonian is written as
\begin{equation}
    H_{\text{sys+field}} = H_{\text{sys}} + \frac{\hbar}{\sqrt{2\pi}}\sum_l \int^\infty_0 d\omega \, \big( -ia_l(\omega)L_l^\dagger (\omega) + \text{h.c.}\big) + \sum_l \int^\infty_0 d\omega \, \hbar \omega a_l^\dagger(\omega)a_l(\omega),
\label{eq:input_output_1}
\end{equation}
 where 
\begin{equation}
    L_l(\omega) = \sqrt{\frac{d_0^2\omega^3 \Delta\Omega}{8\pi^2\epsilon_0 \hbar c^3}} \sum_{j=1}^N \mathbf{d}_j\cdot\hat{\epsilon}\, \sigma^{(j)}_-,
\label{eq:L_dagger_def}
\end{equation}
for a small solid angle mode (Eq. (\ref{eq:small_solid_angle_E_main_text})). For notational simplicity, we denote the mode indices $(m,\lambda)$ by $l$. The dipole moment $\Tilde{\mathbf{d}}_j$ is expressed as a scalar unit dipole $d_0$ multiplying a dimensionless vector dipole moment $\mathbf{d}_j$. In Eq. (\ref{eq:input_output_1}), we have used a rotating wave approximation to drop terms that do not conserve excitation numbers, i.e. $a^\dagger(\omega) L^\dagger(\omega)$ and $a(\omega)L(\omega)$. The constant factor $\frac{\hbar}{\sqrt{2\pi}}$ is chosen for later convenience.  For a paraxial mode (Eq. (\ref{eq:paraxial_Efield_main_text})), the prefactor in Eq. (\ref{eq:L_dagger_def}) is $\sqrt{\frac{d_0^2\omega}{2\epsilon_0 \hbar c}}\Tilde{u}(\mathbf{0})$.
For a polarization mode (Eq. (\ref{eq:polarization_mode_E_main_text})), the prefactor is $\sqrt{\frac{d_0^2\omega^3}{3\pi\epsilon_0 \hbar c^3}}$. In the current work one of the field modes indexed by $l$ in Eq. (\ref{eq:input_output_1}) is the incoming paraxial mode containing the N-photon Fock state pulse. In the following, we shall denote quantities pertaining to the incoming mode by replacing the subscript ``$l$" with subscript ``inc".
\par Since the system is resonant with the field in a narrow frequency range near the pulse center frequency $\omega_0$, only the frequency near $\omega_0$ contributes significantly to the overall dynamics. We make the approximation $L_l(\omega)\approx L_l(\omega_0)$, and simply denote it as $L_l$ hereafter. In order to have a unified approach to describe different ways to decompose the electric field, it is useful to express the prefactor in Eq. (\ref{eq:L_dagger_def}) as $\sqrt{\Gamma_0 \eta}$, where $\Gamma_0= \frac{d_0^2 \omega_0^3}{3\pi\epsilon_0\hbar c^3}$ is the spontaneous emission rate of a unit dipole $d_0$ with transition frequency $\omega_0$ in vacuum, and $\eta$ is a geometric factor depending on the type of the field mode.
For a paraxial mode (Eq. (\ref{eq:paraxial_Efield_main_text})), $\eta=\frac{3\pi c^2}{2\omega_0^2 }|\Tilde{u}(\mathbf{0})|^2$; for a small solid angle mode (Eq. (\ref{eq:small_solid_angle_E_main_text})), $\eta=\frac{3}{8\pi}\Delta\Omega$; for a polarization mode (Eq. (\ref{eq:polarization_mode_E_main_text})), $\eta=1$. 

To summarize, $L_l$ is now written as
\begin{equation}
    L_l = \sqrt{\Gamma_0 \eta_l}\sum_j \mathbf{d}_j \cdot \hat{\epsilon} |g\rangle\langle j|.
\label{eq:L_def_2}
\end{equation}
To account for the dielectric medium, we will replace the spontaneous emission rate in vacuum with the spontaneous emission rate in the empty cavity model with index of refraction $n$, given by
\begin{equation} \label{eq:Gamma_0}
    \Gamma_0=n\left(\frac{3n^2}{2n^2+1}\right)^2 \frac{d_0^2 \omega_0^3}{3\pi\epsilon_0\hbar c^3}
\end{equation}
\cite{Glauber_1991, Knox_2003}. We take $n=1.33$ as the index of refraction of water for the numerical simulations. 
\par $L_l$ can be simplified further by writing
\begin{equation}
    L_l = \sqrt{\Gamma_l}|g\rangle\langle B_l|,
\label{eq:L_def_bright_state}
\end{equation}
with $\Gamma_l$ the effective coupling constant 
\begin{equation}
    \Gamma_l = \Gamma_0 \eta_l \sum_j |\mathbf{d}_j \cdot \hat{\epsilon}|^2
\label{eq:effective_Gamma}
\end{equation}
and $| B_l \rangle$ the normalized bright state
\begin{equation}
    |B_l\rangle = (\sum_j |\mathbf{d}_j \cdot \hat{\epsilon}|^2)^{-1/2}\sum_j \mathbf{d}_j \cdot \hat{\epsilon} |j\rangle.
\end{equation}
As a physical motivation for this notation, we note that the excited state $|B_l\rangle$ spontaneously emits into mode $l$ at the rate $\Gamma_l$. A more detailed discussion of the emission rate, or more generally, the photon flux, is given in Section \ref{sec:flux}.
\par Going into an interaction frame where we rotate out the zeroth order non-interacting Hamiltonian
\begin{equation}
    H_0 = \sum^N_{j=1} \hbar\omega_0 |j\rangle \langle j| + \sum_l \int^\infty_0 d\omega\, \hbar\omega a_l^\dagger(\omega)a_l(\omega),
\label{eq:H_0}
\end{equation}
the interaction frame Hamiltonian becomes
\begin{equation}
    H_{\text{int}}(t) = \underbrace{\bigg(\sum^N_{j=1}\Delta_j|j\rangle\langle j| + \sum_{j\neq k} J_{jk} |j\rangle \langle k| \bigg)}_{H} + \frac{\hbar}{\sqrt{2\pi}}\sum_l \int^\infty_{0} d\omega \, \big(-i a_l(\omega)e^{-i(\omega-\omega_0)t} L_l^\dagger+ \text{h.c.}\big),
\label{eq:input_output_3}
\end{equation}
where $\Delta_j\equiv \epsilon_j-\hbar\omega_0$ is the site energy detuning (see Eq. (\ref{eq:system_Hamiltonian})). The term inside the first parenthesis acts on the system only and does not depend on time. We shall denote this simply as $H$. 
\par
Changing to the relative frequency variable
$\omega' = \omega-\omega_0$ and re-indexing the frequency components $a(\omega)\rightarrow a(\omega-\omega_0)$, Eq. (\ref{eq:input_output_3}) becomes
\begin{equation}
    H_{\text{int}}(t) = H + \frac{\hbar}{\sqrt{2\pi}}\sum_l\int^\infty_{-\infty} d\omega' \, \big(-i a_l(\omega')e^{-i\omega't} L_l^\dagger+ \text{h.c.}\big).
\label{eq:input_output_4}
\end{equation}
Since the dynamics is dominated by the field modes in a narrow bandwidth near resonance ($\omega'\approx0$) and $\omega_0$ is much larger than this bandwidth, we let the lower bound of the integral, $-\omega_0$, go to $-\infty$.

Finally, we define the quantum white noise operator 
\begin{equation}
    a_l(t)\equiv \sqrt{\frac{1}{2\pi}}\int^\infty_{-\infty} d\omega'\,a_l(\omega')e^{i\omega' t},
\end{equation}
a central object in the input-output formalism. With this definition, the field operators $a_l(t)$ satisfy the bosonic commutation relations: $[a_l(t), a_{l'}^\dagger(t')]=\delta_{l,l'}\delta(t-t')$ and $[a_l(t), a_{l'}(t')]=[a_l^\dagger(t), a_{l'}^\dagger(t')]=0$. The system-light interaction (Eq. (\ref{eq:input_output_4})) can now be written in a simple-looking form as
\begin{equation}
    H_{\text{int}}(t) = H +\sum_l \left( -i\hbar a_l(t)L_l^\dagger 
 + i\hbar a_l^\dagger(t)L_l \right).
\label{eq:input_output_5}
\end{equation}
\par
An N-photon Fock state pulse in mode $l$ with temporal profile $\xi(t)$ is defined as
\begin{equation}
    |N_\xi\rangle_l = \frac{1}{\sqrt{N!}}\bigg(\int d\tau\,\xi(\tau)a_l^\dagger(\tau)\bigg)^N|\phi\rangle,
\end{equation}
where $\xi(t)$ is normalized according to $\int d\tau\,|\xi(\tau)|^2=1$ and $|\phi\rangle$ is the vacuum state of the field.
More general N-photon Fock states are possible, as discussed in Ref.~\cite{Baragiola_2012}, but we shall restrict ourselves to this single mode product form here.

\par Another important object in the input-output formalism is the unitary $U(t)$ defined by
\begin{equation}
    \frac{dU(t)}{dt} = -\frac{i}{\hbar} H_{\text{int}}(t) U(t)
\label{eq:QSDE_unitary_1}
\end{equation}
with initial condition $U(0)=1$.
The system+field unitary in the Schrodinger picture, $e^{-\frac{i}{\hbar}H_{\text{sys+field}}t}$, is related to the interaction picture unitary $U(t)$ by
\begin{equation}
    e^{-\frac{i}{\hbar}H_{\text{sys+field}}t} = e^{-\frac{i}{\hbar}H_0 t}U(t)
\label{eq:sys+field_unitary_interation_pic}
\end{equation}
(see Eq. (\ref{eq:H_0})). We will set $\hbar=1$ from now on.

\subsection{Some results in the input-output formalism}
This section summarizes some results in the input-output formalism that will be used later. The operators $a(t)$ and $U(t)$ satisfy the so-called input-output relation \cite{Gardiner_Collett_1985}:
\begin{equation}
U(t')a_l(t)U^\dagger(t') =
\begin{cases}
a_l(t)\quad t'<t\\
a_l(t) + \frac{1}{2}L_l(t) \quad t'=t\\
a_l(t) + L_l(t) \quad t'>t,
\end{cases}
\label{eq:input_output_relation}
\end{equation}
with $L_l(t)\equiv U^\dagger(t) L_l U(t)$ (see appendix \ref{app:input_output_relation}). At time $t'<t$, $a_l(t)$ has not interacted with the system, so time evolution has no effect on it (i.e. $U(t')a_l(t)U^\dagger(t')=a_l(t)$). For this reason, $a_l(t)$ is called the input field.
At time $t'>t$, $a(t)$ has interacted with the system and will not interact with the system anymore. For this reason, the time-evolved operator $U(t')a_l(t)U^\dagger(t')=a_l(t)+L_l(t)\equiv a_{l,\text{out}}(t)$ is called the output field.
Using the Markov property that $a(t)$ only interacts with the system at time $t$, we have thereby expressed the time-evolved field operator, which usually mixes the system and field degrees of freedom in some complicated way, as a simple sum of a pure field operator and a time-evolved system operator. 
\par Left-multiplying Eq. (\ref{eq:input_output_relation}) by $U(t)$, we then obtain the following commutation relation at $t=t'$, 
\begin{equation}
    [a_l(t), U(t)] = \frac{1}{2}L_l U(t).
\label{eq:a_U_commutator}
\end{equation}
Using this commutation relation, we can rewrite Eq. (\ref{eq:QSDE_unitary_1}) (also see Eq. (\ref{eq:input_output_5})) in normal-ordered form as
\begin{equation}
    \frac{d}{dt}U(t) = \big(-iH-\frac{1}{2}\sum_l L_l^\dagger L_l\big)U(t)-\sum_l L_l^\dagger U(t)a_l(t) + \sum_l a_l^\dagger(t)L_l U(t).
\label{eq:QSDE_unitary_3}
\end{equation}
In the following sections, we will see that the commutation relation and the normal-ordered form of the time evolution derivative ensure that we only need to consider the action of normal-ordered field operators acting on field states, greatly simplifying the calculations.

Input-output theory is traditionally presented in terms quantum stochastic differential equations (QSDE) where the input field operators are taken to be differentials, e.g. $d A_{l,t} = a_l(t + dt) - a_l(t)$, analogous to the Wiener process increment $dW_t$ in classical stochastic calculus. As in classical stochastic calculus, integrals over the quantum stochastic differential $\int X_t dA_t$ can be interpreted either as Stratonivich integrals or as Ito integrals, resulting in different values~\cite{gardiner_1983,gardinerzoller_2004}. The Stratonovich integral takes a ``mid-point'' time approximation for the integrand, while the Ito integral takes the initial time approximation for the integrand. To make connections to the QSDE approach, we note that Eq. (\ref{eq:QSDE_unitary_3}) can be written as
\begin{equation}
    dU_t = -iH U_t dt-\sum_l L_l^\dagger U_t \circ dA_{l,t} + \sum_l L_l U_t \circ dA_{l,t}^\dagger ,
\end{equation}
where the symbol $\circ$ indicates that the differential expression is to be integrated in the Stratonovich sense, or it can also be written as
\begin{equation}
    dU_t = \big(-iH-\frac{1}{2}\sum_l L_l^\dagger L_l\big) U_t dt-\sum_l L_l^\dagger U_t \cdot dA_{l,t} + \sum_l L_l U_t \cdot dA_{l,t}^\dagger ,
\label{eq:QSDE_Unitary_Ito}
\end{equation}
where the symbol $\cdot$ indicates that the differential expression is to be integrated in the Ito sense.
 
\par
It was shown in \cite{Gough_2006} that a usual time-ordered product of iterated time integrals that include products of the field operators $a_l(t)$ and $a_l^\dag(t)$ should be interpreted as a quantum Stratonovich integral. Conversely, if the iterated time integrals were instead written in normal-order then that corresponds to a quantum Ito integral.  Thus by writing Eq. (\ref{eq:QSDE_unitary_3}) in normal order, the solution will have the same mathematical properties as a quantum Ito integral (Eq. (\ref{eq:QSDE_Unitary_Ito})), without delving into the mathematics of quantum Ito integration and its modified rules of calculus.

\subsection{Fock State Master Equations}
Let $|N_\xi \rangle$ be the field state where a single photon pulse with temporal profile $\xi(t)$ is in the incoming mode ($l=\text{inc}$), and all other field modes are in the vacuum state. The system density matrix in the interaction picture, $\Tilde{\rho}_{\text{sys}}(t)$, is obtained by evolving the initial state then partially tracing over the field, i.e., 
\begin{equation}
    \Tilde{\rho}_{\text{sys}}(t) = \text{Tr}_{\text{field}}\Big(U(t)\rho_0\otimes |N_\xi\rangle\langle N_\xi|U^\dagger(t)\Big).
\end{equation}
We assume at $t=0$, the system+field state is a factorizable state, $\rho_0\otimes |N_\xi\rangle\langle N_\xi|$. Using Eqs. (\ref{eq:a_U_commutator}) and (\ref{eq:QSDE_unitary_3}), it can be shown that the reduced system dynamics follow a hierarchy of equations \cite{Baragiola_2012} (see appendix \ref{app:deriving_Fock_state_master_equation})
\begin{equation}
\begin{split}
    \frac{d\rho_{m,n}}{dt} =& -i[H,\rho_{m,n}] +\sum_l \mathcal{D}[L_l](\rho_{m,n})\\
    &+\sqrt{m}\xi(t)[\rho_{m-1,n},L_{\text{inc}}^\dagger]+\sqrt{n}\xi^*(t)[L_{\text{inc}}, \rho_{m,n-1}],
\label{eq:Fock_state_master_equation_single_photon}
\end{split}
\end{equation}
where $\mathcal{D}[L]$ is the Lindblad superoperator defined as $\mathcal{D}[L](\rho)\equiv - \frac{1}{2}L^\dagger L \rho - \frac{1}{2}\rho L^\dagger L + L\rho L^\dagger $ and the density matrices $\rho_{m,n}$ are defined as
\begin{equation}
    \rho_{m,n}(t)\equiv \text{Tr}_{\text{field}}\Big( U(t)\rho_0\otimes |m_\xi\rangle\langle n_\xi|U^\dagger(t)\Big).
\label{eq:Fock_auxiliary_rho_def_main_text}
\end{equation}
Given an N-photon input, $\rho_{N,N}(t)=\Tilde{\rho}_{\text{sys}}(t)$ is the physical density matrix that describes the system state. This couples to auxiliary density matrices $\rho_{m,n}$ with lower indices, where $0 \le m,n\leq N$. When solving the equations without phonons, we can take advantage of the fact that $\rho_{m,n} = \rho_{n,m}^\dagger$. At $t=0$, the initial condition is $\rho_{m,n}=\delta_{m,n}\rho_0$. The Fock state master equation (Eq. (\ref{eq:Fock_state_master_equation_single_photon})) was originally derived by Baragiola et. al. \cite{Baragiola_2012} using the more mathematical quantum stochastic differential equations (QSDE). In appendix \ref{app:deriving_Fock_state_master_equation}, we give a more accessible alternative derivation based on ordinary differential equations (ODE). 

\subsection{System plus field pure state}
\label{sec:pure_state_no_phonon}
In the special case of a single photon Fock state input, the Fock state master equations (Eq. (\ref{eq:Fock_state_master_equation_single_photon})) can be solved analytically. However, rather than solving Eq. (\ref{eq:Fock_state_master_equation_single_photon}) directly, a more physically intuitive approach is to write the system+field pure state $|\psi(t)\rangle$ as 
\begin{equation}
    |\psi(t)\rangle = |\beta(t)\rangle |\text{vac}\rangle + |g\rangle \sum_l \int^\infty_{-\infty} d t_r\, \phi_l(t, t_r)a_l^\dagger(t_r)|\text{vac}\rangle,
\label{eq:pure_state_0}
\end{equation}
where $|\beta(t)\rangle$ is an unnormalized system state in the excited subspace, and $\phi_l(t, t_r)$ is an unnormalized ``wavefunction" of the single photon field state in mode $l$ at time $t$. This is in fact the most general form of the system+field pure state when there is exactly one excitation in the system and the field. We can solve for $|\beta(t)\rangle$ and $\phi_l(t, t_r)$ to obtain (see appendix \ref{app:pure_state})
\begin{subequations}
\begin{align}
    |\beta(t)\rangle &= -\int^t_0 d\tau \, \xi(\tau) e^{(-iH-\frac{1}{2}\sum_l L^\dagger_l L_l)(t-\tau)}L^\dagger_{\text{inc}}|g\rangle \label{eq:beta_no_phonon}\\
    \phi_l(t,t_r) &=
    \begin{cases}
    \delta_{l, \text{inc}}\xi(t_r)\quad,t<t_r\\
    \delta_{l, \text{inc}}\xi(t_r)+\frac{1}{2}\langle g|L_l|\beta(t_r)\rangle \quad, t=t_r\\
    \delta_{l, \text{inc}}\xi(t_r)+\langle g|L_l|\beta(t_r)\rangle \quad, t>t_r
    \end{cases}
\end{align}
\end{subequations}
Since $\langle \psi(t)|\psi(t)\rangle=1$, we have the property
\begin{equation}
    \langle \beta(t)|\beta(t)\rangle + \sum_l \int dt_r \, |\phi_l(t, t_r)|^2 = 1.
\end{equation}
Using Eq. (\ref{eq:Fock_auxiliary_rho_def_main_text}), we find that the solution to the single photon Fock state master equation (Eq. (\ref{eq:Fock_state_master_equation_single_photon})) is
\begin{equation}
\begin{split}
    &\rho_{1,1}(t) = |\beta(t)\rangle\langle \beta(t)| + |g\rangle\langle g|\sum_l \int^\infty_{-\infty} dt_r \, |\phi_l(t,t_r)|^2 \\
    & \rho_{1,0}(t) = |\beta(t)\rangle \langle g| \\
    & \rho_{0,0}(t) = |g\rangle\langle g|.
\end{split}
\label{eq:Fock_state_master_equation_solution_no_phonon}
\end{equation}
\par
The total probability of being in the excited state is $\langle \beta(t)|\beta(t)\rangle$. However, since the system energy scale $||H||$ is much larger than the spontaneous emission rate $||\sum_l L^\dagger_l L_l||$, if we are interested in the system behavior at short times we will have $||\sum_l L^\dagger_l L_l||\,t\ll 1$. It is then useful to drop the spontaneous emission terms in Eq. (\ref{eq:beta_no_phonon}) and approximate $|\beta(t)\rangle$ as
\begin{equation}
    |\beta_\xi'(t)\rangle \equiv -\int^t_0 d\tau \, \xi(\tau) e^{-iH(t-\tau)}L^\dagger_{\text{inc}}|g\rangle,
\label{eq:beta_prime_def}
\end{equation}
where for later convenience, we added the subscript $\xi$ to emphasize the dependence of the excited state $|\beta'(t)\rangle$ on the temporal profile $\xi(t)$. 

\subsection{Photon Flux}
\label{sec:flux}
The photon flux of a field mode is defined as the rate at which photons pass through the mode and has the dimension of 1/[time]. At time $t$, the photon flux immediately downstream of the system is given by $a^\dagger_{\text{out}}(t)a_{\text{out}}(t)$ \cite{Baragiola_2012, Loudon_2000_book}.

The expectation value of the photon flux $F_{l}$ in spatial mode $l$ is the trace over the initial state:
\begin{equation}
    F_{l}=\text{Tr}\Big(a^\dagger_{l,\text{out}}(t)a_{l,\text{out}}(t)\rho_0\otimes |N_\xi\rangle\langle N_\xi|\Big).
\end{equation}
Substituting $a_{l,\text{out}}(t)=a_l(t) + L_l(t)$ in Eq. (\ref{eq:input_output_relation}) and following a similar procedure as in appendix \ref{app:deriving_Fock_state_master_equation}, we obtain the following explicit expression for the photon flux into mode $l$
\begin{equation}
    F_{l}= 
    \begin{cases}
    N|\xi(t)|^2 + \big( \sqrt{N}\xi^*(t) \langle L_l \rangle_{N, N-1} + \text{c.c.}\big) + \langle L_l^\dagger L_l \rangle_{N,N} \quad ,\,l=\text{inc}\\
    \langle L_l^\dagger L_l \rangle_{N,N} \quad ,\, \text{otherwise},
    \end{cases}
\label{eq:photon_flux_general}
\end{equation}
where $\langle X \rangle_{n,m}\equiv \text{Tr}(X\rho_{n,m})$. In the expression on the first line for the incoming mode, the first term represents the transmission of the incoming photon and the second term arises from the coherent dynamics between the system and the field. A negative value of the latter term represents the absorption of photons. A positive value for this term corresponds to stimulated emission. It is interesting to note that the second term can be positive also in the case of a single photon input, meaning that a single photon can stimulate its own emission. The final term on the first line for the incoming mode has the same form as the flux expression for the non-incoming modes and represents the spontaneous emission into the particular mode $l$.

\subsection{System-vibration interaction}
\label{sec:sys_vib_interaction}
We model the interaction with the phonon using the shifted harmonic oscillator model, where the excited state potential energy surface consists of harmonic potentials shifted from the ground state potential energy surface \cite{Ishizaki_2010,Ishizaki_2009}. The overall Hamiltonian can be written as 
\begin{equation}
    H_{\text{sys+vib}}=\sum_j \epsilon_j |j\rangle\langle j| + \sum_{j\neq k} J_{jk} |j\rangle\langle k| + H_{\text{vib}} + \sum_j |j\rangle \langle j| u_j,
\label{eq:sys+vib_Hamiltonian_main_text}
\end{equation}
where 
\begin{equation}
    H_{\text{vib}} = \sum_{j,\xi} \frac{p^2_{j,\xi}}{2} + \frac{\omega^2_{j,\xi}q^2_{j,\xi}}{2}
\end{equation}
is the vibrational Hamiltonian in the electronic ground state, and
\begin{equation}
    u_j = \sum_\xi c_{j,\xi} q_{j,\xi}.
\end{equation}
$\xi$ indexes the phonon modes, and $p_{j,\xi}$ and $q_{j,\xi}$ are the momentum and position of the phonon mode $\xi$ on site $j$. $u_j$ is a collective phonon coordinate characterized by the spectral density $J(\omega)$, defined as
\begin{equation}
    J(\omega) = \frac{\pi}{2} \sum_\xi \frac{c_{j,\xi}^2}{\omega_{j,\xi}}\delta(\omega-\omega_{j,\xi}).
\end{equation}
Since the system-vibration coupling term does not involve the system ground state, to a very good approximation, the initial thermal state $\propto e^{-\beta H_{\text{sys+vib}}}$ is a product state between the system ground state $|g\rangle\langle g|$ and the vibrational thermal state $\propto e^{-\beta H_{\text{vib}}}$. For the rest of Section \ref{sec:methods}, we analyze the effect of vibration using two different methods. The first method considers an initial vibrational pure state, and solves for the pure state analogous to Section \ref{sec:pure_state_no_phonon}. Then by averaging the dynamics starting from a thermal distribution of vibrational pure states, we can analyze the behavior given an initial vibrational thermal state. This method can be applied numerically only for a small number of discrete vibration modes, but it gives us useful analytical expressions in the continuum limit. The second method uses the HEOM formalism to trace out the vibration degrees of freedom and represent the effect of a continuum of vibration modes by a set of auxiliary density matrices containing only the system degrees of freedom. 

\subsection{System plus field plus vibration pure state}
\label{sec:pure_state_with_phonon}
Generalizing the analysis presented in Section \ref{sec:pure_state_no_phonon}, one can write down a system+field+vibration pure state as a function of time, given the system+vibration state initialized in the pure product state $|g\rangle\otimes |v\rangle$, where $|v\rangle$ is an arbitrary vibration state. In Section \ref{sec:analysis_on_absorption}, we shall take $|v\rangle$ to be the eigenstate of $H_{\text{vib}}$, the vibrational Hamiltonian in the electronic ground state with energy $E_v$, so that $H_{\text{vib}}|v\rangle = E_v |v\rangle$. The overall pure state $|\psi(t)\rangle$ is written as
\begin{equation}
    |\psi(t)\rangle = |\gamma(t)\rangle|\text{vac}\rangle + |g\rangle \sum_l \int^\infty_{-\infty} dt_r \, |\chi_l(t, t_r)\rangle a^\dagger_l(t_r) |\text{vac}\rangle,
\end{equation}
with
\begin{subequations}
\begin{align}
    |\gamma(t)\rangle &= -\int^t_0 d\tau \, \xi(\tau) e^{(-iH_{\text{sys+vib}}-\frac{1}{2}\sum_l L^\dagger_l L_l)(t-\tau)}L^\dagger_{\text{inc}}|g\rangle e^{-iH_{\text{vib}}\tau}|v\rangle \\
    \chi_l(t,t_r) &=
    \begin{cases}
    \delta_{l, \text{inc}}\xi(t_r)e^{-iH_{\text{vib}}t}|v\rangle\quad,t<t_r\\
    \delta_{l, \text{inc}}\xi(t_r)e^{-iH_{\text{vib}}t}|v\rangle+\frac{1}{2}e^{-iH_{\text{vib}}(t-t_r)}\langle g|L_l|\gamma(t_r)\rangle \quad, t=t_r\\
    \delta_{l, \text{inc}}\xi(t_r)e^{-iH_{\text{vib}}t}|v\rangle+e^{-iH_{\text{vib}}(t-t_r)}\langle g|L_l|\gamma(t_r)\rangle \quad, t>t_r.
    \end{cases}
\end{align}
\end{subequations}
Here $|\gamma(t)\rangle$ is an unnormalized system+vibration state with the system being in the excited subspace, and $\chi_l(t, t_r)$ is an unnormalized vibration state at time $t$. One can check that
\begin{equation}
\begin{split}
    &\rho_{1,1}(t) = |\gamma(t)\rangle\langle \gamma(t)| + |g\rangle\langle g|\sum_l \int^\infty_{-\infty} dt_r \, |\chi_l(t,t_r)\rangle\langle \chi_l(t,t_r)| \\
    & \rho_{1,0}(t) = |\gamma(t)\rangle \langle g|\langle v|e^{iH_{\text{vib}}t} \\
    & \rho_{0,0}(t) = e^{-iH_{\text{vib}}t}|v\rangle|g\rangle\langle g|\langle v|e^{iH_{\text{vib}}t}
\end{split}
\end{equation}
solves the single photon Fock state master equation (Eq. (\ref{eq:Fock_state_master_equation_single_photon})) with vibration. For later convenience, we drop the spontaneous emission terms and define
\begin{equation}
    |\gamma_{\xi,v}'(t)\rangle \equiv -\int^t_0 d\tau \, \xi(\tau) e^{-iH_{\text{sys+vib}}(t-\tau)}L^\dagger_{\text{inc}}|g\rangle e^{-iH_{\text{vib}}\tau}|v\rangle
\label{eq:gamma_prime_def}
\end{equation}
to emphasize the dependence on the temporal profile $\xi(t)$ and the initial vibrational state $|v\rangle$.

\subsection{Hierarchical equations of motion (HEOM)}
\label{section:HEOM}
Another way to treat the vibration effects is to use the HEOM formalism. We let each chromophore couple to a phonon bath with a Drude-Lorentz spectral density
\begin{equation}
    J(\omega) = \frac{2\lambda\gamma\omega}{\omega^2+\gamma^2}.
\label{eq:Drude_Lorentz_main_text}
\end{equation}
The numerical values of the parameters in the spectral density are taken from \cite{Bennett_2013}. Specifically, $\lambda$ is the reorganization energy, taken to be 37 $\text{cm}^{-1}$ for all sites, and  $\gamma$, with physical dimension of frequency, is the decay rate of the phonon correlation function (see appendix \ref{app:deriving_HEOM}), which characterizes the time scale of vibration-induced fluctuations in the electronic excitation energy. For chlorophyll a, we take $\gamma=30\,\text{cm}^{-1}$; for chlorophyll b, we take $\gamma=48\,\text{cm}^{-1}$. Under high temperatures (characterized by $\hbar\gamma /k_B T \ll 1$, where $k_B$ and $T$ are the Boltzmann constant and temperature, respectively), the interaction picture system density matrix (i.e., with $\sum_j \omega_0 |j\rangle\langle j|$ rotated out) follows the hierarchical equations of motion \cite{Tanimura_1989,Ishizaki_2009}
\begin{equation}
    \frac{d}{dt}\rho^{\Vec{n}}(t) = -iH^\times \rho^{\Vec{n}} - (\sum_j n_j \gamma_j) \rho^{\Vec{n}} - \sum_j \lambda_j P_j^\times \rho^{\Vec{n}+\hat{e}_j} + n_j(2k_B T P_j^\times - i\gamma_j P_j^o) \rho^{\Vec{n}-\hat{e}_j},
\label{eq:HEOM_main_text}
\end{equation}
where $P_j\equiv |j\rangle\langle j|$, $A^o B\equiv \{A, B\}$ is the anticommutator superoperator, and $A^\times B\equiv [A, B]$ is the commutator superoperator. $\Vec{n} = (n_1, \cdots, n_N)$ is a vector of N integers, where the element $n_j$ is the hierarchical level of the j\textsuperscript{th} site. $\hat{e}_j\equiv (0, \cdots, 0, 1, 0, \cdots, 0)$ is the ``unit vector" with the j\textsuperscript{th} element equals to 1 and all other elements equal to 0. $\rho^{(\Vec{0})}$ is the physical density matrix, and the other $\rho^{(\Vec{n})}$'s are non-physical auxiliary density matrices that capture the non-Markovian effects of the phonon environment. The initial contition is
\begin{equation}
    \rho^{\Vec{n}}(0) =
    \begin{cases}
    \rho_\text{sys}(0)\quad ,\Vec{n}=\Vec{0}\\
    0 \quad ,\Vec{n}\neq \Vec{0}.
    \end{cases}
\end{equation}
Numerically, a cutoff level $N_\text{cutoff}$ has to be introduced so that only a finite number of auxiliary density matrices with $\sum_j n_j \leq N_\text{cutoff}$ are solved. Given the cutoff, the total number of auxiliary density matrices is $\binom{N+N_{\text{cutoff}}}{N_{\text{cutoff}}}$ \cite{Kreisbeck_2011}. The auxiliary density matrices having $\sum_j n_j = N_\text{cutoff}$ are called the terminators. To capture the effects of one-higher-level set of auxiliary density matrices $\rho^{(\Vec{n})}$, we have developed the following modified equations of motion for the terminators (appendix \ref{app:HEOM_terminator}):
\begin{equation}
\begin{split}
    \frac{d}{dt}\rho^{\Vec{n}}(t) = &-iH^\times \rho^{\Vec{n}} - (\sum_j n_j \gamma_j) \rho^{\Vec{n}} + \sum_j n_j(2k_B T P_j^\times - i\gamma_j P_j^o) \rho^{\Vec{n}-\hat{e}_j} \\
    &-\sum_{j,k} \lambda_j \frac{n_k + \delta_{j,k}}{\gamma_j + \sum_l n_l \gamma_l} P_j^\times (2k_B T P_k^\times -i\gamma_k P_k^o)\rho^{\Vec{n}+\hat{e}_j - \hat{e}_k}.
\end{split}
\label{eq:HEOM_terminator_main_text}
\end{equation}

\subsection{Combining the input-output and HEOM formalisms}
To simultaneously study the effects of the single photon and the phonon bath on the excitonic system, we now combine the input-output and HEOM formalisms. Two different formalisms are needed to treat the effects of these two bosonic baths because of their different properties. The input-output formalism is based on the frequency-independent coupling and the wide-band approximation (see Section \ref{sec:sys-light_interaction_input_output}), which has been used extensively in quantum optics to treat the interaction of matter with the photon field. The coupling to phonons, on the other hand, is frequency-dependent, and therefore cannot be treated with the assumptions of the input-output formalism. The input-output formalism allows us to explicitly calculate properties of the outgoing photon field (see Section \ref{sec:flux}), while the HEOM formalism traces out the bath degrees of freedom. The HEOM formalism is well-suited to treat the coupling to phonons, since the phonon correlation function is Gaussian (see appendix \ref{app:deriving_HEOM}), while the correlation function (e.g. $\langle a^\dagger(t_2)a^\dagger(t_1)\rangle$, $\langle a^\dagger(t_2)a(t_1)\rangle$, etc.) of an N-photon Fock state is not Gaussian. As an aside, we note that in contrast to a Fock state, for a multimode coherent state the correlation function is Gaussian, and for a coherent state input one can in fact treat the interaction with photons using the HEOM formalism.  In addition, because the second cumulant $\langle a(t_2)a^\dagger(t_1)\rangle - \langle a(t_2)\rangle \langle a^\dagger(t_1)\rangle$ is proportional to a delta function for a coherent state, the resulting reduced system dynamics is Markovian and does not involve auxiliary density matrices (see Section \ref{sec:Fock_vs_coherent_system_state}) \cite{wiseman_milburn_2009_book,Herman_2018}.
\par
To combine the input-output and HEOM formalisms, we use Eqs. (\ref{eq:input_output_1}) and (\ref{eq:sys+vib_Hamiltonian_main_text}) to write the full Hamiltonian as
\begin{equation}
    H_\text{total} = H_\text{field} + H_\text{vib} + \sum_j \omega_0 |j\rangle\langle j| + \underbrace{(H_\text{sys} -\sum_j \omega_0 |j\rangle\langle j|)}_{H} + H_\text{sys-field} + H_\text{sys-vib}.
\end{equation}
Here the term inside the parenthesis is the Hamiltonian appearing in the Fock state master equation, and will be denoted simply as $H$. Moving into the interaction picture where we now rotate out $H_\text{field} + H_\text{vib} + \sum_j \omega_0 |j\rangle\langle j|$, the full Hamiltonian becomes
\begin{equation}
    H_{\text{total}} (t) = H + \sum_j |j\rangle\langle j | u_j(t) + \sum_l (-i a_l(t)L_l^\dagger + \text{h.c.}).
\end{equation}
This is to be compared with the interaction picture Hamiltonian for the system+field state alone, Eq. (\ref{eq:input_output_5}). Given a multimode Fock state photon in one spatial mode as the input field, the reduced dynamics in the system+vibration degrees of freedom is then given by the Fock state master equation Eq. (\ref{eq:Fock_state_master_equation_single_photon}), with $H$ replaced by $H + \sum_j |j\rangle\langle j| u_j(t)$. To apply the HEOM formalism to this, we rewrite the Fock state master equation in the block matrix form
\begin{equation}
\begin{split}
    &\frac{d}{dt}
    \begin{pmatrix}
    \rho_{N,N} \\
    \vdots \\
    \rho_{m,n} \\
    \vdots \\
    \rho_{0,0}
    \end{pmatrix}
    = \\
    &\begin{pmatrix}
    -i\sum_j (P_j u_j(t))^\times \rho_{N,N} +\big(-i H^\times  + \sum_l \mathcal{D}[L_l]\big)\rho_{N,N} - \sqrt{N}\xi(t)L^{\dagger\times}_{\text{inc}} \rho_{N-1,N} +\sqrt{N}\xi^*(t)L^\times_{\text{inc}} \rho_{N,N-1} \\
    \vdots \\
    -i\sum_j (P_j u_j(t))^\times \rho_{m,n} +\big(-i H^\times  + \sum_l \mathcal{D}[L_l]\big)\rho_{m,n} - \sqrt{m}\xi(t)L^{\dagger\times}_{\text{inc}} \rho_{m-1,n} +\sqrt{n}\xi^*(t)L^\times_{\text{inc}} \rho_{m,n-1} \\
    \vdots \\
    -i\sum_j (P_j u_j(t))^\times \rho_{0,0} +\big(-i H^\times  + \sum_l \mathcal{D}[L_l]\big)\rho_{0,0}
    \end{pmatrix}.
\end{split}
\label{eq:Fock_state_master_equation_matrix}
\end{equation}
Eq. (\ref{eq:Fock_state_master_equation_matrix}) can be written in a more compact notation as
\begin{equation}
    \frac{d}{dt} \Xi(t) = (\mathcal{V}(t) + \mathcal{W}(t))\Xi(t),
\label{eq:Fock_state_master_equation_matrix_simplified}
\end{equation}
where
\begin{equation}
    \Xi(t) = 
    \begin{pmatrix}
    \rho_{N,N} \\
    \vdots \\
    \rho_{0,0}
    \end{pmatrix},
\end{equation} and $\mathcal{V}(t)$ and $\mathcal{W}(t)$ are linear operators on $\Xi$.
$\mathcal{V}(t)$ is the operator that acts nontrivially on the vibrational degrees of freedom. Its effect on $\Xi$ is given by
\begin{equation}
    \mathcal{V}(t)\Xi = 
    \begin{pmatrix}
    -i\sum_j (P_j u_j(t))^\times \rho_{N,N}\\
    \vdots \\
    -i\sum_j (P_j u_j(t))^\times \rho_{0,0}
    \end{pmatrix}.
\end{equation}
The effect of $\mathcal{W}(t)$ on $\Xi$ is to produce the rest of the terms in Eq. (\ref{eq:Fock_state_master_equation_matrix}). We note that $\mathcal{W}(t)$ acts trivially on the vibrational degrees of freedom. Now, from Eq. (\ref{eq:Fock_state_master_equation_matrix_simplified}), we can write the vector $\chi(t)$ of reduced Fock state auxiliary density matrices on the system, i.e., 
\begin{equation}
    \chi(t) = \text{Tr}_{\text{vib}}(\Xi(t)) = \begin{pmatrix}
    \text{Tr}_{\text{vib}}(\rho_{N,N}) \\
    \vdots \\
    \text{Tr}_{\text{vib}}(\rho_{0,0})
    \end{pmatrix}
\end{equation}
formally as a time-ordered exponential 
\begin{equation}
    \chi(t) = \text{Tr}_{\text{vib}}\Big(\hat{T}\exp \big(\int^t_0 d\tau \, (\mathcal{V}(\tau) + \mathcal{W}(\tau)\big)\rho_{\text{vib,thermal}}\Big) \chi(0),
\end{equation}
where we have used the fact $\Xi(0)=\chi(0)\otimes \rho_{\text{vib,thermal}}$ to pull out $\chi(0)$ from the partial trace. Using the Gaussian property of the vibration correlation function (see appendix \ref{app:deriving_HEOM}), we perform a generalized cumulant expansion \cite{Kubo_1962} on the time-ordered exponential and obtain
\begin{equation}
    \chi(t) = \hat{T} \mathcal{Z} \chi(0),
\end{equation}
where $\mathcal{Z}$ is defined as
\begin{equation}
    \mathcal{Z} = \exp (\int^t_0 dt_1\, \mathcal{W}(t_1) - \sum_j \int^t_0 dt_2 \int^{t_2}_0 dt_1\, \lambda_j e^{-\gamma_j (t_2-t_1)} P_j^\times(t_2) (2k_B T  P_j^\times(t_1) -i \gamma_j P_j^o(t_1)) ).
\end{equation}
The integrand of the double integral is now understood as an operator that applies to every block matrix $\text{Tr}_{\text{vib}}(\rho_{m,n})$ of $\chi$. Following a similar procedure as in appendix \ref{app:deriving_HEOM}, we then obtain the Fock state + HEOM master equation
\begin{equation} \label{eq:Fock+HEOM_mastereq}
    \frac{d}{dt}\chi^{\Vec{n}}(t) = \mathcal{W}(t) \chi^{\Vec{n}} - (\sum_j n_j \gamma_j) \chi^{\Vec{n}} - \sum_j \lambda_j P_j^\times \chi^{\Vec{n}+\hat{e}_j} + n_j(2k_B T P_j^\times - i\gamma_j P_j^o) \chi^{\Vec{n}-\hat{e}_j}.
\end{equation}
Written in terms of individual auxiliary density matrices, this is equivalent to
\begin{equation}
\begin{split}
    \frac{d}{dt} \rho^{\Vec{n}}_{m,n} = &(-iH^\times +\sum_l \mathcal{D}[L_l]) \rho^{\Vec{n}}_{m,n} - \sqrt{m}\xi(t)L^{\dagger\times}_{\text{inc}} \rho^{\Vec{n}}_{m-1,n} + \sqrt{n}\xi^* (t) L^\times_{\text{inc}}\rho^{\Vec{n}}_{m,n-1} \\
    & - (\sum_j n_j \gamma_j)\rho^{\Vec{n}}_{m,n} - \sum_j \lambda_j P_j^\times \rho^{\Vec{n}+\hat{e}_j}_{m,n} + n_j(2k_B T P_j^\times - i\gamma_j P_j^o) \rho^{\Vec{n}-\hat{e}_j}_{m,n},
\end{split}
\label{eq:Fock+HEOM_final}
\end{equation}
with the initial condition
\begin{equation}
    \rho^{\Vec{n}}_{m,n}(0) = 
    \begin{cases}
    \rho_{\text{sys}}(0)\quad , \Vec{n}=\Vec{0} \text{ and } m=n \\
    0\quad, \text{otherwise}.
    \end{cases}
\end{equation}
Note that $\rho^{\Vec{0}}_{N,N}$ is the only physical density matrix. 

The set of equations in Eq.~(\ref{eq:Fock+HEOM_final}) consist of a double hierarchical structure. The supercripted index $\Vec{n}$ indexes the HEOM auxiliary density matrices, and the subscripted index $(m,n)$ indexes the Fock state master equation auxiliary density matrices. The total number of auxiliary density matrices is $\binom{N+N_{\text{cutoff}}}{N_{\text{cutoff}}} (N_{\text{photon}}+1)^2$. Note that in general $\rho^{\Vec{n}}_{m,n}\neq \rho^{\Vec{n}\dagger}_{n,m}$, while the equality holds without HEOM. Since the photon flux operators act trivially on the vibrational degrees of freedom, the expressions for photon fluxes are the same as Eq. (\ref{eq:photon_flux_general}), with the replacement of $\rho_{m,n}$ by $\rho^{\Vec{0}}_{m,n}$.
\par
The double hierarchical structure of Eq. (\ref{eq:Fock+HEOM_final}) makes the computation quite expensive, so we first turn to analytical studies to understand some of its features and consequences in the next two Sections.  Following this, in Section \ref{sec:LHCII_calculation}, we present a numerical simulation using the double hierarchical structure for single photon Fock state absorption and excitonic energy transfer in the LHCII monomer (14-mer) system.

\section{Fock state vs coherent state input}
\label{sec:Fock_vs_coherent}
In this Section we shall examine the relationship between the dynamics under Fock state input photon fields and under coherent state input fields.
We show that unlike coherent state inputs, Fock state inputs do not induce any coherence between excited states with different total number of excitations. If a Fock state input and a coherent state have the same average photon number and the same temporal profile, then when the weak coupling condition $N\Gamma_{\text{inc}}\tau_{\text{pulse}}\ll 1$ (where $N$ is the average photon number, $\Gamma_{\text{inc}}$ is the coupling strength between system and the incoming paraxial mode (see Eq. (\ref{eq:effective_Gamma})), and $\tau_{\text{pulse}}$ is the pulse duration) holds, the system density matrices in the single excitation subspace are the same. Furthermore, the output photon flux is also the same for both Fock and coherent state input fields. We derive these results by first examining the case of a single input photon, then generalizing to the case of N input photons to show that the excited part of the system state is directly proportional to the number of photons. 

\subsection{System state}
\label{sec:Fock_vs_coherent_system_state}
A coherent state with temporal profile $\xi(t_r)$ and coherent amplitude $\alpha$ is written as
\begin{equation}
    |\alpha_\xi\rangle = \exp \Big(\int dt_r \, \alpha(t_r) a^\dagger_{\text{inc}}(t_r) - \alpha^*(t_r) a_{\text{inc}}(t_r)\Big) |\text{vac}\rangle
\end{equation}
where $\alpha(t_r) = \alpha\xi(t_r)$. The average photon number of $|\alpha_\xi\rangle$ is equal to $\int dt_r |\alpha(t_r)|^2=|\alpha|^2$.
Given this coherent state input, the system dynamics is exactly the semiclassical equation plus spontaneous emission
\begin{equation}
    \frac{d}{dt}\rho = -i[H-i\alpha(t)L^\dagger_{\text{inc}}+i\alpha^*(t)L_{\text{inc}}, \rho] + \sum_l L_l \rho L^\dagger_l - \frac{1}{2} L^\dagger_l L_l \rho - \frac{1}{2} \rho L^\dagger_l L_l
\label{eq:coherent_state_master_equation}
\end{equation}
(see appendix \ref{app:coherent_state_master_equation} for a derivation of this based on the input-output formalism). 
Since spontaneous emission occurs on a much longer time scale than the excitonic dynamics of the system, we will ignore spontaneous emission in the following analysis. Performing a second order perturbation (PT2) analysis on the initial state $|g\rangle\langle g|$, the system state can be written in the block matrix form as
\begin{equation}
    \rho_{c}(t) = 
    \begin{pmatrix}
        \big(1-\langle \beta_\alpha'(t)|\beta_\alpha'(t)\rangle\big)|g\rangle\langle g| &  |g\rangle\langle \beta_\alpha'(t)|\\
         & \\
        |\beta_\alpha'(t)\rangle\langle g| & |\beta_\alpha'(t)\rangle\langle \beta_\alpha'(t)|
    \end{pmatrix},
\label{eq:coh_PT2_no_phonon}
\end{equation}
where $|\beta_\alpha'(t)\rangle$ is defined in Eq. (\ref{eq:beta_prime_def}). In the presence of phonons, writing the initial phonon thermal state as a mixture of pure states $\sum_v P_v |v\rangle\langle v|$, the system+vibration state in PT2 is given by
\begin{equation}
    \rho_{c'}(t) =\sum_v P_v
    \begin{pmatrix}
        \big(1-\langle \gamma_{\alpha,v}'(t)|\gamma_{\alpha,v}'(t)\rangle\big)|g\rangle\langle g| &  \text{Tr}_{\text{vib}}\,|g\rangle\langle \gamma_{\alpha,v}'(t)|\\
         & \\
        \text{Tr}_{\text{vib}}\,|\gamma_{\alpha,v}'(t)\rangle\langle g| & \text{Tr}_{\text{vib}}\,|\gamma_{\alpha,v}'(t)\rangle\langle \gamma_{\alpha,v}'(t)|
    \end{pmatrix},
\label{eq:coh_PT2_with_phonon}
\end{equation}
where $|\gamma_{\alpha,v}'(t)\rangle$ is defined in Eq. (\ref{eq:gamma_prime_def}). Details of the second order perturbation calculation are presented in appendix \ref{app:coherent_PT2}.

The perturbative approach works well when product of the perturbation $\alpha(t)L^\dagger_{\text{inc}}$ (or its Hermitian conjugate) and the interaction time is $\ll 1$. The coherent amplitude $\alpha(t)=\alpha \xi(t)$ is on the order of $\sqrt{N/\tau_{\text{pulse}}}$, where $N$ is the average photon number and $\tau_{\text{pulse}}$ is the pulse duration, since $N=|\alpha|^2$ and $\xi(t)$ has the normalization $\int dt |\xi(t)|^2=1$. $L_{\text{inc}}$ is on the order of $\sqrt{\Gamma_{\text{inc}}}$ because $L_{\text{inc}} = \sqrt{\Gamma_{\text{inc}}} |g\rangle\langle B_{\text{inc}}|$ (see Eq. (\ref{eq:L_def_bright_state})). Combining the order of magnitude estimates, we can conclude that the PT2 analysis is accurate when $N\Gamma_{\text{inc}}\tau_{\text{pulse}}\ll 1$. 

\par
As a comparison, given a single photon Fock state input, neglecting spontaneous emission, the system state without the influence of phonons is given exactly by
\begin{equation}
    \rho_{F1}(t) =
    \begin{pmatrix}
        \big(1-\langle \beta_\xi'(t)|\beta_\xi'(t)\rangle\big)|g\rangle\langle g| &  0 \\
         & \\
        0 & |\beta_\xi'(t)\rangle\langle \beta_\xi'(t)|
    \end{pmatrix}
\label{eq:Fock_exact_no_phonon}
\end{equation}
(see Eq. (\ref{eq:Fock_state_master_equation_solution_no_phonon})), and in the presence of phonons it is given by
\begin{equation}
    \rho_{F1'}(t)= \sum_v P_v
    \begin{pmatrix}
        \big(1-\langle \gamma_{\xi,v}'(t)|\gamma_{\xi,v}'(t)\rangle\big)|g\rangle\langle g| & 0\\
         & \\
        0 & \text{Tr}_{\text{vib}}\,|\gamma_{\xi,v}'(t)\rangle\langle \gamma_{\xi,v}'(t)|
    \end{pmatrix}.
\label{eq:Fock_exact_with_phonon}
\end{equation}
\par
Thus with or without phonons, when the Fock state temporal profile $\xi(t)$ is equal to the coherent amplitude $\alpha(t)$, the block diagonal terms of $\rho_{F1}(t)$ turn out to be the same as those of $\rho_{c}(t)$ in this weak coupling situation.
In contrast, the off-diagonal blocks representing the coherence between ground and singly excited states are nonzero in $\rho_{c}(t)$, while these blocks are $0$ in $\rho_{F1}(t)$. The fact that the coherence terms between subspaces of different excitation number are zero for Fock state input fields derives from a much more general observation, namely that:
\textbf{given the system initializes in the electronic ground state, an n-photon Fock state input does not induce any coherence between system subspaces of different electronic excitation number.}
\par
The proof of this statement makes use of the excitation conserving property of the overall Hamiltonian. Defining the total excitation number as the number of photons plus the number of electronic excitations in the system, the total excitation is equal to $n$, the photon number of the input Fock state, at all times, since both the system-field and the system-vibration interactions conserve the total excitation number. Any pure state $|\Psi\rangle$ with $n$ total excitations lives in the subspace
\begin{equation*}
    |\Psi\rangle\in\bigoplus_{m=0}^n \mathcal{S}_m \otimes \mathcal{F}_{n-m},
\end{equation*}
where $\mathcal{S}_m$ is the system m-excitation subspace, and $\mathcal{F}_m$ is the m-photon subspace of the field. Since $\mathcal{F}_m$ and $\mathcal{F}_{m'}$ are orthogonal to each other if $m \neq m'$, the reduced system density matrix $\text{Tr}_{\text{field}}|\Psi\rangle\langle \Psi | = \sum_{m=0}^{n} \rho_m$ is block-diagonal, with $\rho_m$ being nonzero only in the m-excitation block. Any matrix element connecting states with different excitation numbers is identically zero. In the more general case that the system+field+vibration state is a mixture of different pure states with n total excitations, the reduced system density matrix becomes a mixture of block-diagonal matrices, which is still block-diagonal. 

This result has the important implication that when only average quantities of the system are measured, i.e., averages over the system density matrix, identical results are obtained from excitation by a single photon Fock state pulse and a coherent state pulse having the same temporal profile.   
This is verified numerically below for the LHCII monomer, which possesses 14 chlorophyll chromophores. In contrast, system quantities that are conditioned on the outcomes of single photon photon counting experiments~\cite{Li_2021} require a quantum trajectory description of individual single photon experiments for which this equivalence does not hold~\cite{cook_trajectories_2021}.

\subsection{Photon flux}
\label{sec:Fock_vs_coherent_photon_flux}
The photon flux under a coherent state input, also an averaged quantity, is similarly identical to the flux under a single photon Fock state input.
Under a coherent state input, the photon flux in each mode is
\begin{equation}
    F_{l}= 
    \begin{cases}
    |\alpha(t)|^2 + \big(\alpha^*(t) \langle L_l \rangle + \text{c.c.}\big) + \langle L_l^\dagger L_l \rangle \quad ,\,l=\text{inc}\\
    \langle L_l^\dagger L_l\rangle \quad ,\, \text{otherwise},
    \end{cases}
\label{eq:photon_flux_coherent}
\end{equation}
where $\langle X \rangle\equiv \text{Tr}(X\rho(t))$ (see appendix \ref{app:coherent_state_master_equation}). Substituting Eqs. (\ref{eq:coh_PT2_no_phonon}) or (\ref{eq:coh_PT2_with_phonon}) into Eq. (\ref{eq:photon_flux_coherent}), and substituting Eqs. (\ref{eq:Fock_exact_no_phonon}) or (\ref{eq:Fock_exact_with_phonon}) into Eq. (\ref{eq:photon_flux_general}), we see that if the single photon Fock state temporal profile $\xi(t)$ is equal to the coherent amplitude $\alpha(t)$, then the photon fluxes from the single photon Fock state input are the same as the photon fluxes from the coherent state within a PT2 description.

\subsection{N-photon Fock state input}
\label{sec:N_photon_Fock_vs_coh}
\begin{figure}[htbp]
    \centering
    \includegraphics[scale=0.5]{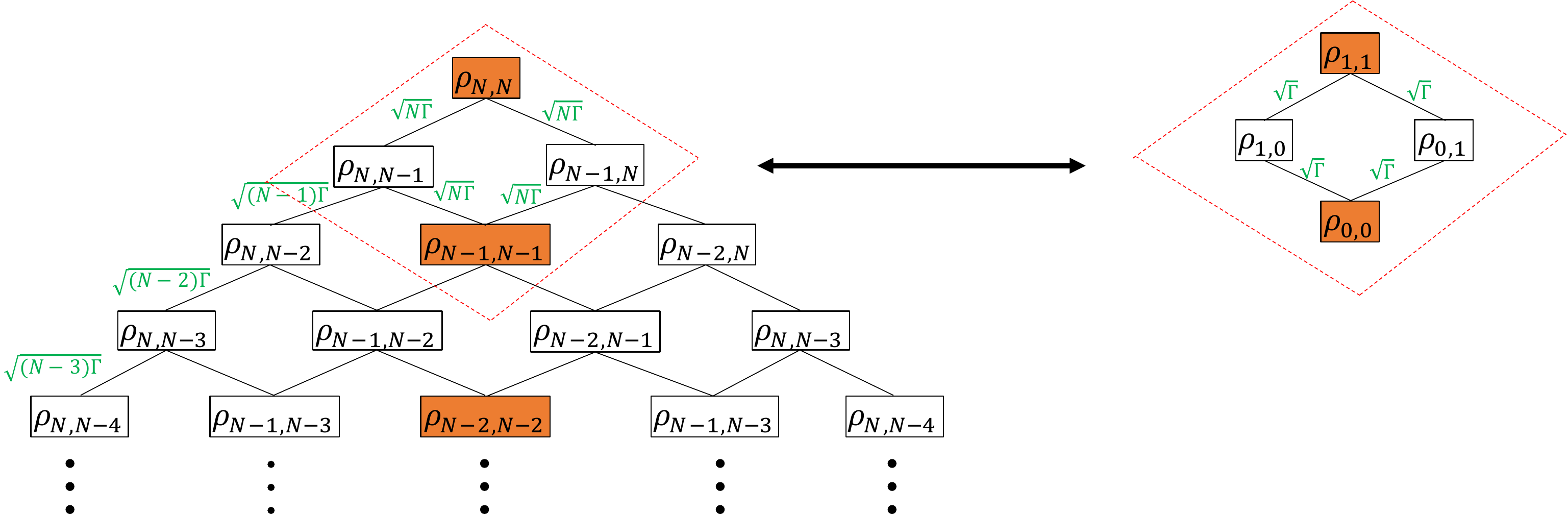}
    \caption{Fock state hierarchy for the (left) N-photon Fock state master equations and (right) single photon Fock state master equations. Each box represents an auxiliary density matrix. Solid orange boxes are the diagonal ``source" terms, and the other boxes are the off-diagonal ``non-source" terms. The ``bonds" between auxiliary density matrices represent the coupling between them. We label the coupling strengths for some of the bonds. To the lowest order in $\Gamma_{\text{inc}}\tau_{\text{pulse}}$, only the top four auxiliary density matrices (enclosed in red dashed lines) contribute to the dynamics of the physical density matrix. The equations for the top four density matrices are the same as those for the single photon master equations, with the replacement of $\Gamma_{\text{inc}}$ by $N\Gamma_{\text{inc}}$.}
    \label{fig:Fock_state_ladder}
\end{figure}

When $N\Gamma_{\text{inc}}\tau_{\text{pulse}}\ll 1$, the excited part of the system density matrix under excitation by an N-photon Fock state is equal to $N$ times that under excitation by a single photon Fock state. To understand this relationship, consider the N-photon Fock state hierarchy, which is shown schematically in Figure (\ref{fig:Fock_state_ladder}). The diagonal density matrices $\rho_{m,m}$, indicated by the solid orange boxes, are initialized in $|g\rangle\langle g|$, and are considered as the ``source" terms of the master equations. The off-diagonal density matrices $\rho_{m,n}$ ($m\neq n$) are initialized in $0$, and are considered as the ``non-source" terms. An auxiliary density matrix $\rho_{m,n}$ couples downward to $\rho_{m-1,n}$ and $\rho_{m,n-1}$, with coupling strength $\sqrt{m\Gamma}$ and $\sqrt{n\Gamma}$, respectively (see Eq. (\ref{eq:Fock_state_master_equation_single_photon})). The couplings are drawn as bonds between auxiliary density matrices. Perturbatively speaking, changes in the physical density matrix $\rho_{N,N}$ are due to its coupling to other ``source" density matrices because they have nonzero initial values. Therefore if $N\Gamma_{\text{inc}}\tau_{\text{pulse}}\ll 1$ (i.e. if the coupling $\sqrt{N\Gamma_{\text{inc}}}\xi(t)$ times the interaction time $\tau_{\text{pulse}}$ is $\ll 1$), the dynamics of $\rho_{N,N}$ is dominated by its 2-bond couplings to $\rho_{N-1,N-1}$. The couplings to other source density matrices require more than 2 bonds, which contribute much less than the 2-bond coupling. As an aside, double excitations in the system require at least 4 bonds, so the probability for such events is much lower than for single excitations. This justifies our restriction to the ground and singly excited states.
\par
Focusing on the four auxiliary density matrices involved in the 2-bond coupling (i.e., $\rho_{N,N}$, $\rho_{N-1,N}$, $\rho_{N,N-1}$, and $\rho_{N-1,N-1}$) and dropping all other auxiliary density matrices, we notice that the four auxiliary density matrices follow the same master equations as the single photon master equations, with the replacement of $\Gamma_{\text{inc}}$ in the single photon master equations by $N\Gamma_{\text{inc}}$.
Since $|\beta'_\xi (t)\rangle$ in Eq. (\ref{eq:Fock_exact_no_phonon}) and $|\gamma'_{\xi,v} (t)\rangle$ in Eq. (\ref{eq:Fock_exact_with_phonon}) are both proportional to $\sqrt{\Gamma_{\text{inc}}}$, , the system state under the excitation of an N-photon Fock state in the absence of phonons is 
\begin{equation}
    \rho_{FN}(t) =
    \begin{pmatrix}
        \big(1-N\langle \beta_\xi'(t)|\beta_\xi'(t)\rangle\big)|g\rangle\langle g| &  0 \\
         & \\
        0 & N|\beta_\xi'(t)\rangle\langle \beta_\xi'(t)|
    \end{pmatrix},
\label{eq:N_Fock_exact_no_phonon}
\end{equation}
to the lowest order in $N\Gamma_{\text{inc}}\tau_{\text{pulse}}$. The corresponding equation in the presence of phonon coupling follows similarly. Thus with or without phonons, the entire single excitation block of the system density matrix (lower right block in Eq.~\ref{eq:N_Fock_exact_no_phonon}), containing both population and coherence terms, is now a factor of $N$ times that derived from excitation by a single photon Fock state. 
\par
Comparing Eq. (\ref{eq:N_Fock_exact_no_phonon}) and its generalization in the presence of phonons to the coherent state results in Eqs. (\ref{eq:coh_PT2_no_phonon}) and (\ref{eq:coh_PT2_with_phonon}), and using the properties $|\beta'_\alpha(t)\rangle = \alpha |\beta'_\xi(t)\rangle$ and $|\gamma'_{\alpha,v}(t)\rangle = \alpha |\gamma'_{\xi,v}(t)\rangle$, we see that the $|\text{ground}\rangle\langle\text{ground}|$ and the $|\text{excited}\rangle\langle\text{excited}|$ blocks of the system density matrix under the excitation by a N-photon Fock state is the same as those under the excitation by a coherent state with the same temporal profile and average photon number. This relationship is verified numerically below for a dimer system under excitations with average 20 photons. It should be emphasized again that this relationship holds when $N\Gamma_{\text{inc}}\tau_{\text{pulse}}\ll 1$. It is well known that in the case of average single photon, when $\Gamma_{\text{inc}}\tau_{\text{pulse}}\sim 1$, a Fock state input and a coherent state input with the same temporal profile can generate very different dynamics in a two-level system \cite{Wang_2011}.

\par

\subsection{Numerical comparison between Fock state vs coherent state input}
To give numerical verifications to the analysis above, we first calculate the system dynamics and photon fluxes of an LHCII monomer (14-mer) system under excitation by a single photon Fock state pulse and by a coherent state with coherent amplitude $\alpha=1$ (average $|\alpha|^2=1$ photon). Details of the LHCII Hamiltonian and transition dipoles are given in~Ref.~\cite{Supplementary}. The temporal profile of both coherent state and Fock state pulses are set to equal to the Gaussian form
\begin{equation}
    \xi(t) = \Big(\frac{\Omega^2}{2\pi}\Big)^{1/4} e^{-\Omega^2 (t-t_0)^2/4}.
\label{eq:Gaussian_temporal_profile}
\end{equation}
The details of the calculation are described in Section \ref{sec:LHCII_calculation}.
\par
\begin{figure}[htbp]
    \centering
    \includegraphics[scale=0.55]{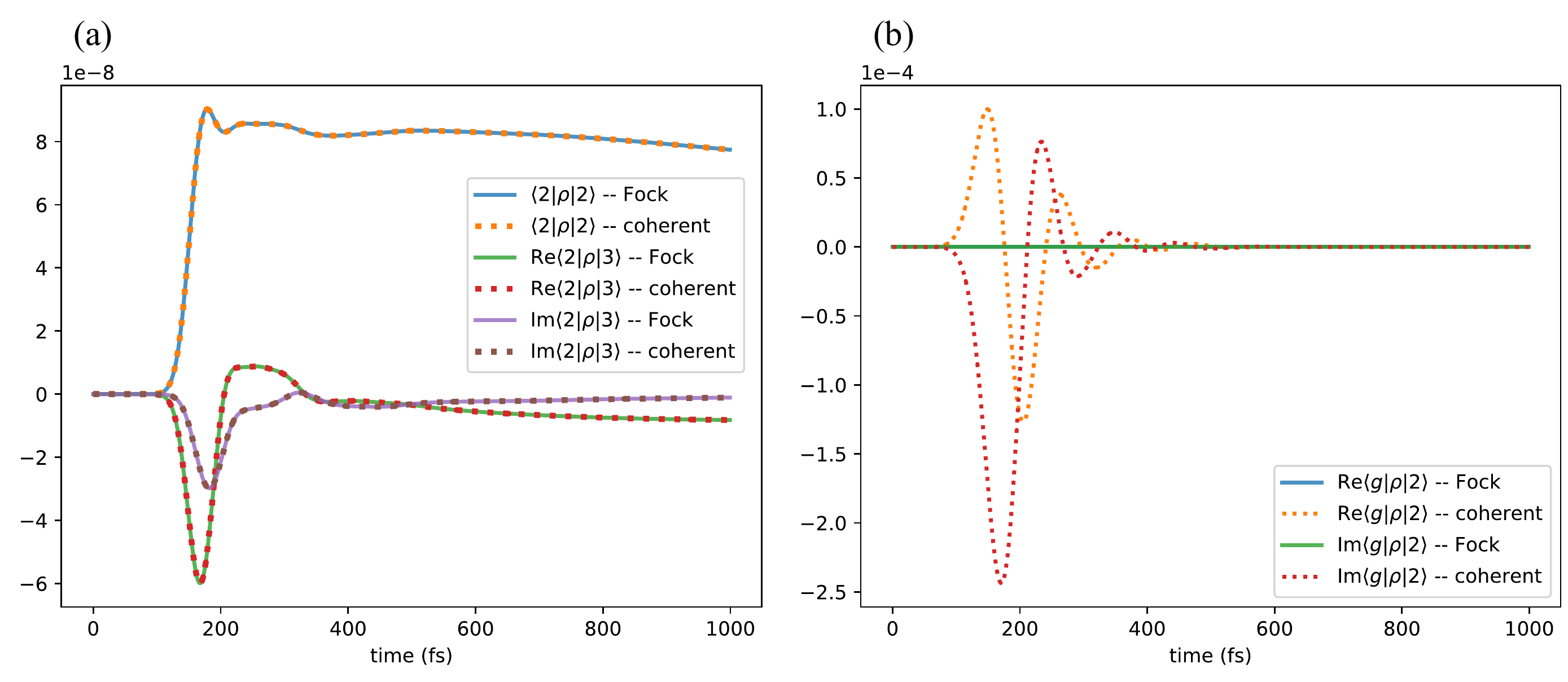}
    \caption{Time dependence of selected excitonic density matrix elements of LHCII under excitation by a single photon Fock state pulse of light and by a coherent state pulse of light containing an average of one photon. The light pulse is centered at $150$ fs. 5 HEOM levels were used for these calculations. (a) Comparison of matrix elements $\langle 2|\rho|2\rangle$ and $\langle 2|\rho|3\rangle$ with indexing referring to the chromophores at sites 2 and 3 of LHCII (see \cite{Supplementary}), showing that density matrix component in the first excitation subspace is almost identical under excitation by a Fock state (solid lines) and by a coherent state of light (dotted lines) with the same average photon number. 
    The differences between the matrix elements under excitation by these two states of light are less than the numerical accuracy of the numerical integrator. 
    (b) Real and imaginary parts of matrix elements between the LHCII ground state and the excited state of chromophore 2, $\langle g|\rho|2\rangle$, under Fock state excitation (solid lines) and under coherent state input light (dotted lines). For this single-photon excitation, the ground-excited coherence terms are identically zero for a Fock state input (solid lines), but non-zero for a coherent state input (dotted lines).}
    \label{fig:Fock_vs_coherent}
\end{figure}

If we characterize the pulse duration $\tau_{\text{pulse}}$ by the inverse bandwidth $1/\Omega\approx 17.7\,\text{fs}$ (see Eq. (\ref{eq:Gaussian_temporal_profile})), then $\Gamma_{\text{inc}}\tau_{\text{pulse}} \approx 4.4\times 10^{-7} \ll 1$.
Therefore we expect the numerical results to show good agreement with the analyses in Sections \ref{sec:Fock_vs_coherent_system_state} and \ref{sec:Fock_vs_coherent_photon_flux}. Figure (\ref{fig:Fock_vs_coherent}a) shows for example the site 2 probabilities, $\langle 2|\rho|2\rangle$, and the coherence term between site 2 and site 3, $\langle 2|\rho|3\rangle$, under Fock state and coherent state inputs. Both the population and the coherence terms are nearly identical under the two input light states, with the relative difference smaller than the numerical accuracy of the numerical integrator (relative tolerance $=10^{-3}$).
The main difference between the two system states is in the ground-excited state coherence. Figure (\ref{fig:Fock_vs_coherent}b) shows for example the coherence term $\langle 0|\rho|2\rangle$ under the two input light. As predicted in Section~\ref{sec:Fock_vs_coherent_system_state} above, $\langle 0|\rho|2\rangle$ is zero for a Fock state input, but nonzero for a coherent state input. 
\par
Similar results are obtained for the photon flux, as demonstrated explicitly in Ref.~\cite{Supplementary}, as expected since these are obtained by averaging over the excited state density matrix (see Eqs. (\ref{eq:photon_flux_general}) and (\ref{eq:photon_flux_coherent})).

\begin{figure}[htbp]
    \centering
    \includegraphics[scale=0.55]{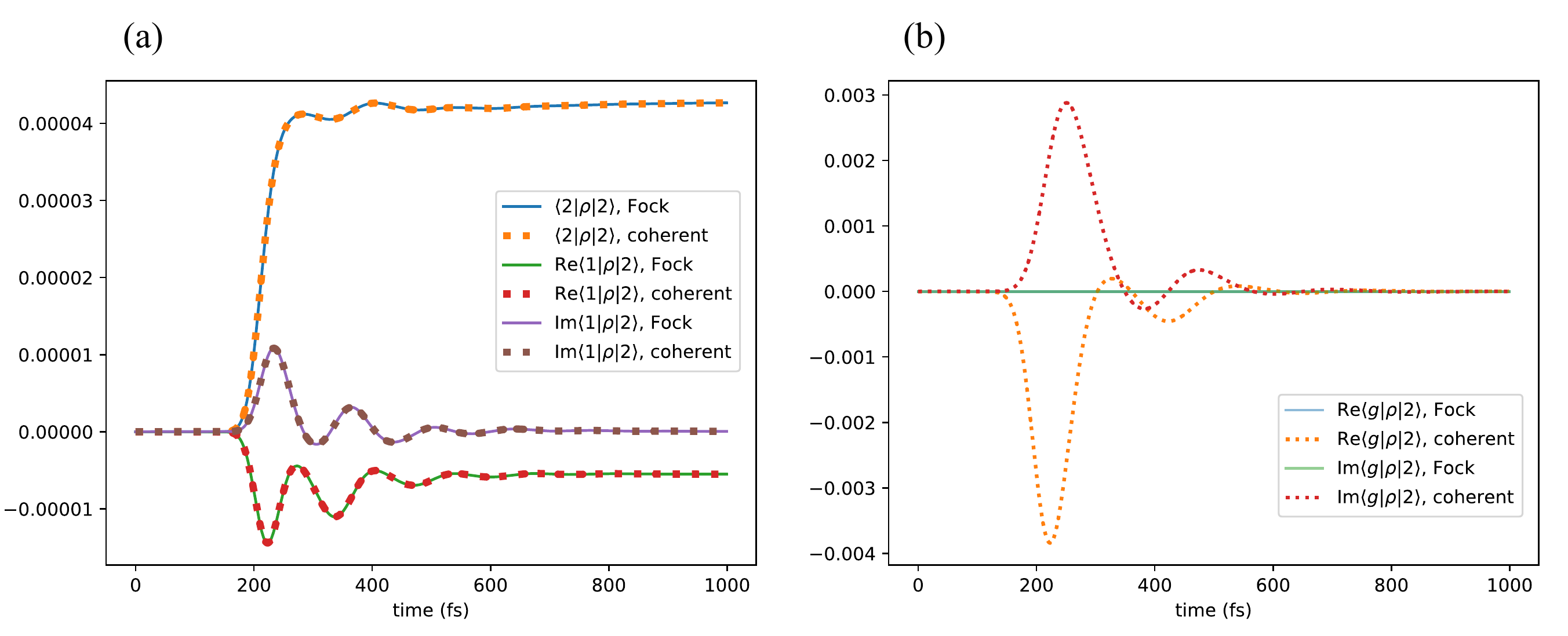}
    \caption{Time dependence of selected excitonic density matrix elements of a dimer under excitation by a 20-photon Fock state pulse of light and by a coherent state pulse of light containing an average of 20 photons. The light pulse is centered at $200$ fs. 5 HEOM levels were used for these calculations. (a) Comparison of matrix elements $\langle 2|\rho|2\rangle$ and $\langle 1|\rho|2\rangle$ in the $|\text{excited}\rangle\langle \text{excited}|$ block under Fock state excitation (solid lines) and under coherent state input light (dotted lines). (b) Comparison of the matrix element $\langle g|\rho|2\rangle$ under Fock state excitation (solid lines) and under coherent state input light (dotted lines). }
    \label{fig:Fock_vs_coh_dimer_20_photons}
\end{figure}
As another example, we consider a dimer system under the excitation of a 20-photon Fock state and under the excitation of a coherent state with $\alpha=\sqrt{20}$, corresponding to an average photon number of $20$. For this example $N\Gamma_{\text{inc}}\tau_{\text{pulse}}\approx 1.8\times 10^{-5}$. Similar to the previous result, Fig. (\ref{fig:Fock_vs_coh_dimer_20_photons}a) shows that both the population and coherence terms in the $|\text{excited}\rangle\langle \text{excited}|$ block of the system density matrix are almost the same under excitation by a 20-photon Fock state and excitation by a coherent state with an average of 20 photons. The difference between the singly excited blocks is again less than the numerical accuracy of the numerical integrator. 
Matrix elements between the ground state and the excited subspace are identically zero under Fock state excitation, and are non-zero under coherent state excitation (see Fig. (\ref{fig:Fock_vs_coh_dimer_20_photons}b)). 
\par
The fact that the coherent state and Fock state inputs give similar results when $N\Gamma_{\text{inc}}\tau_{\text{pulse}}\ll 1$ means that we can in fact simulate the average dynamics under N-photon Fock state excitation by simulating the dynamics under a coherent state and then setting the ground-excited coherence terms to be zero.  This can drastically reduce the computational runtime by avoiding the Fock state hierarchy. For example, for the 20-photon input light fields above, the computational runtime of this Fock state calculation is $\approx 500$ s, while the runtime of the corresponding coherent state calculation is $\approx 1$ s.
\par
\begin{figure}
    \centering
    \includegraphics[scale=0.3]{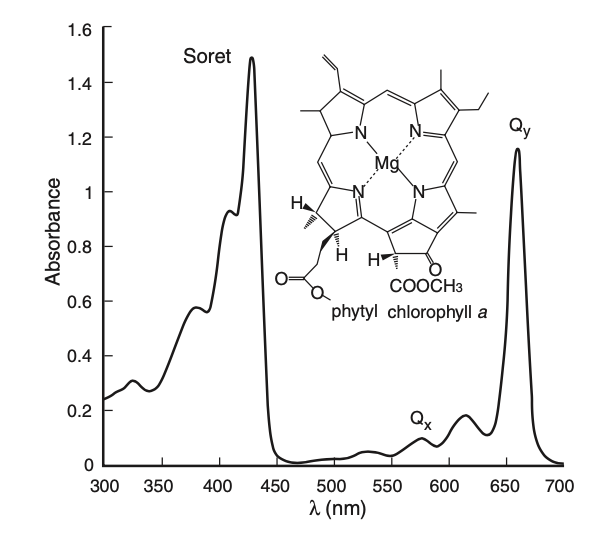}
    \caption{Absorption spectrum of chlorophyll a. Figure taken from \cite{Blankenship}.}
    \label{fig:Chla_abs_spectrum}
\end{figure}
The Q\textsubscript{y} absorption bands of chlorophyll molecules have a bandwidth of $\Omega\approx 400\,\text{cm}^{-1}$\cite{Blankenship} (see Fig. (\ref{fig:Chla_abs_spectrum})). 
If we consider the absorption as a measurement process whereby the chlorophyll molecules detect incoming light at frequencies inside their $Q_y$ bandwidth, then the corresponding duration of optimally detectable pulses is the inverse bandwidth $\tau_{\text{pulse}}=1/\Omega$. This leads to $\Gamma_{\text{inc}}\tau_{\text{pulse}}\approx 4\times 10^{-7}$ for a single chromophore with a transition dipole of 4 Debye, typical of chlorophyll molecules~\cite{Novoderezhkin_2011}. This small value of $\Gamma_{\text{inc}}\tau_{\text{pulse}}$ suggests that our result for the similarity between excitation under a Fock state input and under a coherent state input is indeed applicable to photosynthesis in natural conditions.

\section{Analysis of Absorption}
\label{sec:analysis_on_absorption}
The spontaneous emission time scale $\tau_{\text{emission}}\sim 10 \text{ ns}$ is much longer than the system+vibration time scale $\tau_\text{sys+vib}\sim 10-100$ fs. If we let the pulse duration $\tau_{\text{pulse}}$ be much shorter than $\tau_{\text{emission}}$, we can study the dynamics within short times $t\ll \tau_{\text{emission}}$ without considering the effects of spontaneous emission, since spontaneous emission only reduces the total excitation probability by a small fraction. Within the short time regime, we can define the absorption probability as the total excitation probability $\sum_j \langle j|\rho|j\rangle$ immediately after all but an exponentially small tail of the pulse has passed. After this time the interaction with the phonon bath can redistribute the excitation between the chromophores, but it does not change the total excitation probability. 
\par 
From the solution to the pure state equations without phonons (Eq. (\ref{eq:beta_prime_def})), we know that the absorption probability as a function of time is 
\begin{equation}
    \text{abs. prob.}_\text{no phonon} = \langle \beta_\xi'(t)|\beta_\xi'(t)\rangle.
\label{eq:abs_prob_1}
\end{equation}
To find the absorption probability at time $t$ when almost all of the pulse has passed, we notice that the magnitude of the integrand in Eq. (\ref{eq:beta_prime_def}) is very small outside of the integration bounds $(0,t)$, since the temporal profile $\xi(\tau)$ is localized in this interval. Therefore we can extend the range of integration to $(-\infty, \infty)$. Next, we replace the $L_{\text{inc}}$ in Eq. (\ref{eq:beta_prime_def}) with $\sqrt{\Gamma_{\text{inc}}}|g\rangle\langle B_{\text{inc}}|$ (see Eq. (\ref{eq:L_def_bright_state})), where $\Gamma_{\text{inc}}$ is the effective coupling constant to the incoming paraxial mode and $|B_{\text{inc}}\rangle$ is the normalized bright state corresponding to the incoming mode polarization. Then Eq. (\ref{eq:beta_prime_def}) becomes
\begin{equation}
    |\beta_\xi'\rangle = -\sqrt{\Gamma_{\text{inc}}}\sum_n \int^\infty_{-\infty} d\tau \, \xi(\tau)e^{-i(E_n-E_0) (t-\tau)} \langle n|B_{\text{inc}}\rangle |n\rangle,
\label{eq:abs_prob_3}
\end{equation}
where we inserted a resolution of identity $\sum_n |n\rangle \langle n|=1$ with $n$ indexing the eigenstates of the exciton system Hamiltonian $H_{\text{sys}}$ of Eq (\ref{eq:system_Hamiltonian}). Since the $H$ in the exponential of Eq. (\ref{eq:beta_prime_def}) already has a carrier frequency $E_0=\hbar\omega_0$ rotated out (see Eq. (\ref{eq:H_0})), the eigenvalue of $H$ is the original system eigenenergy $E_n$ minus the carrier frequency $E_0$, hence the factor $E_n-E_0$ appearing in the exponential in Eq. (\ref{eq:abs_prob_3}).
Substituting Eq. (\ref{eq:abs_prob_3}) into Eq. (\ref{eq:abs_prob_1}), the absorption probability without phonon becomes
\begin{equation}
    \text{abs. prob.}_{\text{no phonon}} = \Gamma_{\text{inc}} \sum_n c_n |\Tilde{\xi}(E_n-E_0)|^2,
\label{eq:abs_prob_2}
\end{equation}
where 
\begin{equation}
    c_n = |\langle n|B_{\text{inc}}\rangle|^2
\end{equation}
is the overlap between the system eigenstate and the bright state, and 
\begin{equation}
    \Tilde{\xi}(E) = \int^\infty_{-\infty} d\tau \, \xi(\tau) e^{iE \tau}
\end{equation}
is the Fourier transform of the temporal profile.
Note that $\sum_n c_n = 1$, so the sum in Eq. (\ref{eq:abs_prob_2}) can be thought of as a weighted average of the frequency components of the incoming pulse.
\par
The analysis above can be generalized to take into account the effects of phonons. To distinguish between the analysis without or with phonons, we denote the eigenstate and eigenenergy of $H_{\text{sys}}$ (Eq. (\ref{eq:system_Hamiltonian})) as $|n\rangle$ and $E_n$, respectively. We denote the eigenstate and eigenenergy of $H_{\text{sys+vib}}$ (Eq. (\ref{eq:sys+vib_Hamiltonian_main_text})) as $|\widetilde{n}\rangle$ and $\widetilde{E}_n$, respectively.  
\par First, we assume an initial phonon state $|v\rangle$ that is an eigenstate of $H_{\text{vib}}$, the vibrational Hamiltonain in the ground electronic state, with energy $E_v$. Following the same procedure that we used to obtain Eq. (\ref{eq:abs_prob_3}), we rewrite Eq. (\ref{eq:gamma_prime_def}) as 
\begin{equation}
    |\gamma_{\xi,v}'\rangle = -\sqrt{\Gamma_{\text{inc}}} \sum_n \int^\infty_{-\infty} d\tau \, \xi(\tau)e^{-i(\widetilde{E}_n - E_0)t}e^{i(\widetilde{E}_n-E_0-E_v)\tau}\langle \widetilde{n}|B_{\text{inc}},v\rangle |\widetilde{n}\rangle.
\end{equation}
$|B_{\text{inc}},v\rangle$ denotes the product state $|B_{\text{inc}}\rangle\otimes |v\rangle$. The absorption probability $\langle\gamma_{\xi,v}'| \gamma_{\xi,v}'\rangle$ given a pure initial vibrational state $|v\rangle$ is evaluated as
\begin{equation}
    \text{abs. prob.}_{\text{ pure phonon},v} =  \Gamma_{\text{inc}} \sum_n c'_{n,v} |\Tilde{\xi}(\widetilde{E}_n - E_v-E_0)|^2,
\label{eq:abs_prob_4}
\end{equation}
with $c'_{n,v} = |\langle \widetilde{n}|B_{\text{inc}},v\rangle|^2$. Note that the vibronic eigenstates $|\widetilde{n}\rangle$ can be expanded in the eigenbasis of the shifted harmonic oscillators corresponding to the vibrational Hamiltonian in the excited electronic states. This will result in Franck-Condon vibration overlap factors appearing in the expression of $c'_{n,v}$. However, due to the fact that the vibrational modes are distributed over all sites and that dipole-dipole coupling mixes the excitonic states, one cannot in general decompose $|\widetilde{n}\rangle$ into a simple product of an electronic state and a vibrational state. Therefore we will simply use the symbol $|\widetilde{n}\rangle$ to represent the complicated superposition of different vibrational states in the electronically excited subspace.
\par To obtain the absorption probability given an initial phonon thermal state, note that the thermal state can be treated as a classical mixture of energy eigenstates $|v\rangle$. Therefore the total absorption probability becomes
\begin{equation}
    \text{abs. prob.}_{\text{thermal phonon}} = \sum_{v} P_v\, \text{abs. prob.}_{\text{pure phonon},v} = \Gamma_{\text{inc}} \sum_{\widetilde{n},v} \widetilde{c}_{n,v}|\Tilde{\xi}(\widetilde{E}_n - E_v-E_0)|^2, 
\label{eq:abs_prob_5}
\end{equation}
where $P_v$ is the Boltzmann weight
\begin{equation}
    P_v = \frac{\exp (-E_v/k_B T)}{ \sum_u \exp (-E_u/k_B T)},
\end{equation} and we introduce a thermally weighted squared overlap of the vibronic eigenstate $|\widetilde{n}\rangle$ with the bright state,
\begin{equation} \label{eq:C_nv}
    \widetilde{c}_{n,v} = P_v c'_{n,v} = P_v|\langle \widetilde{n}|B_{\text{inc}},v\rangle|^2.
\end{equation}
Since $\sum_{\widetilde{n},v} \widetilde{c}_{n,v}=1$, Eq. (\ref{eq:abs_prob_5}) can be thought of as a weighted average of $|\Tilde{\xi}(\widetilde{E}_n-E_v-E_0)|^2$.
\par

To understand the factor $\widetilde{E}_n - E_v - E_0$ in Eq. (\ref{eq:abs_prob_5}), we consider the sys+vib eigenenergy $\widetilde{E}_n$ to be in the range $[E_{\text{sys}} + E_v - \mathcal{O}(E_{\text{int}}), E_{\text{sys}} + E_v + \mathcal{O}(E_{\text{int}})]$, where $E_{\text{int}}$ is the energy scale of the system-vibration interaction. Then 
\begin{equation}
    \widetilde{E}_n - E_v - E_0 \in [E_{\text{sys}}- E_0- \mathcal{O}(E_{\text{int}}), E_{\text{sys}}- E_0 + \mathcal{O}(E_{\text{int}})].
\end{equation}
This indicates that the interaction with vibration broadens the photon frequency range that the system can interact with, from exactly $E_{\text{sys}}- E_0$ in Eq. (\ref{eq:abs_prob_2}), to the range $[E_{\text{sys}}- E_0- \mathcal{O}(E_{\text{int}}), E_{\text{sys}}- E_0 + \mathcal{O}(E_{\text{int}})]$ in Eq. (\ref{eq:abs_prob_5}). In the case of zero system-vibration interaction, $E_{\text{int}} = 0$ and Eq. (\ref{eq:abs_prob_5}) reduces to Eq. (\ref{eq:abs_prob_2}). 

\subsection{Absorption probability proportional to system-field coupling}
From Eqs. (\ref{eq:abs_prob_2}) and (\ref{eq:abs_prob_5}), we see that the absorption probability is proportional to $\Gamma_{\text{inc}}$, with or without phonons. The only assumption we have made to arrive at these results is that the spontaneous emission time scale ($1/\Gamma_l\sim\tau_{\text{emission}} \sim 10$ ns) is much longer than both the system+vibration time scale ($\tau_{\text{sys+vib}}\sim 10-100$ fs) and the pulse duration $\tau_{\text{pulse}}$. In other words, $\Gamma_{\text{inc}}$ is the small parameter in the problem.
\par
\begin{figure}[htbp]
    \centering
    \includegraphics[scale=0.6]{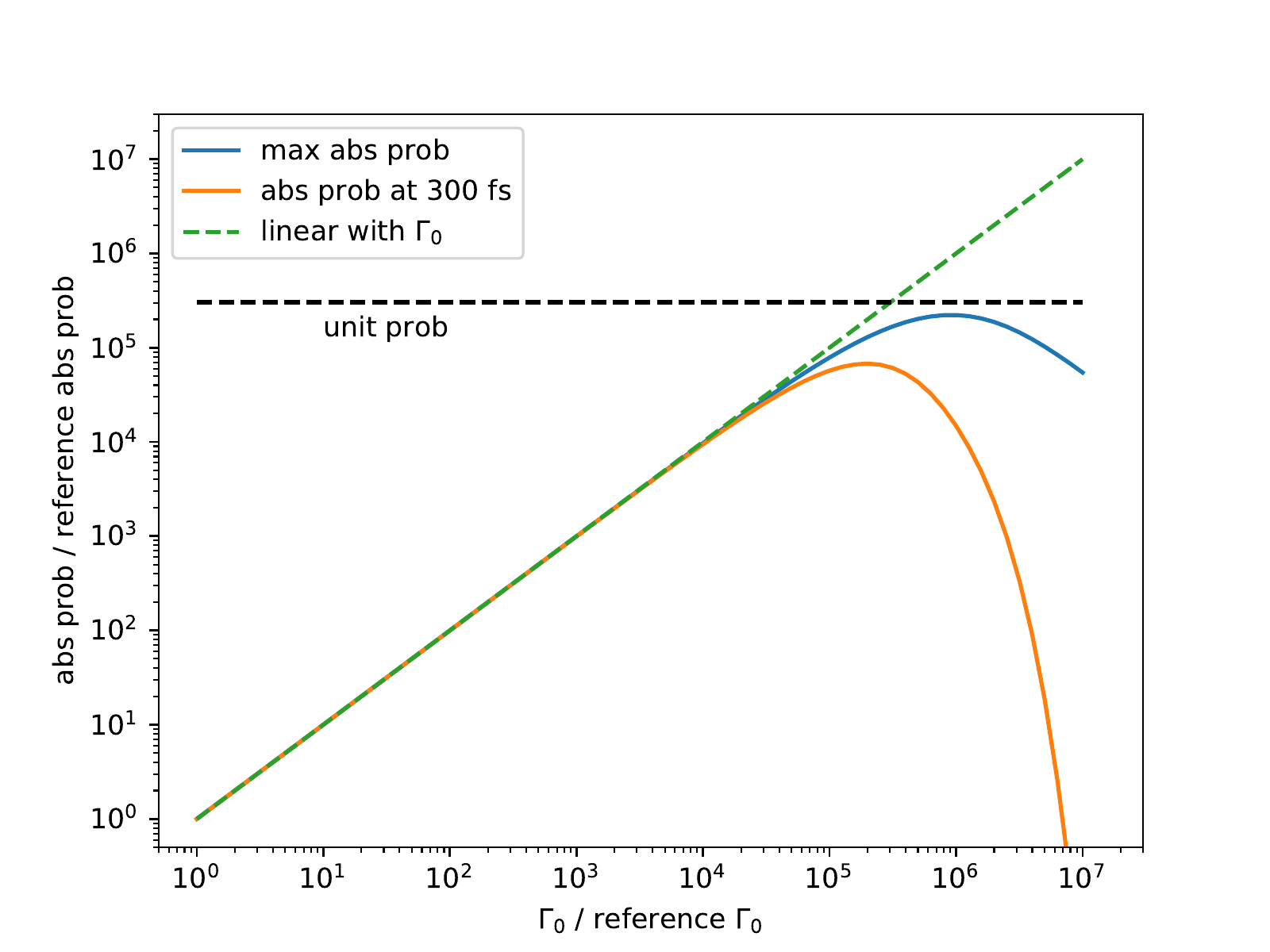}
    \caption{Effect of artificially increasing the system-field coupling strength $\Gamma_0$. Here $\Gamma_0$ is artificially increased by 7 orders of magnitude from the reference physical value, Eq. (\ref{eq:Gamma_0}). The resulting increase in absorption probability is plotted in the vertical axis as the ratio of this to the reference physical absorption probability. The temporal profile of the single photon pulse is a Gaussian centered at 150 fs, with FWHM of 55.5 fs. Two types of absorption probabilities are plotted: maximum absorption probability (blue) and the absorption probability at 300 fs (orange). The absorption probability depends linearly on $\Gamma_0$ across many orders of magnitude until the emission time scale $\tau_{\text{emission}} \sim 1/\Gamma$ becomes comparable to the pulse time scale $\tau_{\text{pulse}}$. At very large $\Gamma_0$, spontaneous emission causes deviation from the linear relationship.  The black dotted line represents unit absorption probability. }
    \label{fig:abs_prob_propto_Gamma}
\end{figure}
\par
Figure (\ref{fig:abs_prob_propto_Gamma}) examines to what extent the linear relationship between the absorption probability and $\Gamma_{\text{inc}}$ holds, or equivalently, to what extent can $\Gamma_{\text{inc}}$ be consider a small parameter. Since $\Gamma_{\text{inc}}=\Gamma_0 \eta \sum_j |\mathbf{d}_j\cdot\hat{\epsilon}|^2$, we artificially increased the unit spontaneous emission rate $\Gamma_0$ by a factor of $1$ to $10^7$ in order to change the value for $\Gamma_{\text{inc}}$. The absorption probability as a function of $\Gamma_0$ was then evaluated numerically for a dimer system with 5 HEOM levels, using a Gaussian pulse with temporal profile $\xi(t)$ (Eq. (\ref{eq:Gaussian_temporal_profile})) centered at 150 fs with a pulse duration of $\tau_{\text{pulse}}=1/\Omega=16.7 \,\text{fs}$. 
The emission time scale $\tau_{\text{emission}}$ is on the order of $1/\Gamma_{\text{inc}} = 18.7\,\text{ns}$. In Figure (\ref{fig:abs_prob_propto_Gamma}) we plot the maximum absorption probability and the absorption probability at 300 fs. We see that the linear relationship between absorption probability and $\Gamma_0$ holds up to $\simeq 10^4$ times the physical value of $\Gamma_0$, indicating that for up to $10^4$ times the physical value of $\Gamma_0$, $\Gamma_{\text{inc}}$ can still be treated as the small parameter in the absorption process. 
At a value $10^5$ times the reference physical $\Gamma_0$ value, the emission time scale $\tau_{\text{emission}} \sim 187\,\text{fs}$ becomes comparable to $\tau_{\text{pulse}}\sim 16.7\,\text{fs}$, so that spontaneous emission occurs before the tail of the Gaussian pulse has passed, making the absorption porbability at 300 fs significantly smaller than the maximum absorption probability. Another reason that the linear relationship cannot hold for large $\Gamma_0$ is of course that the absorption probability cannot exceed one. This limit is plotted as the black dashed line in Figure \ref{fig:abs_prob_propto_Gamma}.

The simple observation that the absorption probability is proportional to $\Gamma_{\text{inc}}=\Gamma_0 \eta \sum_j |\mathbf{d}_j\cdot\hat{\epsilon}|^2$ allows us to quantitatively predict the absorption probability upon varying many different parameters. For example, changing the paraxial beam geometry or the position of the system inside the paraxial spatial mode results in changes in the geometric factor $\eta$. The effect of the dielectric environment is contained in the factor $\Gamma_0$. The effect of light polarization and dipole orientations is described by the factor $\sum_j |\mathbf{d}\cdot \hat{\epsilon}|^2$. We elaborate specifically on the linear dependence on $\sum_j |\mathbf{d}\cdot \hat{\epsilon}|^2$ in the next Section.

\subsection{Light polarization and dipole orientations}
\label{sec:light_polarization_and_dipole_orientation}
To understand the effect of light polarization, it is useful to re-express $\sum_j |\mathbf{d}_j \cdot \hat{\epsilon}|^2$ as the matrix product $\epsilon^\dagger \mathbf{D}^\dagger \mathbf{D} \epsilon$, where $\epsilon$ is a $3\times 1$ unit vector representing the light polarization and $\mathbf{D}$ is an $N\times 3$ matrix with the j-th row being the transition dipole moment of the j-th chromophore. We then perform a singular value decomposition on $\mathbf{D}$ by writing the orthonormal eigenvectors of $\mathbf{D}^\dagger \mathbf{D}$ (or singular vectors of $\mathbf{D}$) as $e_1$, $e_2$, and $e_3$, corresponding to the non-negative eigenvalues $d_1^2$, $d_2^2$, and $d_3^2$, where $d_1$, $d_2$, and $d_3$ are the non-negative singular values.
Without loss of generality, we let $d_1 \geq d_2 \geq d_3 \geq 0$. Light polarized in $e_1$ maximizes $\epsilon^\dagger \mathbf{D}^\dagger \mathbf{D} \epsilon$, therefore maximizing the absorption probability.
Expressing everything in the singular vector basis, given a light polarization $\hat{\epsilon} = a_1 e_1 + a_2 e_2 + a_3 e_3$, then 
\begin{equation}
    \sum_j |\mathbf{d}_j \cdot \hat{\epsilon}|^2 = a_1^2 d_1^2 + a_2^2 d_2^2 + a_3^2 d_3^2.
\label{eq:collective_dipole_SVD}
\end{equation}
\par
To confirm numerically the linear dependence of the absorption probability on $\sum_j |\mathbf{d}_j \cdot \hat{\epsilon}|^2$, we consider single photon excitation of a dimer system coupled to a vibrational bath via the double hierarchy of photon field and HEOM bath. We let each chromophore have a transition dipole moment of 4 Debye, the relevant value for chlorophyll molecules~\cite{Novoderezhkin_2011}. Since any two vectors in 3-dimensional space have a common plane, we only need to consider two singular vectors. The third singular vector is perpendicular to the common plane, and has 0 singular value. On the plane, one of the singular vectors lies in the middle of the two dipoles, corresponding to a singular value of $d_{\text{inner}}=4\sqrt{1+\cos \phi}$, where $\phi$ (satisfying $0\leq\phi\leq \pi$) is the angle between the two dipoles. We call this singular vector the inner singular vector. The other singular vector, labeled as the outer singular vector, is orthogonal to the inner singular vector on the common plane, and corresponds to a singular value of $d_{\text{outer}}=4\sqrt{1-\cos \phi}$. When $0\leq\phi\leq \pi/2$, $d_{\text{inner}} \geq d_{\text{outer}}$, so the inner singular vector maximizes the absorption probability. When $\pi/2 \leq\phi\leq \pi$, $d_{\text{outer}} \geq d_{\text{inner}}$, and the outer singular vector maximizes the absorption probability.
\par
\begin{figure}[htbp]
    \centering
    \includegraphics[scale=0.5]{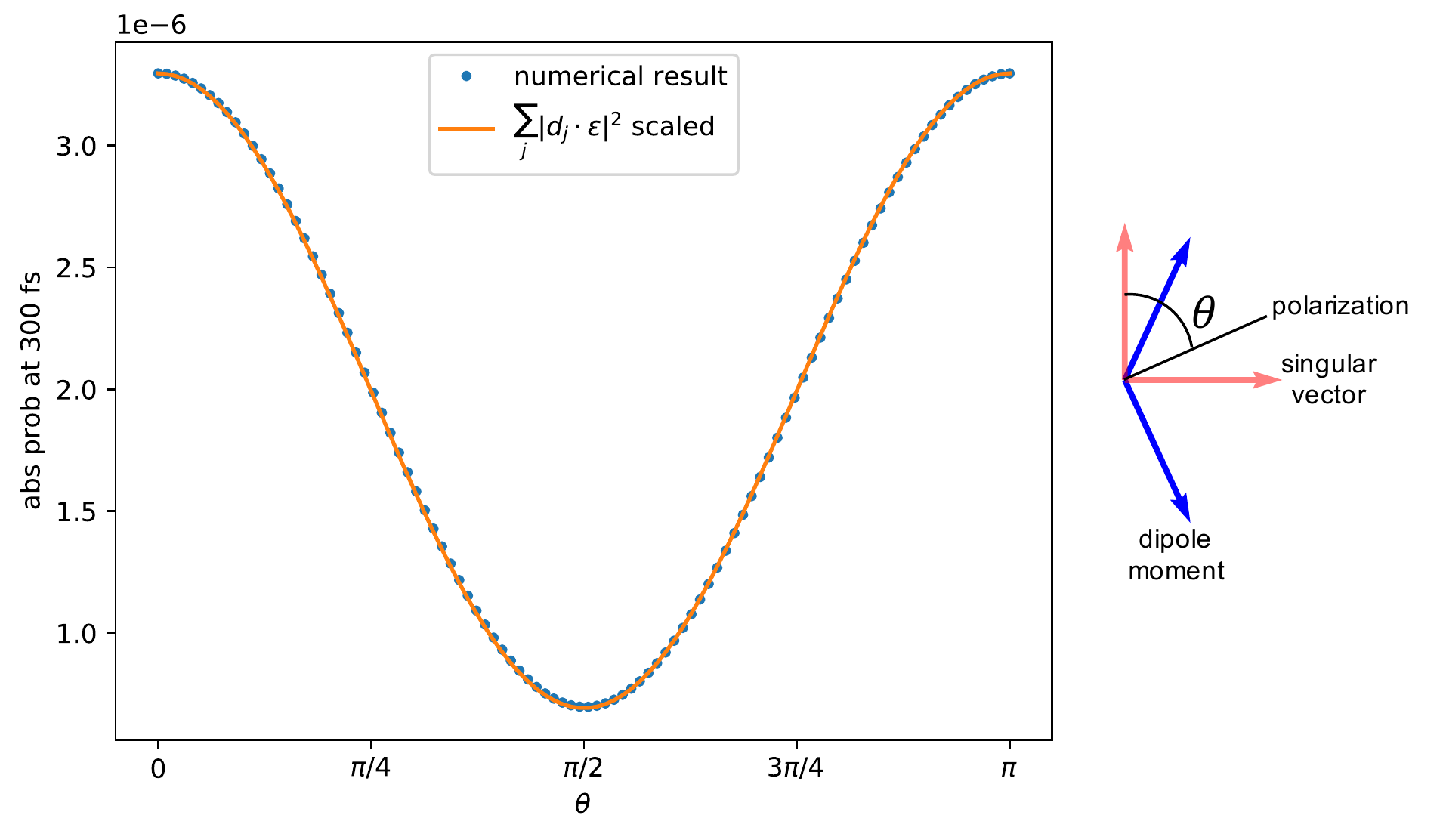}
    \caption{Double hierarchy calculations of single photon absorption for a chromophore dimer system within LHCII, fixing the dipole orientations and varying the light polarization, which is parameterized by $\theta$. The maximum absorption probability is proportional to Eq. (\ref{eq:dipole_SVD_factor}). 5 HEOM levels were included in the calculation. See text and Ref.  \cite{Supplementary} for details of the dimer system.}
    \label{fig:dimer_theta_dependence}
\end{figure}

As a first example, we fix the angle between the two dipole moments to be $\phi\approx 2.28$ rad, corresponding to the dipole orientations of chlorophylls a602 and a603 in LHCII~\cite{Supplementary}. We vary the light polarization along the plane that contains both dipole moments, and parameterize the polarization by the angle $\theta$ to the outer singular vector (see Figure (\ref{fig:dimer_theta_dependence})). Using Eq. (\ref{eq:collective_dipole_SVD}), one can show that
\begin{equation}
    \sum_j |\mathbf{d}_j \cdot \hat{\epsilon}|^2 = 16(1-\cos \phi \cos 2\theta).
\label{eq:dipole_SVD_factor}
\end{equation}
Figure (\ref{fig:dimer_theta_dependence}) shows the results of numerical calculations with the double hierarchy. These show that the maximum absorption probability is indeed proportional to Eq. (\ref{eq:dipole_SVD_factor}) across all $\theta$. Since $\phi$ is between $\pi/2$ and $\pi$, the maximal (minimal) absorption probability occurs at the outer (inner) singular vector.
\par
\begin{figure}[htbp]
    \centering
    \includegraphics[scale=0.5]{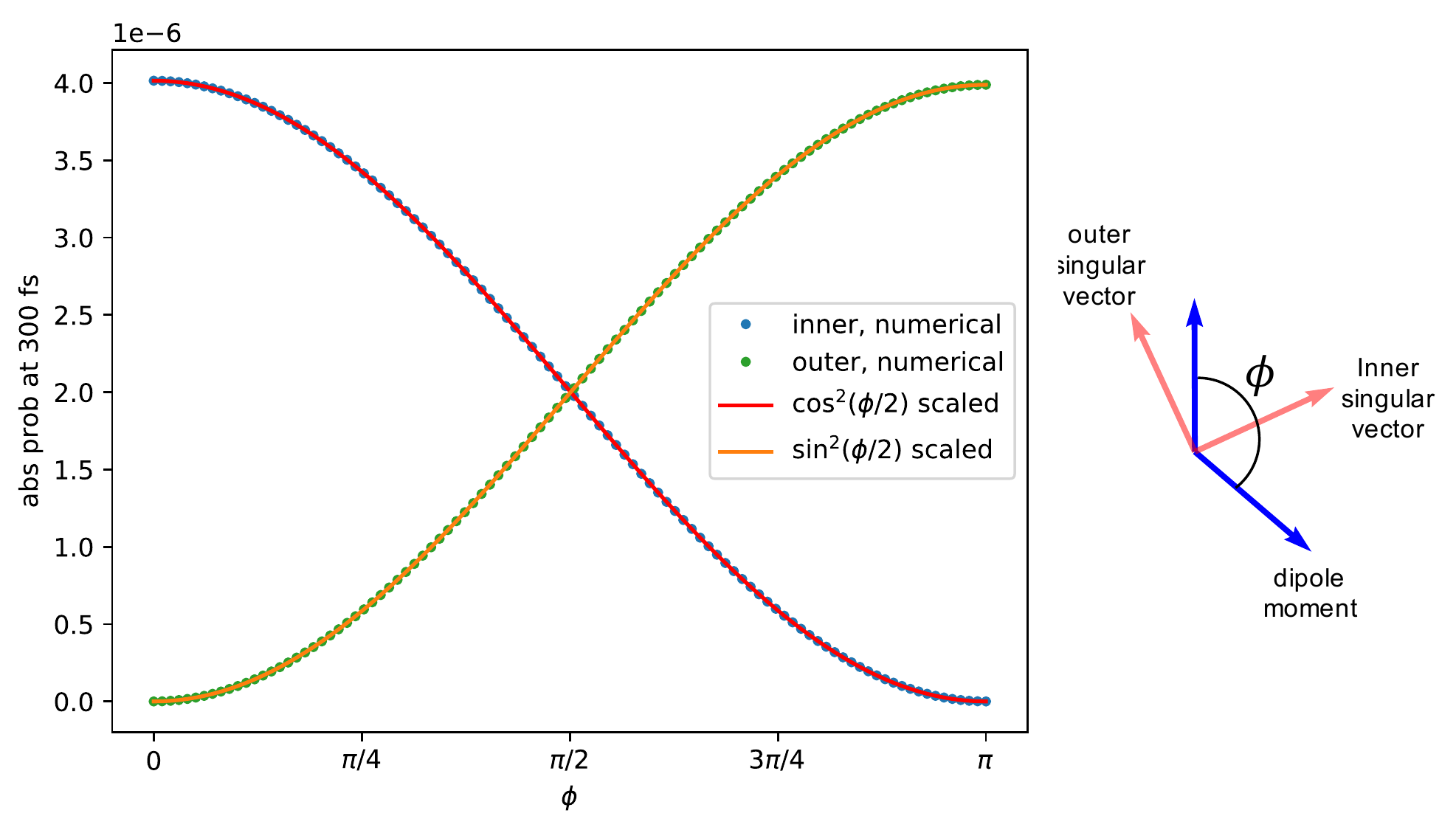}
    \caption{Double hierarchy calculations of single photon absorption for a chromophore dimer system, varying the angle $\phi$ between the dipoles, and letting the light polarization to be either the inner or outer singular vector. The maximum absorption probability is proportional to Eq. (\ref{eq:dipole_SVD_factor}). 5 HEOM levels were included in the calculation. See text and Ref. \cite{Supplementary} for details of the dimer system.}
    \label{fig:dimer_phi_dependence}
\end{figure}

In another example, we fix the polarization to be either the inner singular vector or the outer singular vector, and vary the angle $\phi$ between the dipole moments. Substituting the appropriate $\theta$ into Eq. (\ref{eq:dipole_SVD_factor}), when the polarization is the inner singular vector, $\sum_j |\mathbf{d}_j \cdot \hat{\epsilon}|^2 = 32\cos^2(\phi/2)$, and when the polarization is the outer singular vector, $\sum_j |\mathbf{d}_j \cdot \hat{\epsilon}|^2 = 32\sin^2(\phi/2)$. Figure (\ref{fig:dimer_phi_dependence}) shows the linear dependence of the maximum absorption probability on $\sum_j |\mathbf{d}_j \cdot \hat{\epsilon}|^2$ again. When $\phi=0$, the inner singular vector aligns with both dipoles, maximizing the absorption probability, while the outer singular vector is perpendicular to both dipoles, giving zero absorption probability. The opposite is true when $\phi=\pi$.

\par This type of singular value analysis can be applied to more general chromophore systems to understand the dependence of the absorption probability on the polarization of the incident Fock state photon and the dipole orientations.

\subsection{Orientational average}
Experimentally, the light harvesting systems are typically randomly oriented in solution. Assuming a uniform distribution over all orientations, averaging over all system orientations while fixing the polarization is equivalent to averaging over all polarization directions while fixing the system orientation. Note that rotating the system around the polarization direction does not change $\sum_j |\mathbf{d}_j \cdot \hat{\epsilon}|^2$. Averaging the polarization over all solid angles $\Omega$ , we have
\begin{equation}
    \text{avg} \big(\sum_j |\mathbf{d}_j \cdot \hat{\epsilon}|^2 \big) = \frac{1}{4\pi} \int d\Omega \, \sum_j |\mathbf{d}_j\cdot\hat{\epsilon}(\Omega)|^2 = \frac{1}{3}\sum_j |\mathbf{d}_j|^2,
\end{equation}
which depends only on the magnitude and not on the relative orientation of the dipole moments.

\subsection{Pulse duration dependence of the absorption probability}
To analyze the dependence of absorption probability on pulse duration, we first write a general single photon Fock state photon temporal profile $\xi(t)$ in the scaling form 
\begin{equation}
    \xi(t) = \frac{1}{\sqrt{\tau_{\text{pulse}}}}f(\frac{t-t_0}{\tau_{\text{pulse}}})
\label{eq:temporal_profile_scaling}
\end{equation}
in order to focus on its dependence on the pulse duration $\tau_{\text{pulse}}$. Here $f(x)$ is the scale-invariant pulse shape function, which is dimensionless and square-normalized, i.e., $\int |f(x)|^2 dx = 1$. 
The prefactor $1/\sqrt{\tau_{\text{pulse}}}$ gives $\xi(t)$ the correct dimension and ensures that $\int |\xi(t)|^2\, dt = 1$. We note that $\tau_{\text{pulse}}$ simply characterizes the pulse duration in a general sense, as long as the pulse is reasonably localized in time. Given a pulse shape, one has the freedom to define $\tau_{\text{pulse}}$ up to some $\mathcal{O}(1)$ factors. For example, for the Gaussian temporal profile (Eq. \ref{eq:Gaussian_temporal_profile}), one could define $\tau_{\text{pulse}}=\sqrt{2}/\Omega$ as the standard deviation of $\xi(t)$; alternatively, one could define $\tau_{\text{pulse}}=1/\Omega$ as the standard deviation of $|\xi(t)|^2$.
\par
Using Eq. (\ref{eq:temporal_profile_scaling}), we rewrite the absorption probability without phonons (Eq. (\ref{eq:abs_prob_2})) as
\begin{equation}
    \text{abs. prob.}_{\text{no phonon}} = \Gamma_{\text{inc}} \tau_{\text{pulse}}\sum_n c_n A((E_n - E_0)\tau_{\text{pulse}})
\label{eq:abs_prob_no_phonon_scaling}
\end{equation}
or with (Eq. (\ref{eq:abs_prob_5})) phonons as
\begin{equation}
    \text{abs. prob.}_{\text{thermal phonon}} = \Gamma_{\text{inc}}\tau_{\text{pulse}} \sum_{\widetilde{n},v} \widetilde{c}_{n,v}A((\widetilde{E}_n - E_v - E_0)\tau_{\text{pulse}}),
\label{eq:abs_prob_with_phonon_scaling}
\end{equation}
where we have defined the function
\begin{equation}
    A(E\tau_{\text{pulse}}) = \big|\int dx \, f(x) e^{iE\tau_{\text{pulse}}x}\big|^2.
\label{eq:A(Etau)_def}
\end{equation}
In the following analysis, we assume that the pulse shape function $\xi(t)$ is real, so that $f(t)$ is real and $A(E\tau_{\text{pulse}})$ is a real and even function.

\subsection{Absorption probability in the short pulse regime}
For short pulses, characterized by $\tau_{\text{pulse}}\ll \tau_{\text{sys+vib}}$, $(E_n-E_0)\tau_{\text{pulse}}$ in Eq. (\ref{eq:abs_prob_no_phonon_scaling}) or $(\widetilde{E}_n-E_v-E_0)\tau_{\text{pulse}}$ in Eq. (\ref{eq:abs_prob_with_phonon_scaling}) will be much less than 1.
Furthermore, if $A(E\tau_{\text{pulse}})$ is analytic at $\tau_{\text{pulse}}=0$, then for short pulses, we can Taylor expand it to the second order as
\begin{equation}
    A(E\tau_{\text{pulse}}) = a - b E^2\tau_{\text{pulse}}^2.
\label{eq:A(Etau)_expansion}
\end{equation}
The linear term vanishes because $A(E\tau_{\text{pulse}})$ is an even function. The expansion coefficients $a$ and $b$ are determined by the pulse shape alone and not by the pulse duration.
\par
Substituting Eq. (\ref{eq:A(Etau)_expansion}) into Eqs. (\ref{eq:abs_prob_no_phonon_scaling}) and (\ref{eq:abs_prob_with_phonon_scaling}), we obtain for the short pulse regime,
\begin{equation}
    \text{abs. prob.}_{\text{no phonon}} = \Gamma_{\text{inc}}\tau_{\text{pulse}} \bigg(a - b \tau_{\text{pulse}}^2\sum_n c_n (E_n - E_0)^2\bigg) 
\label{eq:abs_prob_no_phonon_short_pulse}
\end{equation}
and
\begin{equation}
    \text{abs. prob.}_{\text{thermal phonon}} = \Gamma_{\text{inc}}\tau_{\text{pulse}} \bigg(a - b \tau_{\text{pulse}}^2\sum_{\widetilde{n},v} \widetilde{c}_{n,v} (\widetilde{E}_n - E_v - E_0)^2\bigg). 
\label{eq:abs_prob_with_thermal_phonon_short_pulse}
\end{equation}
The term $\sum_n c_n (E_n - E_0)^2$ in Eq. (\ref{eq:abs_prob_no_phonon_short_pulse}) is a bright-state weighted average of the system eigenenergy detunings squared. If the pulse center frequency is set to the average of the system eigenenergies, then this quantity is a skewed variance of the eigenenergies in which eigenstates with more overlap with the bright state have more weight. Similarly, the term $\sum_{\widetilde{n},v} \widetilde{c}_{n,v} (\widetilde{E}_n - E_v - E_0)^2$ in Eq. (\ref{eq:abs_prob_with_thermal_phonon_short_pulse}) is a thermally weighted and bright-state weighted average of $ (\widetilde{E}_n - E_v - E_0)^2$, where more weight is placed on eigenstates (indexed by $\widetilde{n}$) having high overlaps with the bright state and on vibration states (indexed by $v$) with higher Boltzmann weights.
Next, we define an effective energy spread parameter
\begin{equation}
    \Delta^2 = \sum_n c_n (E_n - E_0)^2 = \sum_n |\langle n|B_{\text{inc}}\rangle|^2 (E_n - E_0)^2
\label{eq:Delta_def}
\end{equation}
or 
\begin{equation}
    \Delta^2 = \sum_{\widetilde{n},v} \widetilde{c}_{n,v} (\widetilde{E}_n - E_v - E_0)^2 = \sum_{\widetilde{n},v} P_v |\langle \widetilde{n}|B_{\text{inc}},v\rangle|^2 (\widetilde{E}_n - E_v - E_0)^2,
\label{eq:Delta_def_2}
\end{equation}
depending on whether or not the interaction with phonons is taken into account. Since $\Delta$ characterizes the energy spread of a system, we can identify $1/\Delta$ as a characteristic system+vibration time scale, $\tau_{\text{sys+vib}}$. Substituting Eqs. (\ref{eq:Delta_def}) and (\ref{eq:Delta_def_2}) into Eqs. (\ref{eq:abs_prob_no_phonon_short_pulse}) and (\ref{eq:abs_prob_with_thermal_phonon_short_pulse}), we have a universal expression for the absorption probability in the short pulse regime:
\begin{equation}
    \text{universal short pulse abs. prob.} = \Gamma_{\text{inc}}\tau_{\text{pulse}}(a-b\Delta^2\tau_{\text{pulse}}^2).
\label{eq:short_pulse_general}
\end{equation}
We thus find that despite the complexity of the system and the interaction with phonons, we can describe the absorption probability of the chromophore system in the short pulse regime by only two parameters, $\Gamma_{\text{inc}}$ and $\Delta$.
\par
One important implication of Eq. (\ref{eq:short_pulse_general}) is that a single photon with a delta function temporal profile does not interact with the system, since in the limit $\tau_{\text{pulse}}\rightarrow 0$, the quantity $\text{abs. prob.}=0$. An analogous situation in atomic physics is that to make a $\pi$-pulse, the electric field $E$ times the pulse duration $\tau$ times some constants must be equal to $\pi$ (i.e., $E\tau\sim\pi$, or $E\sim\tau^{-1}$). The photon number, proportional to the energy, in a $\pi$-pulse is proportional to the electric field squared times pulse duration ($E^2\tau$), which is proportional to $\tau^{-1}$. Therefore to make a $\pi$-pulse infinitely short in time requires an infinite number of photons. Hence, a single photon with a delta function temporal profile cannot interact with the system. 
\par
To test the validity of Eq. (\ref{eq:short_pulse_general}), we first consider a Gaussian pulse as in Eq. (\ref{eq:Gaussian_temporal_profile}), and define $\tau_{\text{pulse}} = 1/\Omega$. Comparing Eq. (\ref{eq:Gaussian_temporal_profile}) to Eq. (\ref{eq:temporal_profile_scaling}), yields the scale-invariant pulse shape function
\begin{equation}
    f(x) = \frac{1}{(2\pi)^{1/4}}e^{-x^2/4},
\end{equation}
from which we obtain (using Eq. (\ref{eq:A(Etau)_def}))
\begin{equation}
    A(y) = \sqrt{8\pi} e^{-2y^2}.
\end{equation}
This results in parameter values $a=\sqrt{8\pi}$ and $b = 2\sqrt{8\pi}$ for the Taylor expansion coefficients of $A(y)$ in Eq. (\ref{eq:A(Etau)_expansion}).

We can then evaluate the absorption probability for different systems, with and without phonons, and test Eq.~(\ref{eq:short_pulse_general}) by plotting the scaled absorption probability, abs. prob./$\Gamma_{\text{inc}}\tau_{\text{pulse}}$, as a function of $\tau_{\text{pulse}}$ with the latter given in system-independent units of $1/\Delta$.
For systems without phonons, $\Delta$ is computed directly using Eq. (\ref{eq:Delta_def}) by diagonalizing the system Hamiltonian and calculating the overlaps between the eigenstates and the bright state. For systems with coupling to phonons, $\Delta$ cannot be computed directly, and is obtained instead by fitting Eq. (\ref{eq:short_pulse_general}) to values of the scaled absorption probability calculated using a 5-level HEOM. In each calculation, the pulse is centered at time $t=10\tau_{\text{pulse}}$, the absorption probability is taken as the total excitatoin probability at time $t=20\tau_{\text{pulse}}$.
We calculate data points for shorter and shorter pulses until the fitted $\Delta$ converges. Numerically, we find that the estimated variance of the fitted $\Delta$ decreases to some minimum value and then increases as the pulse shortens. We take the $\Delta$ value with the smallest estimated variance as the $\Delta$ for the system. As a check, applying this fitting procedure to numerical calculations for systems without phonons yields good agreement with the analytically calculated $\Delta$. 
The resulting values of $\Delta$ for various different size systems with and without phonons are tabulated in table (\ref{tab:short_pulse_Delta}).
\begin{table}[h!]
    \centering
    \begin{tabular}{|c|c|c|c|c|c|c|c|}
        \hline
        & monomer& monomer & dimer& dimer & 7-mer & 7-mer & 14-mer \\
        & no & with&  no& with & no & with & no \\
        & phonon& phonon & phonon & phonon & phonon & phonon & phonon \\
        \hline
        $\Delta$ (cm\textsuperscript{-1}) & 50 & 124.1 & 75.3 & 145.2 & 306.7 & 330.8 & 383.1 \\
        \hline
        $1/\Delta$ (fs) & 106.1 & 42.7 & 70.4 & 36.5 & 17.3 & 16.0 & 13.8\\
        \hline
    \end{tabular}
    \caption{$\Delta$ and $1/\Delta$ for various chromophore systems, with or without phonons. For systems without phonons, $\Delta$ was calculated directly using Eq. (\ref{eq:Delta_def}). In the monomer system without phonons, a 50 cm\textsuperscript{-1} detuning was introduced, so $\Delta$ is exactly equal to 50 cm\textsuperscript{-1}. For systems with phonons, $\Delta$ was obtained by numerically fitting to the scaled absorption probability obtained from numerical calculations with a 5-level HEOM at temperature $T=300$ K. }
    \label{tab:short_pulse_Delta}
\end{table}

\begin{figure}[htbp]
    \centering
    \includegraphics[scale=0.6]{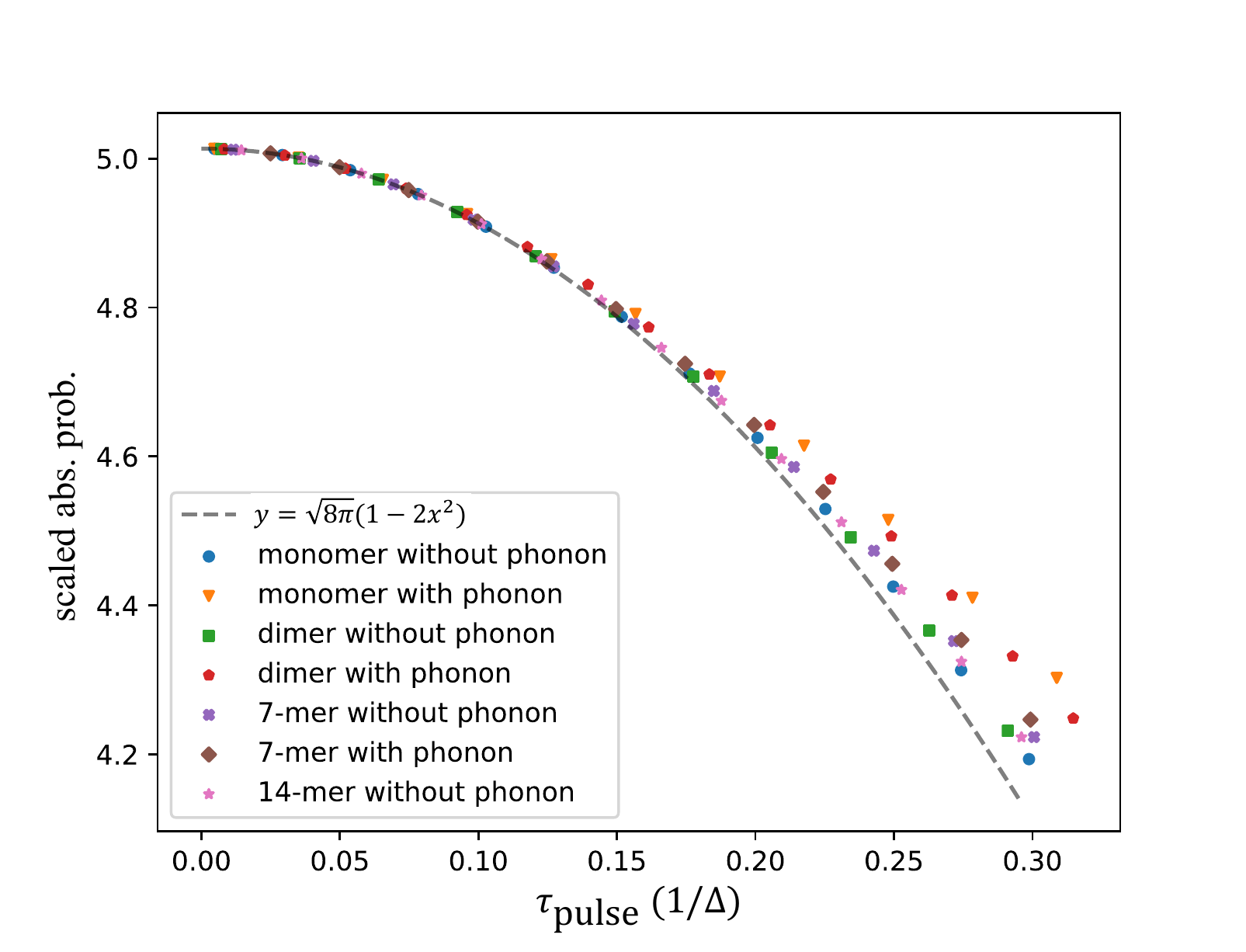}
    \caption{Scaled absorption probability (defined as $\text{abs. prob.}/\Gamma_{\text{inc}}\tau_{\text{pulse}}$) plotted against $\tau_{\text{pulse}}$ for various systems under Gaussian pulses. $\tau_{\text{pulse}}$ is measured in system-dependent time units of $1/\Delta$. The values of $1/\Delta$ are tabulated in table (\ref{tab:short_pulse_Delta}). For short enough pulse duration, the absorption probability follows the analytical expression of Eq. (\ref{eq:short_pulse_general}), shown as the gray dashed line. For systems with phonons, the absorption probability was obtained from numerical calculations with a 5-level HEOM at temperature $T=300$K. To reduce the runtime of the computations, the special HEOM terminator equations (Eq. (\ref{eq:HEOM_terminator_main_text})) were not used here.}
    \label{fig:short_pulse_universality}
\end{figure}

Figure (\ref{fig:short_pulse_universality}) plots the scaled absorption probability, namely abs. prob./$\Gamma_{\text{inc}}\tau_{\text{pulse}}$, as a function of $\tau_{\text{pulse}}$ in units of $1/\Delta$ for various systems, calculated with or without phonons. 
For short enough pulse durations ($\tau_{\text{pulse}} <\approx  0.15/\Delta$ in this case), it is evident that the numerically calculated absorption probabilities match Eq. (\ref{eq:short_pulse_general}) very well for all of the systems considered here.
\par
We further note that as the temperature of the initial phonon state increases, the effective energy spread parameter $\Delta$ also increases. The temperature dependence of $\Delta$ for the dimer system with phonons is shown in Ref.~\cite{Supplementary}.
\par
Since $\Delta$ is independent of pulse shape, having measured it with one pulse form allows us to quantitatively understand the absorption probability under many other pulse shapes. For example, we previously measured $1/\Delta$ = 37.1 fs for the dimer system with phonons under short Gaussian pulses. We can use this information to predict the absorption probability under short square pulses and short exponential pulses by inserting the corresponding expansion coefficients $a$ and $b$. Calculations of these coefficients and plots of the resulting scaling of the short pulse absorption probabilities with $\tau_{\text{pulse}}$ are given in Ref.~\cite{Supplementary}.

\subsection{Absorption Probability in the Long Pulse Regime}
\label{sec:absorption_long_pulse}
The long pulse regime is characterized by $\tau_{\text{sys+vib}} \ll \tau_{\text{pulse}} \ll \tau_{\text{emission}}$, where the pulse duration $\tau_{\text{pulse}}$ is much longer than the chromophore system and vibrational time scale $\tau_{\text{sys+vib}}$, but still much shorter than the emission time scale $\tau_{\text{emission}}$ so that spontaneous emission can be ignored. For many pulse shapes (e.g., Gaussian, square, exponential), $A(E\tau_{\text{pulse}})$ is a localized function peaked at $E\tau_{\text{pulse}}=0$, and $\int dy\, A(y)$ is an $\mathcal{O}(1)$ constant. Therefore, for large $\tau_{\text{pulse}}$, treating $A(E\tau_{\text{pulse}})$ as a function of $E$, we can set
\begin{equation}
    A(E\tau_{\text{pulse}}) = \frac{k}{\tau_{\text{pulse}}}\delta_{\tau_{\text{pulse}}}(E),
\label{eq:long_pulse_A}
\end{equation}
where $k = \int dy\, A(y)$ is a pulse shape dependent $\mathcal{O}(1)$ constant and $\delta_{\tau_{\text{pulse}}}(E)$ is a sharply peaked function with width on the order of $\sim 1/\tau_{\text{pulse}}$. In the limit $\tau_{\text{pulse}}\rightarrow \infty$, $\delta_{\tau_{\text{pulse}}}(E)$ becomes a true delta function $\delta(E)$. Substituting Eq. (\ref{eq:long_pulse_A}) into Eq. (\ref{eq:abs_prob_no_phonon_scaling}), we find
\begin{equation}
    \text{long pulse abs. prob.}_{\text{no phonon}} = \Gamma_{\text{inc}}k \sum_n c_n \delta_{\tau_{\text{pulse}}}(E_n-E_0).
\label{eq:long_pulse_abs_prob_no_phonon}
\end{equation}
The physical explanation for the delta function is that long pulses have very small energy bandwidths, of order $\sim 1/\tau_{\text{pulse}}$, so the center frequency of the pulse needs to be resonant with an eigenenergy in order to have any significant absorption probability. 
\par
If an eigenenergy $E_m$ is resonant with the pulse center frequency, the sum in the absorption probability expression (Eq. (\ref{eq:abs_prob_no_phonon_scaling})) will be dominated by the resonant eigenstate, and we find that
\begin{equation}
    \text{long pulse abs. prob.}_{\text{no phonon, resonant}} = \Gamma_\text{inc}\tau_{\text{pulse}}|\langle m|B_{\text{inc}}\rangle|^2 a,
\label{eq:long_pulse_resonant}
\end{equation}
where $a=A(0)$ is defined by Eq. (\ref{eq:A(Etau)_expansion}). In this case, the absorption probability is proportional to the pulse duration $\tau_{\text{pulse}}$, provided that $\tau_{\text{pulse}}\ll \tau_{\text{emission}}$ so that spontaneous emission can be ignored. If $E_0$ is not resonant with any eigenenergy, then the absorption probability is near zero in the long pulse regime due to the delta function in Eq. (\ref{eq:long_pulse_abs_prob_no_phonon}). Numerical verification of these results is presented below.
\begin{figure}[htbp]
    \centering
    \includegraphics[scale=0.39]{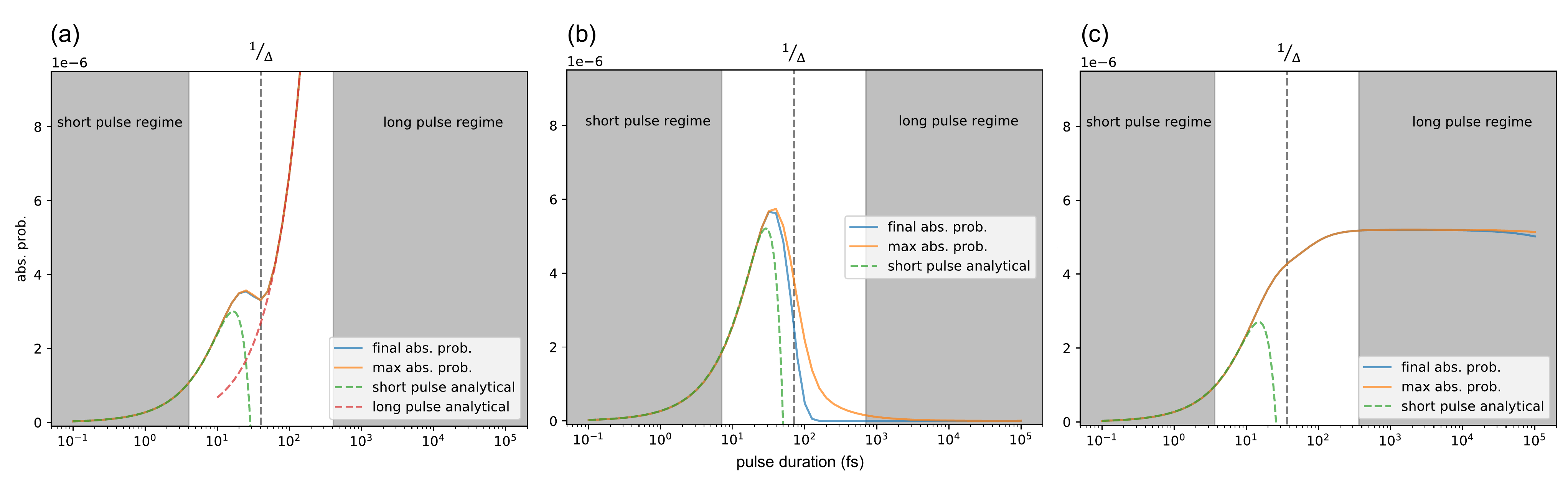}
    \caption{Dependence of absorption probability on pulse duration for a dimer system (a) without phonons and with pulse center frequency resonant with the upper eigenenergy (b) without phonons and with an off-resonant pulse center frequency in the middle of the two eigenenergies (c) with phonons and with an off-resonant pulse center frequency in the middle of the two eigenenergies.}
    \label{fig:short_to_long_pulse}
\end{figure}
\par
Figure (\ref{fig:short_to_long_pulse}) shows the absorption probabilities for a dimer system under a Gaussian pulse in three different scenarios, in each case as a function of pulse duration over a range of six orders of magnitude. 
The Gaussian temporal profile is centered at $10\,\tau_{\text{pulse}},$. Two types of absorption probabilities are measured. The first is the final absorption probability, which is defined as the absorption probability at $20\,\tau_{\text{pulse}}$. The second is the maximum absorption probability, defined as the maximum probability during the time interval $t=0$ to $t=20\,\tau_{\text{pulse}}$. We identify the characteristic system or system+vibration timescale $\tau_{\text{sys+vib}}$ as $1/\Delta$, indicated on the plots by dashed vertical lines. The short and long pulse regimes are identified by $\tau_{\text{pulse}}<0.1/\Delta$ and $\tau_{\text{pulse}}>10/\Delta$, respectively. 
\par
In Figure (\ref{fig:short_to_long_pulse}a), there is no coupling to phonons and the pulse center frequency is set to be resonant to the higher eigenenergy of the $2\times 2$ dimer Hamiltonian. The effective energy spread parameter $\Delta=130.8\,\text{cm}^{-1}$ (see Eq. (\ref{eq:Delta_def})) corresponds to a characteristic time scale of $1/\Delta = 40.6$ fs. In the long pulse regime, the absorption probability follows the linear relationship given by Eq. (\ref{eq:long_pulse_resonant}) up to $\tau_{\text{pulse}}=10^5$ fs. The long pulse absorption probability is not shown due to the small scale of the absorption probability. 
\par
Figure (\ref{fig:short_to_long_pulse}b) considers the same dimer system without phonons, but with the pulse center frequency set to be off-resonant at the average of the two non-degenerate eigenenergies (or equivalently, the average of the site energies). The effective energy spread parameter $\Delta = 75.3\,\text{cm}^{-1}$, and $1/\Delta = 70.4$ fs. Consistent with our analysis, the final absorption probabilities in the long pulse regime are very close to zero and are smaller than the numerical accuracy of the numerical integrator. The maximum absorption probability drops to zero more slowly than the final absorption probability. 
\par
In the presence of phonons, substituting Eq. (\ref{eq:long_pulse_A}) into Eq. (\ref{eq:abs_prob_with_phonon_scaling}), we find
\begin{equation}
    \text{long pulse abs. prob.}_{\text{thermal phonon}} = \Gamma_{\text{inc}}k \sum_{\widetilde{n},v} \widetilde{c}_{n,v}\delta_{\tau_{\text{pulse}}}(\widetilde{E}_n - E_v - E_0),
\label{eq:abs_prob_with_phonon_long_pulse_1}
\end{equation}
where $\widetilde{c}_{n,v}$ are the thermally weighted vibronic overlap with bright state, Eq.~(\ref{eq:C_nv}).
The absorption probability is non-zero if the sum of a vibrational energy $E_v$ and the pulse center frequency $E_0$ is equal to some eigenenergy $\widetilde{E}_n$ of $H_{\text{sys+vib}}$ (Eq. (\ref{eq:sys+vib_Hamiltonian_main_text})). If the vibrational states are dense enough that the spacings between the vibronic energy levels are much less than $1/\tau_{\text{pulse}}$, the width of $\delta_{\tau_{\text{pulse}}}(E)$, we may treat the vibrational energies as a continuum. Defining a coarse-grained version of the bright state overlap function $\widetilde{c}_{n,v}$,
\begin{equation}
    \widetilde{c}(\widetilde{E}_n, E_v) = \sum_{\widetilde{n}',v'} c_{n',v'} \delta(\widetilde{E}_n - \widetilde{E}_{n'})\delta(E_v - E_{v'})
\label{eq:c_function_continuous}
\end{equation}
and writing the sums as integrals
\begin{equation}
    \sum_{\widetilde{n},v} \widetilde{c}_{n,v} \rightarrow \int d\widetilde{E}_n dE_v \, \widetilde{c}(\widetilde{E}_n, E_v),
\end{equation}
Eq. (\ref{eq:abs_prob_with_phonon_long_pulse_1}) becomes
\begin{equation}
    \text{long pulse abs. prob.}_{\text{thermal phonon}} = \Gamma_{\text{inc}}k \int d\widetilde{E}_n dE_v \, \widetilde{c}(\widetilde{E}_n, E_v)\delta_{\tau_{\text{pulse}}}(\widetilde{E}_n - E_v - E_0).
\label{eq:abs_prob_with_phonon_long_pulse_3}
\end{equation}
If the vibrational states are dense enough, it is reasonable to assume that $\widetilde{c}(\widetilde{E}_n, E_v)$ varies slowly on the scale of $1/\tau_{\text{pulse}}$, the width of $\delta_{\tau_{\text{pulse}}}(E)$, since we are in the regime $\tau_{\text{sys+vib}}\ll \tau_{\text{pulse}}$. Hence Eq. (\ref{eq:abs_prob_with_phonon_long_pulse_3}) can be well-approximated by replacing $\delta_{\tau_{\text{pulse}}}(\widetilde{E}_n-E_v-E_0)$ with a true delta function $\delta(\widetilde{E}_n-E_v-E_0)$, and hence
\begin{equation}
    \text{long pulse abs. prob.}_{\text{thermal phonon}} = \Gamma_{\text{inc}}k \int dE_v \, c(\widetilde{E}_n=E_v + E_0, E_v).
\label{eq:abs_prob_with_phonon_long_pulse_2}
\end{equation}
This shows that in the long pulse regime, when the system is coupled to a dense spectrum of phonons, the absorption probability is independent of pulse duration. 
\par
An alternative derivation of the long pulse absorption probability with phonons, presented in appendix \ref{app:long_pulse_phonon}, shows that when $\tau_{\text{pulse}}$ is much longer than the time required for the system to reach a steady state  due to phonon interactions, we have
\begin{equation}
    \text{long pulse abs. prob.}_{\text{thermal phonon}} \sim \frac{\Gamma_{\text{inc}}\tau_{\text{steady}}}{N},
\label{eq:long_pulse_abs_prob_estimate}
\end{equation}
where $\sim$ means on the order of, $N$ is the number of chromophores and $\tau_{\text{steady}}$ is the time scale for the system to approach steady state. Note that this expression does not imply that the absorption probability is inversely proportional to $N$, since $\Gamma_{\text{inc}}$ generally increases as $N$ increases (see Eq. (\ref{eq:effective_Gamma})).
\par
Figure (\ref{fig:short_to_long_pulse}c) shows the absorption probability for the dimer system with phonons (calculated with 5 HEOM levels). The pulse center frequency is chosen to be the average of the two system eigenenergies. Since the coupling to phonons broadens the frequency that the system can interact with, having the pulse center frequency equal to one of the eigenenergies will not result in any qualitative difference. For this example the effective energy spread parameter $\Delta = 145.2\,\text{cm}^{-1}$, and $1/\Delta = 36.5$ fs (see table (\ref{tab:short_pulse_Delta})). Consistent with our analysis, the absorption probability in the long pulse regime is quite independent of $\tau_{\text{pulse}}$ until very large $\tau_{\text{pulse}}$ values, when emission effects become significant. If we simply take $\tau_{\text{steady}}$ in Eq. (\ref{eq:long_pulse_abs_prob_estimate}) to be $1/\gamma$, where $\gamma$ is the HEOM parameter characterizing the phonon dephasing rate, then the order of magnitude estimate $\Gamma_{\text{inc}}\tau_{\text{steady}}/2\approx 5\times 10^{-6}$ is also consistent with the numerical result.
\par
\begin{figure}[htbp]
    \centering
    \includegraphics[scale=0.5]{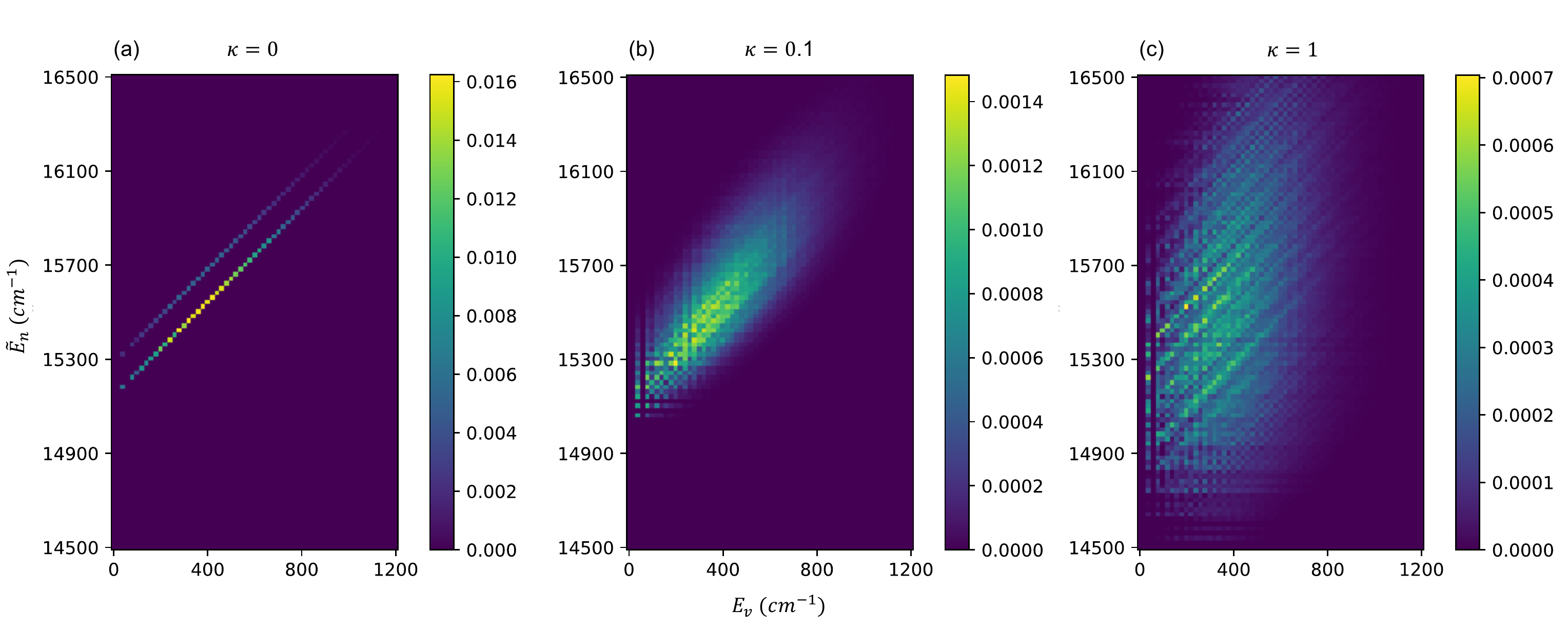}
    \caption{$\widetilde{c}(\widetilde{E}_n,E_v)$ for a dimer system coupled to 4 vibrational modes (2 on each chromophore). Plots (a)-(c) correspond to the exciton-phonon coupling strengths $\kappa=$ $0$, $0.1$, and $1$, respectively.  At $\kappa=0$, the two sharp diagonal peaks indicate that for long pulses, only two frequencies give rise to significant absorption probability. As $\kappa$ increases, the diagonal peaks broaden, and the pulse frequency does not have to be exactly resonant to the system eigenenergy for there to be significant absorption probability.}
    \label{fig:c_function}
\end{figure}

To understand the general behavior of the coarse grained bright state overlap function $c(E_n,E_v)$, we plot this in Figure (\ref{fig:c_function}) for a model dimer system where each chromophore is coupled to two discrete vibrational modes. The total Hamiltonian is
\begin{equation}
    H = H_{\text{sys}} + H_{\text{vib}} + \kappa \sum_{j = 1,2}\sum_{k=1,2} \alpha_{jk} |j\rangle\langle j|(b_{jk}+b^\dagger_{jk}),
\label{eq:diag_1}
\end{equation}
with $H_{\text{sys}}$ expressed in the site basis as
\begin{equation}
    H_{\text{sys}} = \sum_{j=1,2}\epsilon_j |j\rangle\langle j| + J(|1\rangle\langle 2| + |2\rangle\langle 1|),
\end{equation}
and 
\begin{equation}
    H_{\text{vib}} = \sum_{j=1,2}\sum_{k=1,2} \omega_{jk} b^\dagger_{jk}b_{jk},
\label{eq:diag_3}
\end{equation}
with $b_{jk}$ the annihilation operator for the $k$-th vibrational mode coupled to site $j$. 
The numerical values for the parameters in Eqs. (\ref{eq:diag_1})-(\ref{eq:diag_3}) are listed in Ref.~\cite{Supplementary}.
The parameter $\kappa$ sets the overall coupling strength between the excitonic system and the vibrations. 
For ease of calculation, instead of treating $c(E_n,E_v)$ as a function of continuous variables as in Eq. (\ref{eq:c_function_continuous}), we discretized $E_n$ and $E_v$ into 20 cm\textsuperscript{-1} bins and replaced the delta function $\delta(E_1-E_2)$ in Eq. (\ref{eq:c_function_continuous}) by the binning function
\begin{eqnarray}
    \chi(E_1, E_2) = 
    \begin{cases}
    \frac{1}{20\,\text{cm}^{-1}}\quad &E_1\text{ and }E_2\text{ are in the same bin} \\
    0 \quad &\text{otherwise},
    \end{cases}
\end{eqnarray}
where the factor $1/20\,\text{cm}^{-1}$ ensures proper normalization $\int dE_1 \,\chi(E_1, E_2) = 1$. Then the discretized $\widetilde{c}(\widetilde{E}_n,E_v)$ takes the form
\begin{equation}
    \widetilde{c}(\widetilde{E}_n, E_v) = \sum_{\widetilde{n}',v'} \widetilde{c}_{n',v'}\chi(\widetilde{E}_n,\widetilde{E}_n')\chi(E_v,E_v').
\label{eq:c_function_discretized}
\end{equation}
Figure (\ref{fig:c_function}) shows the discretized bright state overlap function $\widetilde{c}(\widetilde{E}_n,E_v)$ for system-vibration coupling values $\kappa=0$, $0.1$, and $1$. 
When there is no system-vibration coupling ($\kappa=0$), the vibronic energy is simply the sum of the system energy and the vibrational energy, i.e., $\widetilde{E}_n = E_n + E_v$. On the other hand, Eq. (\ref{eq:abs_prob_with_phonon_long_pulse_2}) requires that $\int dE_v \,\widetilde{c}(\widetilde{E}_n = E_0 + E_v, E_v)$ be non-zero for there to be a finite absorption probability. Therefore the pulse center frequency $E_0$ has to be equal to a system eigenenergy $E_n$ in order to have nonzero absorption probability.
The two sharp diagonal peaks in panel (a) correspond to the lines $\widetilde{E}_n = E_n + E_v$ for each of the two system energies $E_n$, indicating that for long pulses, the absorption probability is nonzero only when the pulse frequency is equal to one of the two system energies $E_n$. 
As the system-vibration coupling increases, the peaks broaden and their heights decrease, indicating that the pulse center frequency no longer has to be exactly equal to the system energies $E_n$ for there to be nonzero absorption probability and consequently the resonant absorption probability at $E_0=E_n$ decreases.

\section{Analysis of emission}
\label{sec:analysis_on_emission}
In the analytical studies of the previous section we have ignored the effect of spontaneous emission at long times. We now make use of the separation of time scales between the sys+vib dynamics and the spontaneous emission to analyze the long time emission behavior.
\subsection{Uniform exponential decay of excited states in the presence of phonons}
\label{sec:long_time}
The HEOM reaches a steady state on the time scale of $\tau_{\text{steady}}\sim 100$ fs, which is much shorter than the spontaneous emission time scale $\tau_{\text{emission}}\sim 10$ ns. Therefore at a sufficiently long time after the pulse has passed, specfically, when $t-\tau_{\text{pulse}}\gg \tau_{\text{steady}}$, the chromophore system should reach a quasi-steady state with respect to the phonon bath that decays slowly to the ground state due to spontaneous emission. (By quasi-steady state we mean that the chromophore system is in steady state with regard to the phonon bath, but not yet with regard to the photon bath.) We claim that under an N-photon Fock state input, the long time system state takes the form
\begin{equation}
    \rho(t) = |g\rangle\langle g| + b e^{-\Gamma_{\text{long time}}t} \big(\rho_{\text{st}}-|g\rangle \langle g| + \mathcal{O}(\epsilon)\big) ,
\label{eq:observation2_1}
\end{equation}
where the decay rate $\Gamma_{\text{long time}}$ is given by
\begin{equation}
    \Gamma_{\text{long time}} = (1+\mathcal{O}(\epsilon))\sum_l \text{Tr} (L^\dagger_l L_l \rho_{\text{st}}).
\label{eq:observation2_2}
\end{equation}
Here $\rho_{\text{st}}$ is the normalized HEOM steady state in the excited subspace, $\epsilon\sim\tau_{\text{sys+vib}}/\tau_{\text{emission}}$ is a small parameter, and $b$ is a constant to be determined. A detailed derivation of Eqs. (\ref{eq:observation2_1}) and (\ref{eq:observation2_2}) is provided in appendix \ref{app:proof_obs2}.
These equations show that in the quasi-steady state the excited part of the system decays to the ground state following a single exponential whose decay rate is equal to the total emission rate (see Eq. (\ref{eq:photon_flux_general})).
\par
The constant $b$ has some arbitrariness in it, in the sense that $b$ depends on where the time $t=0$ is defined. However, once we fix the $t=0$ point in time, we can determine the constant $b$ numerically by first integrating $\rho(t)$ to some final time $t_f$ that is sufficiently long enough for the system to reach a quasi-steady state. According to Eq. (\ref{eq:observation2_1}), the total excitation probability to the lowest order is then given by $b \exp(-\Gamma_{\text{long time}}t_f)$. Therefore we take 
\begin{equation}
    b = (\text{total excitation prob. at }t_f)e^{\Gamma_{\text{long time}}t_f}.
\end{equation}
The normalized steady state $\rho_{\text{st}}$ in Eqs. (\ref{eq:observation2_1}) and (\ref{eq:observation2_2}) can be evaluated independently of the Fock state hierarchy by propagating the HEOM with an initial state in the excited subspace to long enough time.
\par
This long time behavior implies that we only need to solve the Fock state + HEOM equations (Eqs. (\ref{eq:Fock+HEOM_mastereq}) - (\ref{eq:Fock+HEOM_final})) numerically until the system reaches the quasi-steady state with respect with the phonons. This happens within several mulitples of $\tau_{\text{steady}}$ after the pulse has passed. From this point onwards, the system will behave according to Eq. (\ref{eq:observation2_1}), with the total emission rate given by Eq.(\ref{eq:observation2_2}). 
\par
\begin{figure}[htbp]
    \centering
    \includegraphics[scale=0.6]{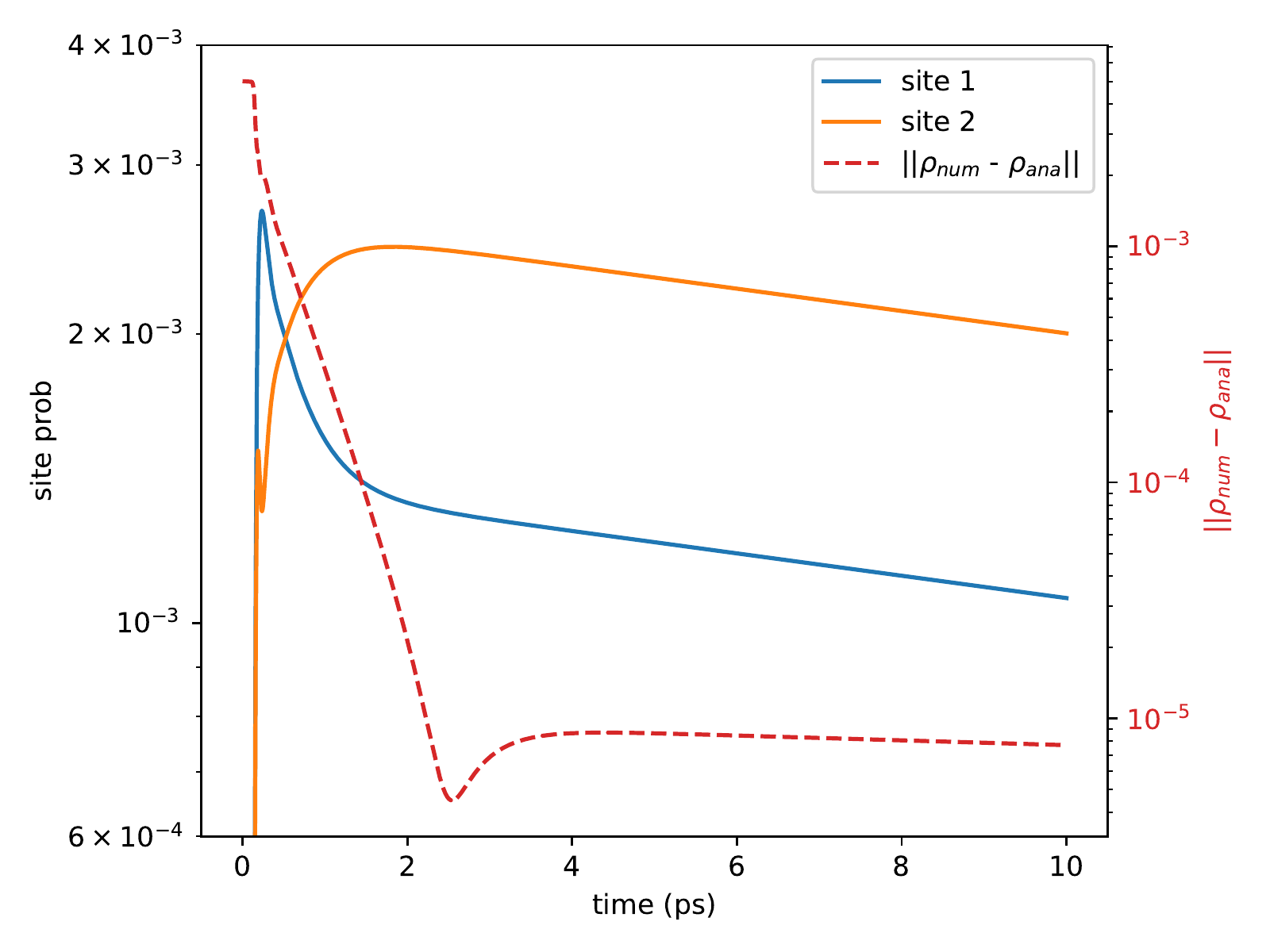}
    \caption{Double hierarchy calculations of the absorption probability for a selected dimer system in LHCII as a function of time. Solid blue and red lines (using the left axis) show excitation probabilities on chromophore sites 1 and 2, respectively, in a log scale as a function of time. 
    The dashed line shows the difference (measured by the Euclidean distance, scale on the right axis) between the numerical density matrix obtained by integrating the double hierarchy of Eq. (\ref{eq:Fock+HEOM_final}) 
    and the lowest order analytical expression of Eq. (\ref{eq:observation2_1}).
    For these calculations the value of $\Gamma_0$ was increased by 1000 relative to the physical value in order to see emission in a reasonable amount of numerical integration time. The quasi-steady state $\rho_{\text{st}}$ is found by propagating the HEOM with initial condition $\rho(0)=|1\rangle\langle 1|$ up to 20 ps, at which time each element of the density matrix remains constant up to 12 digits after the decimal point.}
    \label{fig:dimer_obs2}
\end{figure}

Figure (\ref{fig:dimer_obs2}) shows the excitation probabilities for each chromophore site on a log scale as a function of time for a dimer system. The difference between the density matrix values obtained from numerical integration of the double hierarchy of equations, Eq. (\ref{eq:Fock+HEOM_final}), and those obtained from the lowest order long time analytical expression is quantified by the Euclidean distance
\begin{equation}
    ||\rho_{\text{num}}-\rho_{\text{ana}}|| = \sqrt{\sum_{i,j} |(\rho_{\text{num}})_{ij}- (\rho_{\text{ana}})_{ij}|^2},
\end{equation}
and is plotted in Figure (\ref{fig:dimer_obs2}). Note that we have increased the reference value of spontaneous emission rate $\Gamma_0$ by a factor of 1000 over the physically relevant value in order to see the spontaneous emission within a reasonable amount of numerical integration time. Figure (\ref{fig:dimer_obs2}) shows that in the quasi-steady state after $\sim$ 4 ps, the numerical difference $||\rho_{\text{num}}-\rho_{\text{ana}}|| \sim 10^{-5}$ is about 2 orders of magnitude smaller than the absorption probability ($\sim 10^{-3}$), indicating good convergence. Comparing the $\sim10^{-3}$ probabilities with Eq. (\ref{eq:observation2_1}) indicates that for this system $b\sim 10^{-3}$ and the small parameter $\epsilon \sim 10^{-2}$. 
The numerically fitted value of $\Gamma_{\text{long time}}$ ($2.678\times 10^{-2}\, \text{ps}$, fitted from 8 to 10 ps) is in similarly good agreement with the value $\Gamma_{\text{long time}}$ ($2.685\times 10^{-2} \,\text{ps}^{-1}$) obtained from Eq. (\ref{eq:observation2_2}), with a relative difference on the order of $\epsilon\sim 10^{-2}$, in accordance with Eq. (\ref{eq:observation2_2}).
(Note that if we had not increased $\Gamma_0$ 1000 times, $\epsilon$ would be on the order of $10^{-5}$, making the analytical expression even more accurate.)

\subsection{Collective vs Independent Emission}
After the incoming pulse has passed, the total emission rate of the chromophoric system is 
\begin{equation}
    R_{\text{coll}} = \sum_{l\in \{x,y,z\}} \text{Tr}(L^\dagger_l L_l \rho) = \Gamma_0 \sum_{j,k} \mathbf{d}_j^* \cdot \mathbf{d}_k \langle j|\rho|k\rangle,
\label{eq:collective_emi_1}
\end{equation}
(see Eq. (\ref{eq:photon_flux_general})), where
\begin{equation}
    L_x = \sqrt{\Gamma_0}\sum_j \mathbf{d}_j^* \cdot \hat{x}|g\rangle \langle j|
\label{eq:collective_emi_2}
\end{equation}
and similarly for $L_y$ and $L_z$ (see Eq. (\ref{eq:L_dagger_def}) and its description). As mentioned in Section \ref{sec:sys-light_interaction_input_output}, $\Gamma_0$ is the unit spontaneous emission rate and $\mathbf{d}_j$ the unitless transition dipole moment of site j. The x, y, or z-polarized component of the field couples the ground state to a collective bright state $ \sum_j \mathbf{d}_j \cdot \hat{x}_l |j\rangle$.
Eqs. (\ref{eq:collective_emi_1}) and (\ref{eq:collective_emi_2}) constitute the correct description of emission, which is intrinsically collective \cite{Lehmberg_1970,Gross_Haroche_1982_review, Spano_Mukamel_1989}.
\par It is sometimes assumed that the chromophores spontaneously emit independently of one another \cite{Herman_2018}. In this case, the total emission rate would be given by
\begin{equation}
    R_{\text{indep}} = \sum_{j=1}^N \text{Tr}(L^\dagger_j L_j \rho) = \sum_{j=1}^N \Gamma_0 |\mathbf{d}_j|^2 \langle j|\rho|j\rangle ,
\label{eq:indep_emi_1}
\end{equation}
with
\begin{equation}
    L_j = \sqrt{\Gamma_0}|\mathbf{d}_j|\,|g\rangle\langle j|  . 
\label{eq:indep_emi_2}
\end{equation}
Here $\Gamma_0 |\mathbf{d}_j|^2$ is the spontaneous emission rate of chromophore j. We refer to the spontaneous emission described by Eqs. (\ref{eq:indep_emi_1}) - (\ref{eq:indep_emi_2} as independent emission.
\par
The difference between collective and independent emission rates
\begin{equation}
    R_{\text{coll}} - R_{\text{indep}} = \Gamma_0\sum_{j \neq k} \mathbf{d}_{j}^* \cdot \mathbf{d}_{k} \langle j|\rho|k\rangle,
\label{eq:collective_independent_difference}
\end{equation}
derives from the coherence between different excited states, i.e., from the off-diagonal terms in the density matrix, as well as from the non-orthogonality of the different transition dipoles. In an idealized system where all individual chromophores have the same emission rates $\Gamma$, then $R_{\text{indep}}=\Gamma$, but $R_{\text{coll}}$ can vary between $0$ and $N\Gamma$, depending on the dipole orientations and the extent of excitonic coherence between chromophores. 
\par The long time emission behavior is dominated by the HEOM steady state $\rho_{\text{st}}$ (see Eq. (\ref{eq:observation2_1})). Therefore to observe the difference between collective and independent emission at long times, we need only to substitute $\rho_{\text{st}}$ into Eqs. (\ref{eq:collective_emi_1}), (\ref{eq:indep_emi_1}), and (\ref{eq:collective_independent_difference}). Since $\rho_{\text{st}}\propto \text{Tr}_{\text{vib}}(e^{-\beta H_{\text{sys+vib}}})$ \cite{Tanimura_2020}, we find that $\rho_{\text{st}}\propto e^{-\beta H_{\text{sys}}}$ to the lowest order of the system-vibration coupling. 
\begin{figure}[htbp]
    \centering
    \includegraphics[scale=0.6]{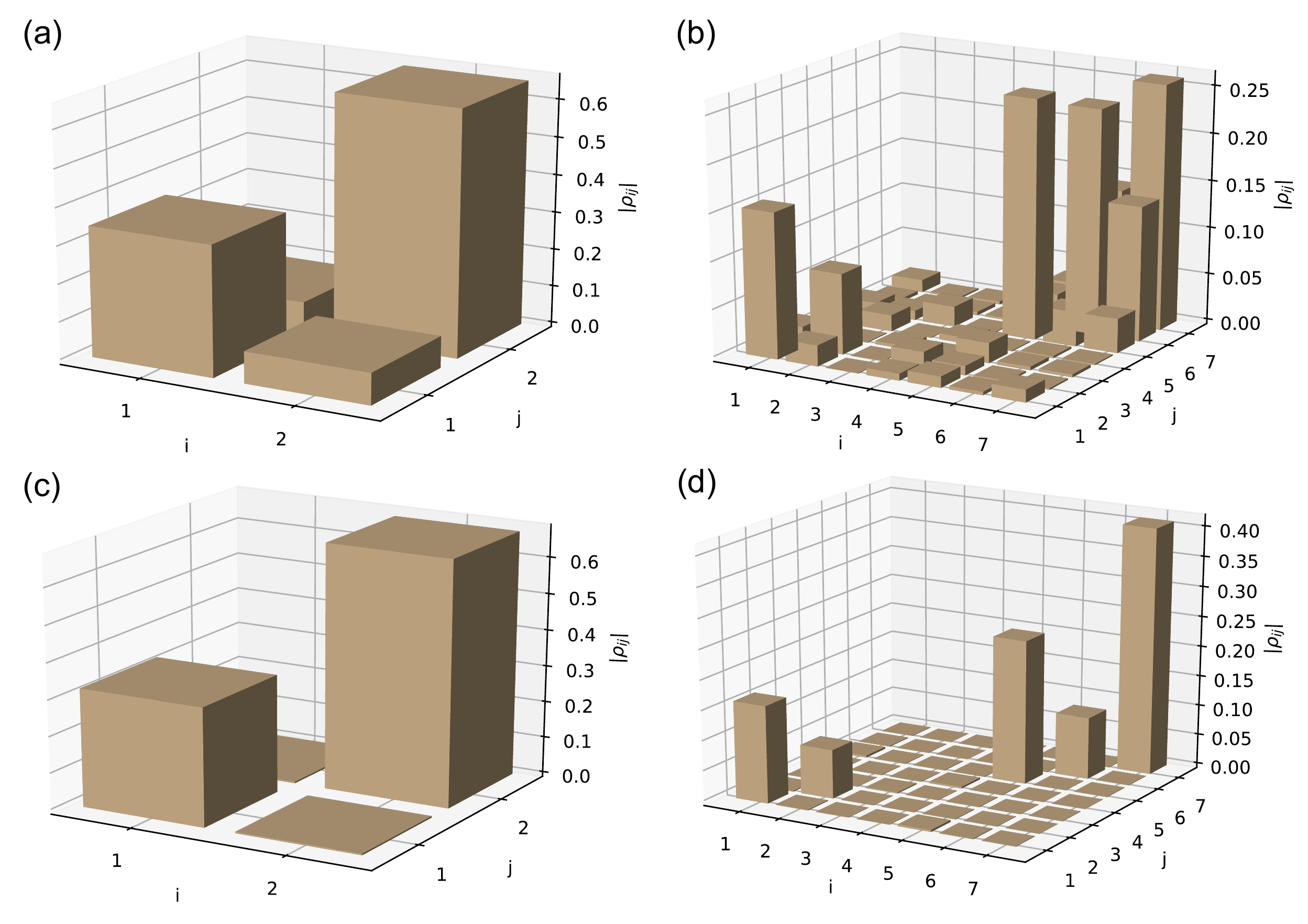}
    \caption{Magnitudes of matrix elements of the HEOM steady state density matrix $\rho_{\text{st}}$ for a dimer and a 7-mer of chromophores in LHCII. (a), (b): dimer and 7-mer system, respectively, in the site basis. (c), (d):  dimer and 7-mer system system, respectively on the energy basis. In the energy eigenbasis, the steady state is almost diagonal. In the site basis, coherences between different sites are present but they are generally smaller than the population terms, consistent with weakly delocalized (Frenkel) excitons.}
    \label{fig:steadystate_rhos}
\end{figure}
Figure (\ref{fig:steadystate_rhos})) shows the result of numerical calculations of the HEOM steady state for a selected dimer and an 7-mer system in LHCII.  These show that $\rho_{\text{st}}$ is close to being diagonal in the energy eigenbasis, and that in the site basis, the off-diagonal coherence terms are generally smaller than the diagonal population terms. This behavior is consistent with the Frenkel character of excitons in LHCII, which show delocalization over a small number of sites (2-3).
Due to the random orientation of the dipoles and the smallness of the coherence between different sites, there is only a modest difference between collective and independent emission rates.
In our LHCII examples, the dimer system has $R_{\text{indep}} = 3.23\times 10^{-2}\,\text{ns}^{-1}$ and $R_{\text{coll}} = 3.59\times 10^{-2}\,\text{ns}^{-1}$; the 7-mer system has $R_{\text{indep}} = 3.28\times 10^{-2}\,\text{ns}^{-1}$ and $R_{\text{coll}} = 4.54\times 10^{-2}\,\text{ns}^{-1}$. 
These small differences are consistent with experimental results for LHCII trimers that show very little enhancement due to collective emission \cite{van_Amerongen_2002}. 
In contrast, bacterial LHI and LHII complexes, which have relatively ordered structures and dipole orientations, exhibit collective emission with enhancement factors of 3-4 over independent emission \cite{van_Grondelle_1997}, while 
bacterial chlorosomes show less enhancement than might be expected from their initial delocalization lengths~\cite{savikhin1998excitation,yakovlev2002exciton,prokhorenko2000exciton} because of exciton relaxation and dynamical disorder~\cite{malina_2021}. 

\section{Double hierarchy solutions for LHCII with calculation of photon fluxes}
\label{sec:LHCII_calculation}
\begin{figure}[htbp]
    \centering
    \includegraphics[scale=0.5]{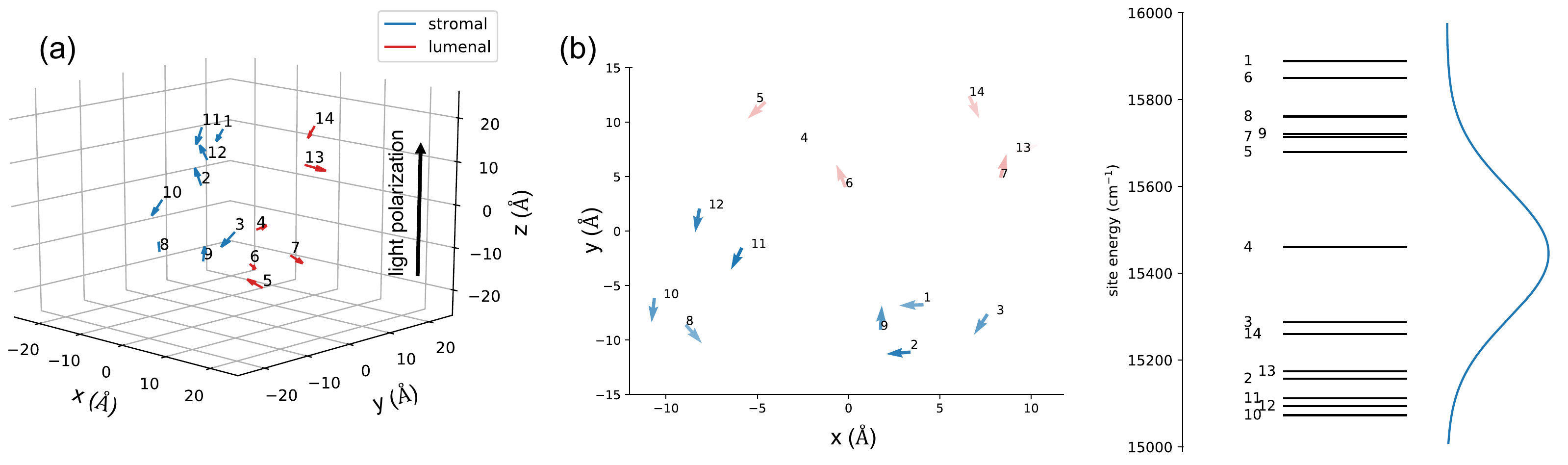}
    \caption{Transition dipole moments and excitonic energy levels of LHCII.  (a) Relative positions and transition dipole moments of the 14 chlorophylls in a LHCII monomer. Blue arrows refer to chlorophylls on the stromal side, red arrows to chlorophylls on the lumenal side. The coordinate system is defined by the singular vector basis of the dipole matrix $\mathbf{D}$ (Section \ref{sec:light_polarization_and_dipole_orientation}). With this convention, the z axis lies approximately in the plane separating the stroma and lumen, and   polarization along $z$ is found to maximize the total absorption probability.
    (b) LHCII chlorophyll transition dipole moments projected into the x-y plane. The intensity of the color of each arrow indicates the extent of overlap between the bright state and the chlorophyll at that site (i.e., $|\langle j|B_{\text{inc}}\rangle|^2$), which is proportional to the square of the z-component of the dipole momen, $|\mathbf{d}_j\cdot \hat{z}|^2$). (c) Site energies of the 14 chlorophylls in a LHCII monomer. The blue curve on the right represents the frequency distribution ($\omega_0 + |\int dt \, \xi(t) e^{i\omega t}|^2$) of the Gaussian single photon pulse.}
    \label{fig:LHCII_setup}
\end{figure}

In this Section we present a numerical solution of the double hierarchy for the Fock state + HEOM master equations (Eq. (\ref{eq:Fock+HEOM_final})) for an LHCII monomer (14-mer) system. The incoming pulse has a Gaussian temporal profile
\begin{equation}
    \xi(t) = \Big(\frac{\Omega^2}{2\pi}\Big)^{1/4} e^{-\Omega^2 (t-t_0)^2/4},
\label{eq:Gaussian_temporal_profile_2}
\end{equation}
where $\Omega$ is the frequency bandwidth, here chosen to be the standard deviation of the site energies ($\Omega\approx (17.7\, \text{fs})^{-1}$ or $299 \,\text{cm}^{-1}$). $t_0$ is the time of the center of the pulse, chosen to be late enough in time (150 fs) that no appreciable tail of the Gaussian pulse is present before $t=0$. The  center frequency is set to be the average site energy ($\omega_0\approx$15445 $\text{cm}^{-1}$) (see Figure (\ref{fig:LHCII_setup})). An experimentally reasonable value for the incoming paraxial beam geometric factor is $\eta=0.11$. In order to maximize the total absorption probability, the incoming field polarization was chosen to be the singular vector of the dipole matrix $\mathbf{D}$ with the largest singular value (see Section \ref{sec:light_polarization_and_dipole_orientation}). We set this singular vector to point along the +z-axis, which is found to lie approximately in the plane of the thylakoid membrane separating the stroma and lumen of a thylakoid disc (see Figure (\ref{fig:LHCII_setup})). The other two orthogonal singular vectors are set to point along the x- and y-axes. 5 HEOM levels were included in the calculation. To check the error due to truncating the HEOM levels, we also performed the calculation with 4 HEOM levels and compare the results below. For any time-dependent quantity $f_n(t)$ obtained from n-level HEOM calculations, we determine an estimated error bar $f_n(t) \pm \text{err}(t)$, where $\text{err}(t)$ is the moving average of $|f_n(t)-f_{n-1}(t)|$ in the $t\pm 20\,\text{fs}$ window.

\par
\begin{figure}[htbp]
    \centering
    \includegraphics[scale=0.6]{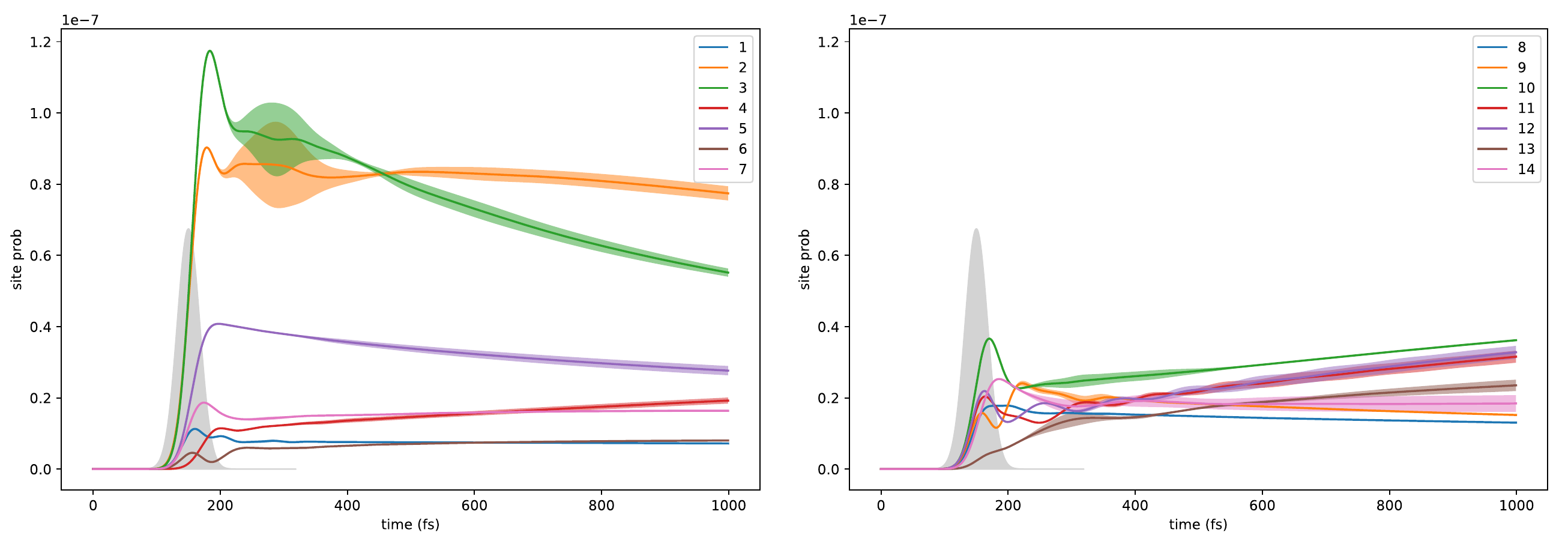}
    \caption{Numerical calculations of the absorption dynamics and subsequent excitonic energy transport for the LHCII monomer, a 14-mer system, using the double hierarchy for the Fock state + HEOM master equation. The 14 site probabilities as functions of time are plotted. The error due to finite HEOM levels are indicated by the colored regions around the solid lines. Gray region represent the Gaussian temporal profile squared $|\xi(t)|^2$ (see Eq. (\ref{eq:Gaussian_temporal_profile})), which has the normalization $\int |\xi(t)|^2 dt = 1$.}
    \label{fig:LHCII_sites}
\end{figure}

Figure (\ref{fig:LHCII_sites}) shows the calculated site probabilities as functions of time. The site probabilities rise initially as the incoming pulse passes. These show oscillations over the following few hundreds of fs, due to the excitonic dipole-dipole couplings and to the interaction with phonons, as discussed in many recent studies~\cite{Ishizaki_2009,Kundu_2020}.
At later times the system slowly approaches a quasi-steady state due to the dephasing interaction with phonons. This is not a true steady state because the excited state site probabilities decay over the ns spontaneous emission time scale (see Section \ref{sec:analysis_on_emission}). In the absence of exciton-phonon interactions, the site probabilities will continue to oscillate due to excitonic coherence until the ns time scale spontaneous emission removes the excitation.
\par
\begin{figure}[htbp]
    \centering
    \includegraphics[scale=0.5]{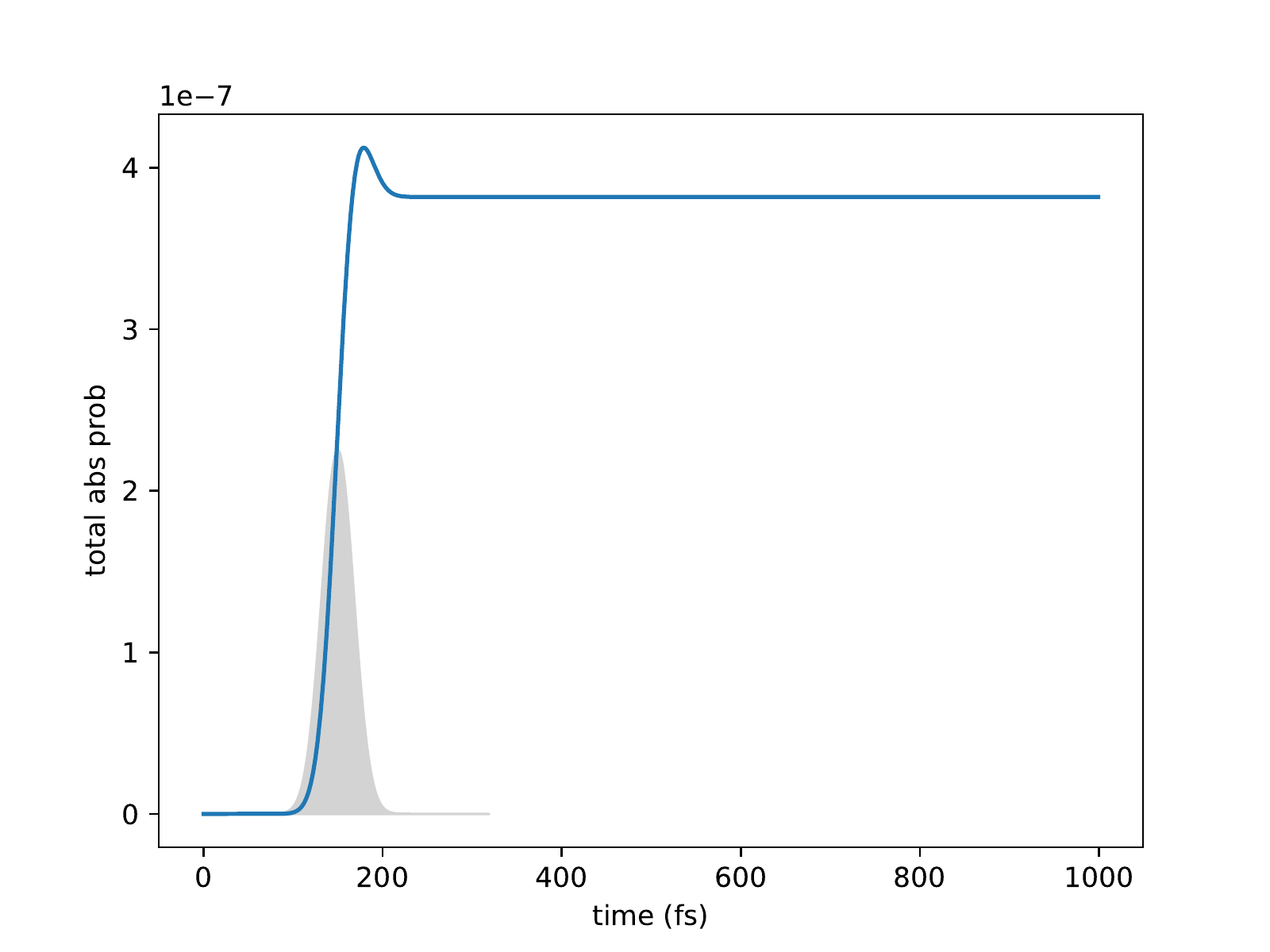}
    \caption{Total absorption probability for LHCII on interaction with a single Fock state photon pulse, defined as the sum over all excitation probabilities for individual sites, plotted as a function of time. The HEOM error bar here (estimated from the difference between calculations with 4 and 5 levels of the HEOM hierarchy) is smaller than the width of the curve. After the pulse, the total absorption probability remains nearly constant at around $4\times 10^{-7}$ and will decay very slowly to zero on a ns time scale due to spontaneous emission.}
    \label{fig:LHCII_tot_abs_prob}
\end{figure}

 
Figure (\ref{fig:LHCII_tot_abs_prob}) now shows the total excitation probability, defined as the sum of all local excitation probabilities on individual sites. Following an initial rise over the duration of the pulse, this remains nearly constant at around $4\times 10^{-7}$ after the pulse has passed over the time scale of 1 ps. The extremely low absorption probability is due to the exceedingly small magnitude of the system-light coupling $L_{\text{inc}} = \sqrt{\Gamma_{\text{inc}}} |g\rangle \langle B_{\text{inc}}|$. This is a generic feature of natural light harvesting systems~\cite{Herman_2018} which can also be related to the exceedingly slow rate of spontaneous emission (cf. Eq. (\ref{eq:photon_flux_general})) relative to the system and system-phonon time scales.

Since the pulse bandwidth was taken here to be the same as the standard deviation of the site energies (Figure \ref{fig:LHCII_setup}(c)), the pulse is neither in the short nor long pulse regime. However, if we apply the long pulse result that the absorption probability is on the order of $\Gamma_{\text{inc}}\tau_{\text{steady}}/N$ (see Eq. (\ref{eq:long_pulse_abs_prob_estimate})), and taking for simplicity $\tau_{\text{steady}} = 1/\gamma\approx 150\,\text{fs}$, we arrive at an absorption probability on the order of $\sim 3\times 10^{-7}$, consistent with the numerically observed absorption probability even though we are in the intermediate pulse regime. 
\begin{figure}[htbp]
    \centering
    \includegraphics[scale=0.55]{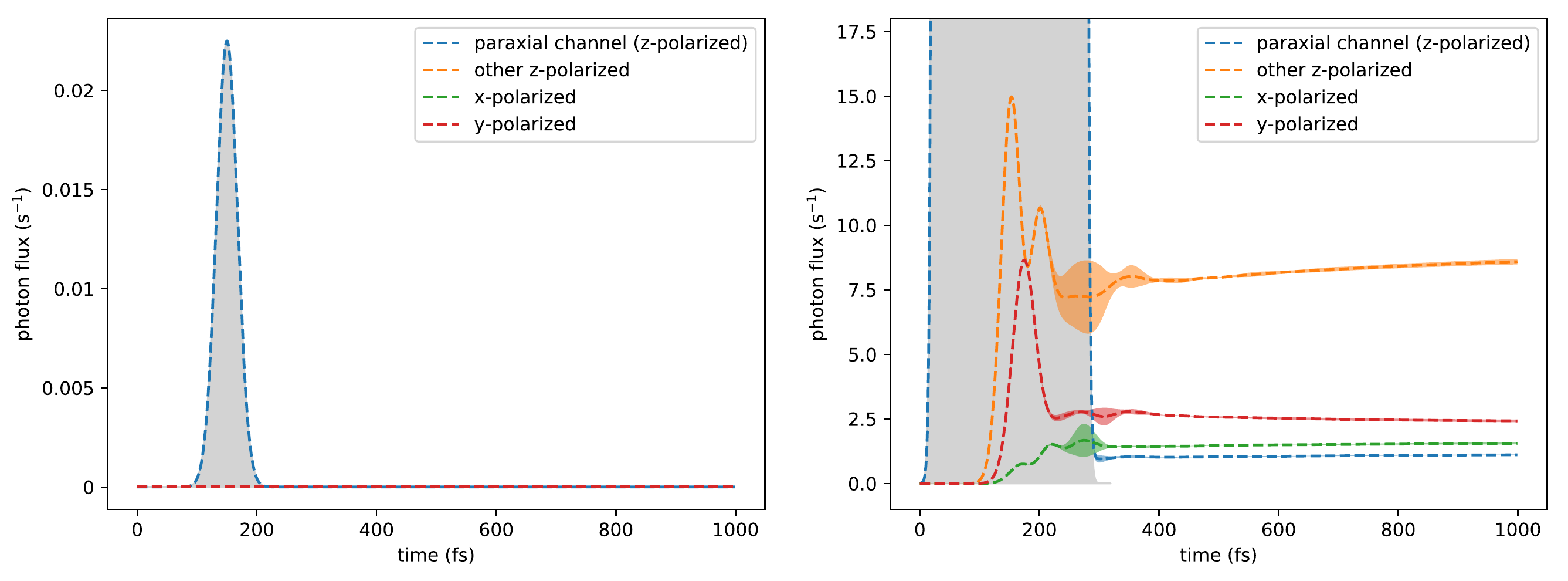}
    \caption{Outgoing photon flux in four different channels following excitation of LHCII by a single photon Fock state pulse.
    Left panel: Due to the small value of the chromophore system-light interaction, the input photon temporal profile squared $|\xi(t)|^2$ (gray region) closely overlaps with the flux in the paraxial channel (blue dashed line)
    Right: Zooming in on the photon flux by about 12 orders of magnitude. Notice that the photon flux is now measured in $\text{s}^{-1}$ instead of $\text{fs}^{-1}$. The small difference between the input photon temporal profile and the outgoing flux in the paraxial channel is now evident.}
    \label{fig:LHCII_flux}
\end{figure}
\par
To analyze the change in photon flux due to absorption and emission by LHCII, the photon fluxes are partitioned into four channels. These are (1) the incoming paraxial channel, z-polarized, with a geometric factor of $\eta=0.11$ (corresponding to a detection area of $7.3\, \%$ of the $4\pi$ solid angle, see discussion above Eq. (\ref{eq:L_def_2})), (2) all other z-polarized light not captured by the incoming channel. (3) all x-polarized light, and (4) all y-polarized light. The system-light coupling operator $L_l$ for each channel is given by Eq. (\ref{eq:L_def_2}), where the incoming channel has a geometric factor $\eta=0.11$, the other z-polarized mode has $\eta=1-0.11=0.89$, and the x- and y-polarized channels have $\eta=1$. The photon fluxes in all four of these channels are plotted as functions of time (see Eq. (\ref{eq:photon_flux_general})) in Figure (\ref{fig:LHCII_flux}). Due to the very small system-light coupling, most of the amplitude of the incoming single photon pulse does not excite the system and appears as outgoing flux in the incoming channel (see Figure (\ref{fig:LHCII_flux})). The fluxes in the other channels, as well as that in the incoming channel after the pulse has passed, are on the order of $\text{s}^{-1}$, around 13 orders of magnitude smaller than the incoming flux before the pulse has passed the LHCII. The very small value of these fluxes after incidence of the single photon is due to the combined effects of the low, $\sim 10^{-7}$, total excitation probability, the weak emission rate, $\Gamma\sim (10\,\text{ns})^{-1}$, and the limited overlap between the system state and the bright state of each channel.

\section{Conclusion}
\label{sec:conclusion}
In this work we have combined an input-output formalism for optical fields with the HEOM formalism for phonon baths to study the excitonic dynamics of photosynthetic light harvesting systems interacting with N-photon Fock state pulses under the influence of coupled phonon degrees of freedom. This combined formalism results in a double hierarchy of equations of motion that need to be solved to obtain the excitonic density matrix. We demonstrated the numerical use of this double hierarchy for single photon absorption and excitonic energy transfer by the LHCII light harvesting complex, possessing 14 chlorophyll chromophores. Under the condition that the system-light coupling is very weak, as for natural light harvesting systems, we developed a number of useful analytic results that can also be applied to larger systems. These include (1) the dependence of the absorption probability on light polarization, dipole orientation, and pulse duration in the limits of short and long pulses, (2) the time evolution of the chromophore system at long times due to spontaneous emission, and (3) the close relationship between the dynamics under Fock state pulses and under coherent state pulses.
\par
To study the absorption behavior, by neglecting the long time spontaneous emission, we could derive expressions for the system state and consequently for the absorption probability. Expressing the temporal profile of the pulse in a scaling form, we were then able to analyze the dependence of absorption probability on pulse duration. 
In the short pulse regime, by defining a system-dependent energy spread parameter $\Delta$ that characterizes the system+vibration time scale ($\tau_{\text{sys+vib}}\sim 1/\Delta$), we found a universal behavior for the absorption probability across all chromophoric systems up to at least the 14 chromophore LHCII system, as well as for different pulse shapes. In the long pulse regime, the absorption probability no longer shows universal behavior and needs to be treated in a case-by-case basis. Taking a chromophore dimer system as an example, we analyzed the different single photon long pulse absorption behavior in three different cases: resonant absorption without phonon coupling, off-resonant absorption without phonon coupling, and off-resonant absorption with phonon coupling. In particular, when phonon coupling is present, the long pulse absorption probability becomes independent of the pulse duration. 
\par 
To study the chromophore system states at long times, we used the fact that the HEOM has a steady state to show that the chromophore system possesses a quasi-steady state, where it reaches a steady state with respect to the phonon bath but has not reached a steady state with respect to the photon bath due to the slow spontaneous emission. This enabled us to understand the chromophore system dynamics in the ps to ns timescale, where numerical integration of the double hierarchy becomes expensive. Furthermore, we used this result to analyze the difference between independent and collective emission as a function of the degree of orientational order and of excitonic coherence. 
We found that for subsystems of LHCII the difference between collective and independent emission is small, implying no significant collective effects, consistent with experimental results for LHCII trimers \cite{van_Amerongen_2002} and expectations based on the non-uniform dipole orientations and the weak extent of coherence between different sites in the excitonic states for LHCII.
\par
An important outcome of this work is the implication of the comparison of the light absorption by photosynthetic systems under excitation by Fock states of light with excitation by coherent states of light.  
For weak system-field couplings $N\Gamma_{\text{inc}}\tau_{\text{pulse}}\ll 1$ (meaning small photon numbers $N$, weak field-chromophore complex coupling constants $\Gamma_{\text{inc}}$, and short pulse durations $\tau_{\text{pulse}}$), we showed that excitation by a coherent state yields the same excited state density matrix, i.e., both populations and coherences, as does excitation by a Fock state with the same temporal profile and average photon number.
This implies that simulating the excitonic dynamics under a short coherent state pulse with an average of $N$ photons, then setting the coherence between ground and excited states to be zero gives an operationally equivalent simulation of the excitonic dynamics under an $N$-photon Fock state excitation. This equivalence holds both with or without phonons. Using physically relevant values of parameters, we showed this equivalence numerically for $N=1$ and $N=20$ photons.\par
This equivalence result and the analysis of absorption probabilities in the limits of short and long pulses reveal a useful complementarity between coherent state and Fock state studies.
For N-photon Fock state studies, coherent states can be used to numerically simulate the more computationally expensive Fock state calculations. On the other hand, the excitation number-conserving property of single photon Fock states has provided us with clues to solve the N-photon Fock state master equations analytically in the physically relevant weak coupling limit. 
This analytical understanding of the absorption probability applies not only to N-photon Fock states, but also to coherent states due to the equivalence in the weak coupling limit. We see that analysis of both Fock and coherent state excitation is valuable for understanding the dynamics of light absorption by light harvesting systems in the weak coupling regime that is relevant to natural photosynthesis in vivo.
 
Finally, we note that the analysis in this work applies to the average state dynamics, relevant to an ensemble of light harvesting systems and an ensemble of experiments with single photons, in which only the output flux of photons is measured.  For consideration of individual experiments with detection of single emitted photons, we can apply a quantum trajectory picture, as described in Ref.~\cite{cook_trajectories_2021}, and obtain additional information about the dynamics of the light harvesting system conditioned upon observation of individual fluorescent photons.  In this interesting situation an incident single photon Fock state and an incident coherent state with an average of one photon no longer give equivalent results~\cite{cook_trajectories_2021}.

\section*{Acknowledgements}
This work was supported by the Photosynthetic Systems program of U.S. Department of Energy, Office of Science, Basic Energy Sciences, within the Division of Chemical Sciences, Geosciences, and Biosciences, under Award No. DESC0019728.

\appendix

\section{Quantizing a paraxial mode}
\label{app:paraxial_quantization}
The vector potential of a paraxial beam propagating in the $+z$ direction takes the form
\begin{equation}
    \mathbf{A}_{\text{para}}(\mathbf{x}, t) = \mathbf{u}(\mathbf{x}, t)e^{ik_0 z - i\omega_0 t} + \text{c.c.},
\label{eq:vector_potential_original}
\end{equation}
where $\mathbf{u}(\mathbf{x}, t)$ is a slowly varying envelope function, and $k_0$ and $\omega_0=ck_0$ are the carrier wavevector and frequency, respectively. The slowly varying envelope is characterized by the conditions
\begin{equation}
\begin{cases}
|\frac{\partial^2 \mathbf{u}}{\partial t^2}|\ll \omega_0|\frac{\partial \mathbf{u}}{\partial t}|\ll \omega_0^2 |\mathbf{u}|\\
|\frac{\partial^2 \mathbf{u}}{\partial z^2}|\ll k_0|\frac{\partial \mathbf{u}}{\partial z}|\ll k_0^2 |\mathbf{u}|.
\end{cases}
\end{equation}
Under these conditions, the wave equation $\nabla^2 \mathbf{A} = \frac{1}{c^2}\frac{\partial^2 \mathbf{A}}{\partial t^2}$ reduces to the paraxial wave equation
\begin{equation}
    \frac{\nabla_{\perp}^2}{2k_0}\mathbf{u} +i (\frac{\partial \mathbf{u}}{\partial z} + \frac{1}{c}\frac{\partial \mathbf{u}}{\partial t})=0,
\end{equation}
which is first order in $z$ and $t$. Therefore we can eliminate one variable and write 
\begin{equation}
    \mathbf{u}(\mathbf{x},t)=f(t_r)\Tilde{\mathbf{u}}(x,y,z),
\label{eq:paraxial_eliminating_variable}
\end{equation}
where $f$ is an arbitrary function and $t_r\equiv t-z/c$ is the retarded time. $\Tilde{\mathbf{u}}(\mathbf{x})$ satisfies the Schr\"{o}dinger-like equation
\begin{equation}
    -i\frac{\partial \Tilde{\mathbf{u}}}{\partial z} = \frac{\nabla^2_\perp}{2k_0}\Tilde{\mathbf{u}}.
\end{equation}
We normalize $\Tilde{\mathbf{u}}$ according to
\begin{equation}
    \int dx dy\,|\Tilde{\mathbf{u}}(x, y, z)|^2 = 1.
\end{equation}
If we fix the spatial mode $\Tilde{\mathbf{u}}(\mathbf{x})$, the field degree of freedom is in the arbitrariness of $f(t_r)$. We express the Fourier components of $f(t_r)$ as
\begin{equation}
    f(t_r)e^{-i\omega_0 t_r} = \frac{1}{\sqrt{L}}\sum_q \underbrace{\phi_q e^{-icqt}}_{\phi_q(t)} e^{iqz},
\label{eq:1d_field_fourier}
\end{equation}
where $q$ takes the values $2\pi n/L$, $n=0,\pm 1, \pm 2, \cdots$. $L$ will be taken to infinity at the end of calculation. Due to the paraxial approximation, $\phi_q$ is localized around $q=k_0$. Substituting Eqs. (\ref{eq:1d_field_fourier}) and (\ref{eq:paraxial_eliminating_variable}) into Eq. (\ref{eq:vector_potential_original}), we have
\begin{equation}
\begin{split}
    \mathbf{A}_{\text{para}}(\mathbf{x}, t) &= \frac{1}{\sqrt{L}} \sum_q \Tilde{\mathbf{u}}(\mathbf{x})\phi_q(t) e^{iqz} + \text{c.c.} \\
    &= \frac{1}{\sqrt{L}} \sum_q [\Tilde{\mathbf{u}}(\mathbf{x})\phi_q(t) + \Tilde{\mathbf{u}}^*(\mathbf{x})\phi_{-q}^*(t)] e^{iqz}.
\end{split}
\end{equation}
From the last equality, we see that the quantity in the square bracket is simply the Fourier-transformed vector potential $\mathbf{A}_q$. Because $\mathbf{A}(\mathbf{x})$ is real-valued, $\mathbf{A}_q=\mathbf{A}_{-q}^*$, so $\{\mathbf{A}_q|q>0\}$ completely specifies $\mathbf{A}(\mathbf{x})$. Since $\{\phi_q\}$ (containing both positive and negative $q$'s) contains twice as many parameters as $\{\mathbf{A}_q|q>0\}$, $\{\phi_q\}$ is a redundant description of the vector potential. To remedy this issue, we set $\phi_q=0,\,\, \forall q<0$. So
\begin{equation}
    \mathbf{A}_{\text{para}}(\mathbf{x}, t) = \frac{1}{\sqrt{L}} \sum_{q>0} \Tilde{\mathbf{u}}(\mathbf{x})\phi_q(t) e^{iqz} + \text{c.c.},
\label{eq:paraxial_A_concise}
\end{equation}
where the sum only ranges over positive $q$'s. The free field electromagnetic Lagrangian in the Coulomb gauge is \cite{Cohen-Tannoudji_photons_and_atoms, Steck_online_book}
\begin{equation}
    \mathcal{L} = \frac{\epsilon_0}{2}\int d^3x\,(\frac{\partial\mathbf{A}}{\partial t})^2 - c^2(\nabla \times \mathbf{A})^2.
\label{eq:EM_Lagrangian}
\end{equation}

Using Eq. (\ref{eq:paraxial_A_concise}), the first term in the Lagrangian is evaluated as
\begin{equation}
\begin{split}
    \int d^3x\, (\frac{\partial\mathbf{A}}{\partial t})^2 &= \frac{1}{L} \int d^3x\,\sum_{q1, q2 >0} \big[\Tilde{\mathbf{u}}^2(\mathbf{x})\Dot{\phi}_{q_1}\Dot{\phi}_{q_2}e^{i(q_1+q_2)z} + \text{c.c.}\big] + 2|\Tilde{\mathbf{u}}(\mathbf{x})|^2\Dot{\phi}_{q_1}\Dot{\phi}^*_{q_2}e^{i(q_1-q_2)z}\\
    &\approx2\sum_q |\Dot{\phi}_q|^2,
\end{split}
\end{equation}
where $\Dot{\phi}$ denotes the time derivative $d\phi/dt$. We dropped the term in the square bracket here because performing the spatial integral gives $\int dz(\int dx dy \Tilde{\mathbf{u}}^2)e^{i(q_1+q_2)z}$. And since the spatial mode is slowly varying in $z$, this integral is non-vanishing only when $q_1+q_2\approx 0$.
However, the paraxial approximation asserts that $\Dot{\phi}_q$ is non-vanishing only when $q\approx k_0$. In this respect, the quantity in the bracket is non-vanishing only when $q_1+q_2\approx 2k_0$, which is incompatible with $q_1+q_2\approx 0$. Therefore the term in the bracket is always small and can be dropped.

In the second term of the Lagrangian,
\begin{equation}
    \nabla\times\mathbf{A} = \frac{1}{\sqrt{L}} \sum_{q>0} (\nabla\times\Tilde{\mathbf{u}} + iq\hat{z}\times\Tilde{\mathbf{u}})\phi_q e^{iqz} + \text{c.c.}
\end{equation}
For the TEM\textsubscript{00} Gaussian beam considered here, $\nabla \times \Tilde{\mathbf{u}}$ is identically zero. For other beam modes $\nabla \times \Tilde{\mathbf{u}}$ is generally nonzero, but it is small compared to the next term $iq\hat{z}\times\Tilde{\mathbf{u}}$, because the sum $\sum_{q>0}$ is dominated by contributions from $q\approx k_0$ and in general $|\nabla\times\Tilde{\mathbf{u}}|$ is much smaller than $|k_0 \Tilde{\mathbf{u}}|$.
The spatial integral of $(\nabla\times\mathbf{A})^2$ follows similarly as above. 

The Lagrangian now takes the form of a collection of harmonic oscillators
\begin{equation}
    \mathcal{L} = \epsilon_0 \sum_{q>0} |\Dot{\phi}_q|^2 - \omega_q^2|\phi_q|^2,
\end{equation}
where $\omega_q\equiv cq$. We first quantize the real and imaginary parts of $\phi_q$ using a set of bosonic operators $\Tilde{a}^{(\text{Re})}_q$, and $\Tilde{a}^{(\text{Im})}_q$, so that the real part of $\phi_q$ is quantized as
\begin{equation}
    \text{Re}\, \phi_q = \sqrt{\frac{\hbar}{4\epsilon_0 \omega_q}} (\Tilde{a}_q^{(\text{Re})} + \Tilde{a}_q^{(\text{Re})\dagger})
\end{equation}
and the imaginary part of $\phi_q$ is quantized as
\begin{equation}
    \text{Im}\, \phi_q = \sqrt{\frac{\hbar}{4\epsilon_0 \omega_q}} (\Tilde{a}_q^{(\text{Im})} + \Tilde{a}_q^{(\text{Im})\dagger})
\end{equation}

Then we transform the ``$\Tilde{a}_q$'' operators into the ``$a_q$'' operators using $a^{(r)}_q = (\Tilde{a}^{(\text{Re})}_q + i\Tilde{a}^{(\text{Im})}_q)/\sqrt{2}$ and $a^{(l)}_q = (\Tilde{a}^{(\text{Re})}_q - i\Tilde{a}^{(\text{Im})}_q)/\sqrt{2}$. The quantized version of the complex-valued $\phi_q$ becomes \cite{Peskin_Schroeder} 
\begin{equation}
    \phi_q = \text{Re}\,\phi_q + i \text{Im}\,\phi_q = \sqrt{\frac{\hbar}{2\epsilon_0\omega_q}}(a^{(r)}_q + a^{(l)\dagger}_q),
\label{eq:quantized_phiq}
\end{equation}
where $a^{(r)}_q$ and $a^{(l)}_q$ are two sets of bosonic operators with commutation relations $[a^{(\nu)}_q, a^{(\nu')\dagger}_{q'}]=\delta_{\nu, \nu'}\delta_{q,q'}$ and $[a^{(\nu)}_q, a^{(\nu')}_{q'}]=[a^{(\nu)\dagger}_q, a^{(\nu')\dagger}_{q'}]=0$. The field Hamiltonian is 
\begin{equation}
    H=\sum_{q>0} \hbar \omega_q (a^{(r)\dagger}_q a^{(r)}_q + a^{(l)\dagger}_q a^{(l)}_q).
\end{equation}
Substituting Eq. (\ref{eq:quantized_phiq}) into Eq. (\ref{eq:paraxial_A_concise}), we arrive at
\begin{equation}
    \mathbf{A}_{\text{para}}(\mathbf{x},t) = \sum_{q>0} \sqrt{\frac{\hbar}{2\epsilon_0 \omega_qL}}\Tilde{\mathbf{u}}\,(\mathbf{x})\left(a^{(r)}_q e^{-i\omega_q t}+a^{(l)^\dagger}_q e^{i\omega_q t}\right)\, e^{iqz} +\text{h.c.}
\end{equation}
Here we see that the transformed operators $a^{(r)}_q$'s and $a^{(l)}_q$'s correspond to right- and left-traveling waves, respectively. The reason that left-traveling waves appear in our quantization is because if we considered left-traveling waves in Eq. (\ref{eq:vector_potential_original}), we would have obtain the exactly same Lagrangian. Since we are only considering the right-traveling waves in this work, we will discard the $a^{(l)}_q$'s and drop the superscript $(r)$.
To obtain the continuum limit $L\rightarrow\infty$, we make the replacements $\sum_{q>0}\rightarrow L/(2\pi c) \int_0^\infty d\omega$ and $a_q\rightarrow \sqrt{2\pi c/L}a(\omega)$, to ensure the commutation relation $[a(\omega),a^\dagger(\omega')]=\delta(\omega-\omega')$.
Finally, using the relation $\mathbf{E} = -\partial\mathbf{A}/\partial t$, we have
\begin{equation}
    \mathbf{E}_{\text{para}}(\mathbf{x},t) =\int_0^\infty d\omega\, \sqrt{\frac{\hbar\omega}{4\pi\epsilon_0 c}} (i\Tilde{\mathbf{u}}(\mathbf{x})a(\omega)e^{-i\omega t_r} + \text{h.c.}).
\end{equation}
Since the size of the chromophoric system is much smaller than the wavelength of visible light, we can apply the dipole approximation and set the light-matter interaction to the form $-\mathbf{d}\cdot\mathbf{E}(\mathbf{x_0})$, where $\mathbf{d}$ is the dipole moment operator of the chromophoric system, and $\mathbf{x_0}$ is the position of this, which is considered as fixed. Without loss of generality, let $\mathbf{x_0}=\mathbf{0}$, and let $\Tilde{\mathbf{u}}(\mathbf{0})$ be real-valued. We then rewrite the electric field of the paraxial mode at location $\mathbf{0}$ as
\begin{equation}
    \mathbf{E}_{\text{para}}(\mathbf{x}= \mathbf{0}, t) =\int_0^\infty d\omega\, \sqrt{\frac{\hbar\omega}{4\pi\epsilon_0 c}}\Tilde{\mathbf{u}}(\mathbf{0}) (ia(\omega)e^{-i\omega t_r} + \text{h.c.}).
\label{eq:paraxial_Efield}
\end{equation}

\section{Decomposing the electric field into a finite sum of 1D fields - small solid angle modes}
\label{app:small_angle_quantization}
The general form of the quantized electric field is \cite{Loudon_2000_book}
\begin{equation}
    \mathbf{E}(\mathbf{x},t) = \int \frac{d^3 k}{(2\pi)^{3/2}} \sum_\lambda \sqrt{\frac{\hbar c|\mathbf{k}|}{2\epsilon_0 }} (ia(\mathbf{k},\lambda)e^{i\mathbf{k}\cdot \mathbf{x}-ic|\mathbf{k}|t}\hat{\epsilon}(\mathbf{k}, \lambda) + \text{h.c.}),
\label{eq:full_quantized_E}
\end{equation}
where we integrate over the 3-dimensional wavevector $\mathbf{k}$ and $\lambda$ indexes the two possible polarization for each $\mathbf{k}$. $\hat{\epsilon}(\mathbf{k}, \lambda)$ is the unit polarization vector corresponding to the mode $(\mathbf{k}, \lambda)$.
From Eq. (\ref{eq:full_quantized_E}), we can partition the integral over all solid angle in $\int d^3 k = \int dk\, k^2 \int d\Omega$ into a sum over integrals over small sections of solid angle $\{\Omega_1, \Omega_2, \cdots\}$. Within a small section of solid angle $\Omega_m$, we can approximate the polarization vectors as two constant unit vectors, $\hat{\epsilon}_{m,1}$ and $\hat{\epsilon}_{m,2}$, for the two polarizations. We can then rewrite the electric field (Eq. (\ref{eq:full_quantized_E})) at position $\mathbf{x}=\mathbf{0}$ as 
\begin{equation}
    \mathbf{E}(\mathbf{x}=\mathbf{0},t) = \sum_m \sum_\lambda \int d\omega\,\frac{\omega^2}{c^3}  \sqrt{\frac{\hbar \omega}{16\pi^3\epsilon_0 }} \int_{\Omega_m}d\Omega\,(ia(\mathbf{k},\lambda)e^{-i\omega t}\hat{\epsilon}_{j,\lambda} + \text{h.c.}).
\end{equation}
Now we define 
\begin{equation}
    a_{m,\lambda}(\omega)\equiv \sqrt{\frac{\omega^2}{\Delta\Omega_m c^3}}\int_{\Omega_m} d\Omega\, a(\mathbf{k}, \lambda),
\label{eq:small_solid_angle_a}
\end{equation}
where $\Delta\Omega_m$ is the ``area" of the solid angle section $\Omega_m$, or $\Delta\Omega_m \equiv \int_{\Omega_m} d\Omega$. The field operators $a_{m,\lambda}(\omega)$'s are defined in this way so that they satisfy the boson commutation relation: $[a_{m,\lambda}(\omega), a^\dagger_{m',\lambda'}(\omega')]=\delta_{m,m'}\delta_{\lambda,\lambda'}\delta(\omega-\omega')$ and $[a_{m,\lambda}(\omega), a_{m',\lambda'}(\omega')]=[a^\dagger_{m,\lambda}(\omega), a^\dagger_{m',\lambda'}(\omega')]=0$. Now we have decomposed the electric field into many one-dimensional fields $\mathbf{E}(t)=\sum_{m,\lambda} \mathbf{E}_{m,\lambda}(t)$, where
\begin{equation}
    \mathbf{E}_{m,\lambda}(t) = \int_0^\infty d\omega \, \sqrt{\frac{\hbar\omega^3\Delta\Omega_m}{16\pi^3\epsilon_0 c^3}}(ia_{m,\lambda}(\omega)e^{-i\omega t}\hat{\epsilon}_{m,\lambda} + \text{h.c.}).
\label{eq:small_solid_angle_E}
\end{equation}

\section{Orthonormal decomposition of the free field modes}
\label{app:orthonormal_decomposition_free_field}
The free field Hamiltonian is
\begin{equation}
    H = \sum_\lambda \int d^3\mathbf{k} \, \hbar c |\mathbf{k}| a_\lambda^\dagger(k) a_\lambda (k),
\end{equation}
where $\mathbf{k}$ indexes the wavevector, $\lambda$ indexes the two polarizations given a wavevector, and $a_\lambda(k)$ is the annihilation operator of field mode indexed by $(k,\lambda)$. In spherical coordinates and with the change of variable $\omega = c |\mathbf{k}|$,
\begin{equation}
    H = \int_0^\infty d\omega\,\hbar\omega \frac{\omega^2}{c^3} \sum_\lambda \int d\Omega \, a_\lambda^\dagger(\omega, \Omega) a_\lambda(\omega, \Omega),
\end{equation}
where $\Omega$ is the solid angle. Suppose there is a complete orthonormal set of functions $g_l(\lambda, \Omega)$ such that
\begin{equation}
    \sum_\lambda \int d\Omega\, g_{l_2}^*(\lambda, \Omega)g_{l_1}(\lambda, \Omega) = \delta_{l_2, l_1} 
\end{equation}
and \cite{Jackson}
\begin{equation}
    \sum_l g_l^*(\lambda_2, \Omega_2) g_l(\lambda_1, \Omega_1) = \delta_{\lambda_2,\lambda_1}\delta(\Omega_2 - \Omega_1),
\end{equation}
then by defining 
\begin{equation}
    a_l(\omega) \equiv \sqrt{\frac{\omega^2}{c^3}} \sum_\lambda \int d\Omega \, g_l(\lambda, \Omega)a_\lambda(\omega, \Omega),
\end{equation}
we have
\begin{equation}
    H = \sum_l \int^\infty_0 d\omega\, \hbar\omega a_l^\dagger(\omega)a_l(\omega).
\end{equation}
Furthermore, the operators $a_l(\omega)$ satisfy the bosonic commutation relations:
$[a_l(\omega),a^\dagger_{l'}(\omega')]=\delta_{l,l'}\delta(\omega-\omega')$ and 
$[a_l(\omega),a_{l'}(\omega')] =[a^\dagger_l(\omega),a^\dagger_{l'}(\omega')]=0$.
\par The electric field decompositions described in appendices \ref{app:small_angle_quantization} and \ref{app:polarization_quantization} are in fact examples of choosing orthonormal but incomplete set of $g_l$'s (see Eqs. (\ref{eq:small_solid_angle_a}) and (\ref{eq:polarization_mode_a})). The set of functions can be made complete by adding infinitely many more orthonormal $g_l$'s.

\section{Decomposing the electric field into a finite sum of 1D fields - polarization modes}
\label{app:polarization_quantization}
Define 
\begin{equation}
    a_x(\omega)\equiv \sqrt{\frac{3\omega^2}{8\pi c^3}} \int d\Omega \sum_\lambda \hat{x}\cdot \hat{\epsilon}(\mathbf{k}, \lambda)a(\mathbf{k},\lambda),
\label{eq:polarization_mode_a}
\end{equation}
where $\omega>0$, $\int d\Omega$ is the integral over all solid angle, and $\mathbf{k}$ is the wavevector corresponding to $\omega$ and $\Omega$. We also apply similar definitions for $a_y(\omega)$ and $a_z(\omega)$. One can directly check that $a_\mu(\omega)\,,\mu\in\{x,y,z\}$ satisfy the boson commutation relations: $[a_\mu(\omega),a^\dagger_{\mu'}(\omega')]=\delta_{\mu,\mu'}\delta(\omega-\omega')$ and $[a_\mu(\omega),a_{\mu'}(\omega')]=[a^\dagger_\mu(\omega),a^\dagger_{\mu'}(\omega')]=0$. The x-component of the electric field at position $\mathbf{x}=\mathbf{0}$ can be written in polarization modes as
\begin{equation}
    E_x(t)=\int^\infty_0 d\omega \, \sqrt{\frac{\hbar\omega^3}{6\pi^2 \epsilon_0 c^3}}(ia_x(\omega)e^{-i\omega t} + \text{h.c.}),
\label{eq:polarization_mode_E}
\end{equation}
and similarly for $E_y$ and $E_z$.

\section{Deriving the input-output relation (Eq. (\ref{eq:input_output_relation}))}
\label{app:input_output_relation}
Using Eqs. (\ref{eq:input_output_5}) and (\ref{eq:QSDE_unitary_1}), we have the time derivative
\begin{equation}
\begin{split}
    \frac{d}{dt'} U^\dagger(t')a_l(t)U(t') &= -i U^\dagger(t')[a_l(t), H_{\text{int}}(t')]U(t') \\
    &= L_l(t')\delta(t-t').
\end{split}
\end{equation}
Solving this equation, we obtain the input-output relation
\begin{equation}
U(t')a_l(t)U^\dagger(t') =
\begin{cases}
a_l(t)\quad t'<t\\
a_l(t) + \frac{1}{2}L_l(t) \quad t'=t\\
a_l(t) + L_l(t) \quad t'>t.
\end{cases}
\label{eq:input_output_relation_appendix}
\end{equation}
In the case $t=t'$, the factor of $1/2$ arises from ``cutting" the delta function in half and dropping the imaginary principal value part \cite{Gardiner_Collett_1985, Gough_2006}. If one works in the language of quantum stochastic differential equations, the case of $t=t'$ requires more careful treatments, as it depends on whether Ito or Stratonovich integration is used \cite{Baragiola_2012, Combes_2017_review, Rob_thesis}. For our purpose, working in the language of ordinary differential equations and cutting the delta function in half allows us to derive the results correctly and self-consistently.

\section{Deriving the Fock state master equation}
\label{app:deriving_Fock_state_master_equation}
For generality, we shall derive the Fock state master equation for a system interacting with an N-photon Fock state in a spatial mode. A normalized N-photon Fock state in spatial mode $l=\text{inc}$ is specified by
\begin{equation}
    |N_\xi\rangle = \frac{1}{\sqrt{N!}} \Big[\int d\tau\,\xi(\tau)a_{\text{inc}}^\dagger(\tau)\Big]^N|\phi\rangle,
\end{equation}
where $\xi(t)$ is the normalized temporal profile of the N-photon Fock state satisfying $\int d\tau\,|\xi(\tau)|^2=1$. Suppose the initial system+field state is factorizable, i.e., $\rho_{\text{sys+field}}=\rho_0\otimes |N_\xi\rangle\langle N_\xi|$. The system state at time $t$ is obtained from a partial trace over the field. The system state $\Tilde{\rho}_{\text{sys}}(t)$ in the interaction picture is then
\begin{equation}
    \Tilde{\rho}_{\text{sys}}(t) = \text{Tr}_{\text{field}}\Big(U(t)\rho_0\otimes |N_\xi\rangle\langle N_\xi|U^\dagger(t)\Big).
\end{equation}
Using Eq. (\ref{eq:QSDE_unitary_3}) to take the time derivative of $\Tilde{\rho}_{\text{sys}}$, we find
\begin{equation}
\begin{split}
    \frac{d\Tilde{\rho}_{\text{sys}}}{dt} = &\big(-iH-\frac{1}{2}\sum_l L_l^\dagger L_l\big)\text{Tr}_{\text{field}}\Big(U(t)\rho_0\otimes |N_\xi\rangle\langle N_\xi|U^\dagger(t)\Big) -\sum_l L_l^\dagger\text{Tr}_{\text{field}}\Big( U(t)a_l(t)\rho_0\otimes |N_\xi\rangle\langle N_\xi|U^\dagger(t)\Big) \\
    &+\sum_l\text{Tr}_{\text{field}}\Big(a_l^\dagger(t) L_l U(t)\rho_0\otimes |N_\xi\rangle\langle N_\xi|U^\dagger(t)\Big)
    +\text{Tr}_{\text{field}}\Big(U(t)\rho_0\otimes |N_\xi\rangle\langle N_\xi|U^\dagger(t)\Big)\big(iH-\frac{1}{2}\sum_l L_l^\dagger L_l\big) \\
    &-\sum_l \text{Tr}_{\text{field}}\Big(U(t)\rho_0\otimes |N_\xi\rangle\langle N_\xi|a_l^\dagger(t)U^\dagger(t)\Big)L_l
    +\sum_l \text{Tr}_{\text{field}}\Big(U(t)\rho_0\otimes |N_\xi\rangle\langle N_\xi|U^\dagger(t)L_l^\dagger a_l(t)\Big)
\end{split}
\label{eq:Fock_master_derivation_1}
\end{equation}
The second (and similarly the fifth) term on the right hand side can be simplified using the identity
\begin{equation}
    a_l(t)|N_\xi\rangle =
    \begin{cases}
    \sqrt{N} \xi(t) |(N-1)_\xi\rangle\, , \quad l=\text{inc} \\
    0 \quad \text{otherwise}.
    \end{cases}
\end{equation}
Hence when $l=\text{inc}$, the partial trace in the second term becomes
\begin{equation}
    \text{Tr}_{\text{field}}\Big( U(t)a_{\text{inc}}(t)\rho_0\otimes |N_\xi\rangle\langle N_\xi|U^\dagger(t)\Big)=\sqrt{N}\xi(t)\text{Tr}_{\text{field}}\Big( U(t)\rho_0\otimes |(N-1)_\xi\rangle\langle N_\xi|U^\dagger(t)\Big).
\label{eq:input_output_trace_trick_1}
\end{equation}
The sixth (and similarly the third) term can be simplified by applying the cyclic property of trace and the commutation relation Eq. (\ref{eq:a_U_commutator}). For example, when $l=\text{inc}$,
\begin{equation}
\begin{split}
    &\text{Tr}_{\text{field}}\Big(U(t)\rho_0\otimes |N_\xi\rangle\langle N_\xi|U^\dagger(t)L_{\text{inc}}^\dagger a_{\text{inc}}(t)\Big)\\
    &=\text{Tr}_{\text{field}}\Big(a_{\text{inc}}(t)U(t)\rho_0\otimes |N_\xi\rangle\langle N_\xi|U^\dagger(t) \Big)L_{\text{inc}}^\dagger\\
    &=\text{Tr}_{\text{field}}\Big(U(t)a_{\text{inc}}(t)\rho_0\otimes |N_\xi\rangle\langle N_\xi|U^\dagger(t) \Big)L_{\text{inc}}^\dagger+\frac{1}{2}L_{\text{inc}}\text{Tr}_{\text{field}}\Big(U(t)\rho_0\otimes |N_\xi\rangle\langle N_\xi|U^\dagger(t) \Big)L_{\text{inc}}^\dagger.
\label{eq:input_output_trace_trick_2}
\end{split}
\end{equation}
If we now define 
\begin{equation}
    \rho_{m,n}(t)\equiv \text{Tr}_{\text{field}}\Big( U(t)\rho_0\otimes |m_\xi\rangle\langle n_\xi|U^\dagger(t)\Big),
\label{eq:Fock_auxiliary_rho_def}
\end{equation}
then following a similar procedure as above, we obtain the full Fock state master equation
\begin{equation}
\begin{split}
    \frac{d\rho_{m,n}}{dt} =& -i[H,\rho_{m,n}] +\sum_l \mathcal{D}[L_l](\rho_{m,n})\\
    &+\sqrt{m}\xi(t)[\rho_{m-1,n},L_{\text{inc}}^\dagger]+\sqrt{n}\xi^*(t)[L_{\text{inc}}, \rho_{m,n-1}],
\label{eq:Fock_state_master_equation_full}
\end{split}
\end{equation}
with $\rho_{N,N}(t)=\Tilde{\rho}_{\text{sys}}(t)$ and $\mathcal{D}[L]$ is the Lindblad superoperator defined as $\mathcal{D}[L](\rho)\equiv - \frac{1}{2}L^\dagger L \rho - \frac{1}{2}\rho L^\dagger L + L\rho L^\dagger $.  
Here $\rho_{N,N}(t)$ is the physical density matrix that describes the system state given an N-photon Fock state input. $\rho_{N,N}$ couples to other auxiliary density matrices corresponding to smaller number of photons, with lower indices down to $\rho_{0,0}$. Therefore, we need to solve for a hierarchy of $(N+1)^2$ coupled density matrix equations. In the absence of phonon coupling, we can use the property $\rho_{m,n} = \rho_{n,m}^\dagger$ to reduce the number of density matrices to solve for to $(N+1)(N+2)/2$. The initial value $\rho_{m,n}=\delta_{m,n}\,\rho_0$ is obtained from Eq. (\ref{eq:Fock_auxiliary_rho_def}). 

\section{Single photon system+field pure state}
\label{app:pure_state}
To derive the pure state equations, first notice that the most general system + field pure state $|\psi(t)\rangle$ with one excitation takes the form
\begin{equation}
    |\psi(t)\rangle = |\beta(t)\rangle |\text{vac}\rangle + |g\rangle \sum_l \int^\infty_{-\infty} d t_r\, \phi_l(t, t_r)a_l^\dagger(t_r)|\text{vac}\rangle,
\label{eq:pure_state_1}
\end{equation}
where $|\beta(t)\rangle$ is an unnormalized system state in the excited subspace, and $\phi_l(t, t_r)$ is an unnormalized ``wave function'' of the single photon field state in mode $l$ at time $t$. 
Taking the time derivative of Eq. (\ref{eq:pure_state_1}), we have
\begin{equation}
    \frac{d}{dt}|\psi(t)\rangle = \frac{d}{dt}|\beta(t)\rangle |\text{vac}\rangle + |g\rangle \sum_l \int^\infty_{-\infty} dt_r\, \frac{\partial}{\partial t} \phi_l(t,t_r)a_l^\dagger(t_r)|\text{vac}\rangle.
\label{eq:pure_state_2}
\end{equation}
On the other hand, we can write the time derivative as
\begin{equation}
    \frac{d}{dt}|\psi(t)\rangle = \frac{d}{dt}U(t) \bigg[ |g\rangle \int^{\infty}_{-\infty} dt_r\, \xi(t_r)a_{\text{inc}}^\dagger(t_r)|\text{vac}\rangle \bigg].
\end{equation}
Using Eqs. (\ref{eq:QSDE_unitary_3}) and (\ref{eq:pure_state_1}), this becomes
\begin{equation}
\begin{split}
    \frac{d}{dt}|\psi(t)\rangle &= (-iH-\frac{1}{2}\sum_l L^\dagger_l L_l)|\psi(t)\rangle - L^\dagger_{\text{inc}}U(t)\xi(t)|g\rangle|\text{vac}\rangle + \sum_l a^\dagger_l(t)L_l |\psi(t)\rangle \\
    &=\bigg[(-iH-\frac{1}{2}\sum_l L^\dagger_l L_l)|\beta(t)\rangle - \xi(t)L^\dagger_{\text{inc}}|g\rangle \bigg]|\text{vac}\rangle + |g\rangle\sum_l \langle g|L_l|\beta(t)\rangle a^\dagger_l(t)|\text{vac}\rangle.
\end{split}
\label{eq:pure_state_3}
\end{equation}
Comparing Eq. (\ref{eq:pure_state_2}) to (\ref{eq:pure_state_3}), we have
\begin{subequations}
\begin{align}
    \frac{d}{dt}|\beta(t)\rangle & = (-iH-\frac{1}{2}\sum_l L^\dagger_l L_l)|\beta(t)\rangle - \xi(t)L^\dagger_{\text{inc}}|g\rangle\\
    \frac{\partial}{\partial t} \phi_l(t, t_r) & = \langle g|L_l|\beta(t)\rangle\delta(t-t_r).
\end{align}
\label{eq:pure_state_4}
\end{subequations}
The solution to Eq. (\ref{eq:pure_state_4}) is
\begin{subequations}
\begin{align}
    |\beta(t)\rangle &= -\int^t_0 d\tau \, \xi(\tau) e^{(-iH-\frac{1}{2}\sum_l L^\dagger_l L_l)(t-\tau)}L^\dagger_{\text{inc}}|g\rangle \\
    \phi_l(t,t_r) &=
    \begin{cases}
    \delta_{l, \text{inc}}\xi(t_r),\quad \quad \quad t<t_r\\
    \delta_{l, \text{inc}}\xi(t_r)+\frac{1}{2}\langle g|L_l|\beta(t_r)\rangle, \quad t=t_r\\
    \delta_{l, \text{inc}}\xi(t_r)+\langle g|L_l|\beta(t_r)\rangle, \quad t>t_r.
    \end{cases}
\end{align}
\end{subequations}

\section{Deriving the HEOM using generalized cumulant expansion}
\label{app:deriving_HEOM}
In this appendix and the next, we will keep track of factors of $\hbar$ explicitly, so that the results can be applied more easily in numerical studies where $\hbar$ is not set to 1.
To model the interaction with phonons, we first employ the Born-Oppenheimer approximation to separate electronic and nuclear degrees of freedom. Each chromophore is coupled to a set of nuclear coordinates, and the nuclear coordinates of different chromophores are independent of each other.  Next, we use the harmonic approximation to describe the nuclear Hamiltonian near the potential energy minumum as a set of harmonic oscillators. Let the nuclear Hamiltonian for the electronic ground state be
\begin{equation}
    H_{\text{vib,g}} = \sum_\xi \frac{p_\xi^2 }{2} + \frac{\omega_\xi^2 q_\xi^2}{2},
\label{eq:ground_nuclear_H}
\end{equation}
where $\xi$ indexes the normal mode (phonon) coordinates, $\omega_\xi$ is the normal mode frequency, $q_\xi$ and $p_\xi$ are the mass-normalized normal mode coordinate and its conjugate momentum. We set the minimum of the ground state potential energy surface to have zero potential energy. We assume the nuclear Hamiltonian for the excited state is described by the same set of normal mode coordinates and that it takes the usual form of shifted harmonic oscillators
\begin{equation}
    H_{\text{vib,e}} = E_0 + \sum_\xi \frac{p_\xi^2 }{2} + \frac{\omega_\xi^2 (q_\xi+d_\xi)^2}{2},
\label{eq:excited_nuclear_H}
\end{equation}
where $E_0$ is the minimum energy of the excited state potential energy surface, and $d_\xi$ is the coordinate shift of normal mode $\xi$. $H_{\text{vib,e}}$ can be re-expressed as
\begin{equation}
    H_{\text{vib,e}} = \epsilon + H_{\text{vib,g}} + u,
\label{eq:H_vib_e_simplified}
\end{equation}
where 
\begin{equation}
    \epsilon = E_{0,j} + \sum_\xi \omega_{\xi}^2 d_{\xi}^2/2
\label{eq:Franck_Condon_energy}
\end{equation}
is the energy of the vertical transition from the ground state minimum, and
\begin{equation}
    u = \sum_\xi \omega_{\xi}^2 d_{\xi}q_{\xi}
\label{eq:collective_phonon_operator}
\end{equation}
is a linear combination of phonon coordinates.
In the continuum limit, the coupling to phonon coordinates can be described by the spectral density $J(\omega)$, defined as
\begin{equation}
    J(\omega) = \sum_\xi   \frac{\pi }{2\omega_\xi}(\omega_\xi^2 d_\xi)^2 \delta(\omega - \omega_\xi)
\end{equation}
The second term on the right-hand side of Eq. (\ref{eq:Franck_Condon_energy}) is the reorganization energy $\lambda$, which is related to the spectral density by
\begin{equation}
    \lambda = \frac{1}{\pi}\int^\infty_0 d\omega \, \frac{J(\omega)}{\omega}.
\label{eq:reorganization_energy}
\end{equation}

The overall system+vibration Hamiltonian in the 0- and 1-electronic excitation subspace is
\begin{equation}
    H_{\text{sys+vib}}=|g\rangle \langle g| \sum_k H_{\text{vib,g}}^{(k)} + \sum_j |j\rangle\langle j| \big( H_{\text{vib,e}}^{(j)}+ \sum_{k\neq j} H_{\text{vib,g}}^{(k)}\big) + \sum_{j\neq k} J_{jk} |j\rangle \langle k|,
\label{eq:sys+vib_Hamiltonian}
\end{equation}
where the last term on the right hand side describes the dipole-dipole interaction between the singly-excited states. We ignore the small effect of phonons on the dipole-dipole interaction~\cite{renger2012normal}. Note that the nuclear Hamiltonian $H_{\text{vib}}^{(j)}$ for different chromophore sites can have different normal modes, displacements $d_\xi$, and energy shifts $E_0$. Using Eq. (\ref{eq:H_vib_e_simplified}), we can simplify Eq. (\ref{eq:sys+vib_Hamiltonian}) as
\begin{equation}
    H_{\text{sys+vib}}=\underbrace{\sum_j \epsilon_j |j\rangle\langle j| + \sum_{j \neq k} J_{jk}|j\rangle\langle k|}_{H_{\text{sys}}} + \underbrace{\sum_k H_{\text{vib,g}}^{(k)}}_{H_{\text{vib}}} + \underbrace{\sum_j |j\rangle\langle j| u_j}_{H_{\text{sys-vib}}}.
\label{eq:sys+vib_Hamiltonian_2}
\end{equation}
We have separated the system+vibration Hamiltonian here into a system part $H_{\text{sys}}$, a vibration part $H_{\text{vib}}$, and a system-vibration interaction part $H_{\text{sys-vib}}$. Note that $H_{\text{sys}}$ takes exactly the same form as Eq. (\ref{eq:system_Hamiltonian}).

\par 
To pave the way for combining the Fock state master equation and the HEOM, we present a derivation of the HEOM based on the generalized cumulant expansion \cite{Kubo_1962}. First, $H_{\text{vib}}$ is rotated out of the system+vibration Hamiltonian, and we write the interaction Hamiltonian as
\begin{equation}
    H_{\text{I}}(t) = H_{\text{sys}} + \sum_j |j\rangle \langle j| u_j(t),
\label{eq:sys-vib_Hamiltonian}
\end{equation}
where $u_j(t)\equiv \exp (iH_{\text{vib}}t) u_j \exp (-iH_{\text{vib}}t)$. An important property of $u_j(t)$ is the Wick's property
\begin{equation}
    \big\langle \hat{T} u_{j_{2n}}(t_{2n})u_{j_{2n-1}}(t_{2n-1})\cdots u_{j_2}(t_2)u_{j_1}(t_1)\big\rangle = \sum_{\text{a.p.p.}} \prod_{k,l} \big\langle \hat{T}u_{j_k}(t_k) u_{j_l}(t_l) \big\rangle,
\label{eq:Wick}
\end{equation}
where $\hat{T}$ is the time-ordering operator, and the angled bracket $\langle X \rangle\equiv\text{Tr}(\rho_\text{thermal} X)$ denotes averaging with a thermal state. The sum on the right hand side is over all possible pairings $(k,l)$ of the 2n operators. Averaging over an odd number of operators, we have $\langle \hat{T} u_{j_{2n-1}}(t_{2n-1})\cdots u_{j_2}(t_2)u_{j_1}(t_1)\rangle=0$. Therefore, under thermal averaging, $u_j(t)$ behaves like a mean-zero Gaussian random process. Note that $\langle \hat{T}u_{j_k}(t_k) u_{j_l}(t_l) \rangle$ is non-zero only when $j_k=j_l$, meaning these correspond to the phonon operator on the same site, because of the assumption that phonons in different sites are independent.
Substituting $q_{\xi}=\sqrt{\hbar/2\omega_{\xi}}(a_{\xi}+a^\dagger_{\xi})$ into Eq. (\ref{eq:collective_phonon_operator}), we find that the two-point correlation function of phonon operators on the same site is
\begin{equation}
    \langle u_j(t_2)u_j(t_1)\rangle = \frac{\hbar}{\pi}\int^\infty_0 d\omega \, J_j(\omega)\big[ \coth(\frac{\beta\hbar\omega}{2})\cos(\omega\tau) -i \sin(\omega\tau)\big],
\label{eq:2-point_correlation_1}
\end{equation}
where $\tau\equiv t_2-t_1$ and $\beta=1/k_B T$ is the inverse temperature. 
We assume the spectral density takes the Drude-Lorentz form
\begin{equation}
    J_j(\omega) = \frac{2\lambda_j\gamma_j\omega}{\omega^2+\gamma_j^2},
\label{eq:Drude_Lorentz}
\end{equation}
corresponding to the overdamped Browninan oscillator model \cite{Ishizaki_2009, Mukamel_1995_book}, where $\gamma$ is the exponential decay rate of the imaginary part of the correlation function. It is interesting to note that if we require the imaginary part of the correlation function (proportional to the linear response of phonons) be an exponential decay with decay rate $\gamma$, and that Eq. (\ref{eq:reorganization_energy}) be satisfied, then the spectral density has to take the Drude-Lorentz form in Eq. (\ref{eq:Drude_Lorentz}).
In modeling photosynthetic systems, typically $\beta\hbar\gamma <1$, and we approximate $\coth(\beta\hbar\omega/2)$ in Eq. (\ref{eq:2-point_correlation_1}) as $2k_B T/\hbar\omega$. Under this high-temperature approximation and using Eq. (\ref{eq:Drude_Lorentz}), Eq. (\ref{eq:2-point_correlation_1}) becomes
\begin{equation}
    \langle u_j(t_2)u_j(t_1)\rangle = \lambda_j e^{-\gamma_j |\tau|}(2k_B T-i\hbar\gamma_j).
\label{eq:2-point_correlation_2}
\end{equation}
The time evolution of the system+vibration density matrix can be expressed as a superoperator acting on the initial system+vibration density matrix
\begin{equation}
    \rho_{\text{sys+vib}}(t) = \hat{T}\exp \bigg( \int^t_0 d\tau \, -\frac{i}{\hbar} H_I^\times (\tau) \bigg) \rho_{\text{sys+vib}}(0).r
\end{equation}
The exponential is time-ordered, and $H_I^\times(\tau) \rho \equiv [H_I (\tau), \rho]$ is the commutator. Assuming an initial factorized state $\rho_{\text{sys+vib}}(0)=\rho_{\text{sys}}(0) \otimes \rho_{\text{vib, thermal}}$, with the vibrational state in thermal equilibrium, the reduced system state is then obtained as the partial trace of the time-evolved system+vibration state
\begin{equation}
\begin{split}
    \rho_{\text{sys}}(t) &= \text{Tr}_{\text{vib}}\Big(\big(\hat{T}\exp\int^t_0 d\tau\, -\frac{i}{\hbar}H_I^\times (\tau)\big) \rho_{\text{sys+vib}}(0) \Big) \\
    &= \Big\langle \hat{T}\exp\int^t_0 d\tau\, -\frac{i}{\hbar}H_I^\times (\tau)\Big\rangle \rho_{\text{sys}}(0).
\end{split}
\label{eq:sys+vib_rho_MGF}
\end{equation}
In the second line, the angled bracket $\langle \cdots \rangle$ means averaging with the vibration thermal state: this maps a superoperator acting on the system+vibration Liouville space to a superoperator acting on the system Liouville space. Performing a generalized cumulant expansion on Eq. (\ref{eq:sys+vib_rho_MGF}),
\begin{equation}
    \rho_\text{sys}(t) = \hat{T}\exp\Big( -\frac{i}{\hbar}\int^t_0 dt_1\,\langle H_I^\times (t_1)\rangle - \frac{1}{\hbar^2} \int^t_0 dt_2 \int^{t_2}_0 dt_1\, \langle H_I^\times (t_2) H_I^\times (t_1)\rangle - \langle H_I^\times(t_2)\rangle \langle H_I^\times (t_1)\rangle \Big)\rho_\text{sys}(0).
\label{eq:cumulant_expansion_1}
\end{equation}
All higher cumulant terms vanish identically because of Wick's property (Eq. (\ref{eq:Wick})). Evaluating the cumulant averages using Eq. (\ref{eq:2-point_correlation_1}), we find
\begin{subequations}
\begin{align}
    &\langle H_I^\times (t_1)\rangle = H_\text{sys}^\times(t_1) \\
    & \langle H_I^\times (t_2)H_I^\times (t_1)\rangle - \langle H_I^\times (t_2) \rangle \langle H_I^\times (t_1)\rangle= \sum_j \lambda_j e^{-\gamma_j |\tau|} P_j^\times(t_2) (2k_B T  P_j^\times(t_1) -i\hbar \gamma_j P_j^o(t_1)),
\end{align}
\end{subequations}
where $P_j\equiv |j\rangle\langle j|$ and $A^o B\equiv \{A, B\}$ is the anticommutator superoperator. The superoperators $H_\text{sys}^\times$, $P_j^\times$, and $P_j^o$ do not depend on time. However, they are still indexed by time so that they can be properly time-ordered inside the time-ordering operator. Now Eq. (\ref{eq:cumulant_expansion_1}) can be rewritten as 
\begin{equation}
    \rho_\text{sys}(t) = \hat{T} \mathcal{Z}\rho_\text{sys}(0),
\end{equation}
where
\begin{equation}
    \mathcal{Z} = \exp\Big( -\frac{i}{\hbar}\int^t_0 dt_1\,H_\text{sys}(t_1) - \frac{1}{\hbar^2} \sum_j \int^t_0 dt_2 \int^{t_2}_0 dt_1\, \lambda_j e^{-\gamma_j (t_2-t_1)} P_j^\times(t_2) (2k_B T  P_j^\times(t_1) -i\hbar \gamma_j P_j^o(t_1)) \Big).
\end{equation}
We now further define
\begin{equation}
    \mathcal{Y}_j \equiv \frac{1}{\hbar}\int^t_0 dt_1\, e^{-\gamma_j (t-t_1)}(2k_B T P_j^\times (t_1)-i\hbar\gamma_j P_j^o (t_1)),
\label{eq:HEOM_Y_operator}
\end{equation}
as well as the auxiliary density matrices
\begin{equation}
    \rho^{\Vec{n}}(t) \equiv \hat{T}(\prod_j \mathcal{Y}_j^{n_j})\mathcal{Z} \rho_\text{sys}(0),
\label{eq:auxiliary_rho}
\end{equation}
with $\Vec{n} = (n_1, \cdots, n_N)$ is a list of N integers where each integer $n_j$ corresponds to a site. The factor $1/\hbar$ in $\mathcal{Y}$ makes $\mathcal{Y}$ dimensionless and ensures that all auxiliary density matrices have the same dimension. Notice that $\rho^{\Vec{0}}$ is the physical density matrix, and that at time $t=0$,
\begin{equation}
    \rho^{\Vec{n}}(0) =
    \begin{cases}
    \rho_\text{sys}(0)\quad ,\Vec{n}=\Vec{0}\\
    0 \quad ,\Vec{n}\neq \Vec{0}.
    \end{cases}
\end{equation}
We can then obtain the HEOM by taking the time derivative of $\rho^{\Vec{n}}(t)$ (Eq. (\ref{eq:auxiliary_rho})) to arrive at
\begin{equation}
    \frac{d}{dt}\rho^{\Vec{n}}(t) = -\frac{i}{\hbar}H_{\text{sys}}^\times \rho^{\Vec{n}} - (\sum_j n_j \gamma_j) \rho^{\Vec{n}} - \sum_j \frac{\lambda_j}{\hbar} P_j^\times \rho^{\Vec{n}+\hat{e}_j} + n_j(\frac{2k_B T}{\hbar} P_j^\times - i\gamma_j P_j^o) \rho^{\Vec{n}-\hat{e}_j},
\label{eq:HEOM}
\end{equation}
where $\hat{e}_j\equiv (0, \cdots, 0, 1, 0, \cdots, 0)$ is the ``unit vector" with the j\textsuperscript{th} element equals to 1 and all other elements equal to 0. 

Note that we began the derivation in the interaction picture where we rotated out $H_\text{vib}$, but because the auxiliary density matrices contain only the system degrees of freedom, the rotation has no effect on the auxiliary density matrices and one can interpret the HEOM (Eq.(\ref{eq:HEOM})) as being in the Schrodinger picture. Numerically, a cutoff level $N_\text{cutoff}$ has to be introduced, so that only the auxiliary density matrices with $\sum_j n_j \leq N_\text{cutoff}$ are solved. The total number of auxiliary density matrices is $\binom{N+N_{\text{cutoff}}}{N_{\text{cutoff}}}$. 

\section{HEOM terminator equations}
\label{app:HEOM_terminator}
To capture the effect of the auxiliary density matrices $\rho^{\Vec{n}+\hat{e}_j}$ that are one level beyond the terminators, we first write their time derivatives as
\begin{equation}
    \frac{d}{dt}\rho^{\Vec{n}+\hat{e}_j}(t) = -\frac{i}{\hbar}H_{\text{sys}}^\times \rho^{\Vec{n}+\hat{e}_j} - (\gamma_j + \sum_k n_k \gamma_k) \rho^{\Vec{n}+\hat{e}_j} + \sum_k (n_k+\delta_{j,k})(\frac{2k_B T}{\hbar} P_j^\times - i\gamma_j P_j^o) \rho^{\Vec{n}+\hat{e}_j-\hat{e}_k},
\label{eq:HEOM_terminator_1}
\end{equation}
where we have dropped the terms involving auxiliary density matrices that are two levels beyond the terminators. If the cutoff level is high enough such that $(\gamma_j+\sum_k n_k \gamma_k)$ is much larger than the scale of $H_{\text{sys}}$, then Hamiltonian term in Eq. (\ref{eq:HEOM_terminator_1}) can be dropped. Then solving Eq. (\ref{eq:HEOM_terminator_1}) formally, we have
\begin{equation}
    \rho^{\Vec{n}+\hat{e}_j}(t) = \int^t_0 d\tau \, e^{-(\gamma_j+\sum_k n_k \gamma_k)(t-\tau)}\sum_k (n_k+\delta_{j,k})(\frac{2k_B T}{\hbar} P_j^\times - i\gamma_j P_j^o) \rho^{\Vec{n}+\hat{e}_j-\hat{e}_k}(\tau).
\label{eq:HEOM_terminator_2}
\end{equation}
Approximating $e^{-(\gamma_j+\sum_k n_k \gamma_k)|t-\tau|}$ as $2\delta(t-\tau)/(\gamma_j+\sum_k n_k \gamma_k)$, Eq. (\ref{eq:HEOM_terminator_2}) becomes
\begin{equation}
    \rho^{\Vec{n}+\hat{e}_j}(t) = \sum_k \frac{n_k+\delta_{j,k}}{\gamma_j+\sum_k n_k \gamma_k}(\frac{2k_B T}{\hbar} P_j^\times - i\gamma_j P_j^o) \rho^{\Vec{n}+\hat{e}_j-\hat{e}_k}(t).
\label{eq:HEOM_terminator_3}
\end{equation}
Substituting Eq. (\ref{eq:HEOM_terminator_3}) into Eq. (\ref{eq:HEOM}) for the terminators, the time derivatives of the terminators can now be written explicitly as 
\begin{equation}
\begin{split}
    \frac{d}{dt}\rho^{\Vec{n}}(t) = &-\frac{i}{\hbar}H_{\text{sys}}^\times \rho^{\Vec{n}} - (\sum_j n_j \gamma_j) \rho^{\Vec{n}} + \sum_j n_j(\frac{2k_B T}{\hbar} P_j^\times - i\gamma_j P_j^o) \rho^{\Vec{n}-\hat{e}_j} \\
    &-\sum_{j,k} \frac{\lambda_j}{\hbar} \frac{n_k + \delta_{j,k}}{\gamma_j + \sum_l n_l \gamma_l} P_j^\times (\frac{2k_B T}{\hbar} P_k^\times -i\gamma_k P_k^o)\rho^{\Vec{n}+\hat{e}_j - \hat{e}_k}.
\end{split}
\end{equation}

\section{Coherent state master equation and photon flux}
\label{app:coherent_state_master_equation}
A coherent state with coherent amplitude  $\alpha$ and temporal profile $\xi(t_r)$ is given by
\begin{equation}
    |\alpha_\xi\rangle = \exp \Big(\int dt_r \, \alpha(t_r) a^\dagger_{\text{inc}}(t_r) - \alpha^*(t_r) a_{\text{inc}}(t_r)\Big) |\text{vac}\rangle ,
\end{equation}
where $\alpha(t_r)=\alpha \xi(t_r)$.
This is an eigenstate of the annihilation operator $a(t)$ for all $t$, i.e.,
\begin{equation}
    a(t)|\alpha_\xi\rangle = \alpha(t)|\alpha_\xi\rangle.
\end{equation}
The system state $\rho(t)$ in the interaction frame (see Eqs. (\ref{eq:H_0}) and (\ref{eq:input_output_3})) is given by
\begin{equation}
    \rho(t) = \text{Tr}_{\text{field}}\bigg( U(t)\rho(0)\otimes |\alpha\rangle\langle \alpha| U^\dagger(t) \bigg).
\end{equation}
Using Eq. (\ref{eq:QSDE_unitary_3}), we have
\begin{equation}
\begin{split}
    \frac{d}{dt}\rho(t) = & -i[H,\rho(t)] + \frac{1}{2}\sum_l \Big( -L^\dagger_l L_l \rho(t) - \rho(t)L^\dagger_l L_l\Big) - \alpha(t) L^\dagger_{\text{inc}}\rho(t) - \alpha^*(t)\rho(t)L_{\text{inc}}  \\
    & + \sum_l L_l\text{Tr}_{\text{field}}\bigg( U(t)\rho(0)\otimes |\alpha(t_r)\rangle\langle \alpha(t_r)| U^\dagger(t)a_l^\dagger(t) \bigg)\\
    &\qquad \quad+ \text{Tr}_{\text{field}}\bigg( a_l(t)U(t)\rho(0)\otimes |\alpha(t_r)\rangle\langle \alpha(t_r)| U^\dagger(t) \bigg) L_l^\dagger.
\end{split}
\end{equation}
Using the commutation relation Eq. (\ref{eq:a_U_commutator}) to simplify the partial traces yields
\begin{equation}
    \frac{d}{dt}\rho = -i[H-i\alpha(t)L^\dagger_{\text{inc}}+i\alpha^*(t)L_{\text{inc}}, \rho] + \sum_l L_l \rho L^\dagger_l - \frac{1}{2} L^\dagger_l L_l \rho - \frac{1}{2} \rho L^\dagger_l L_l.
\end{equation}
We identify a time dependent classical electric field $\mathbf{E}(t)$ as the expectation value of the coherent state, i.e.,
\begin{equation}
    \mathbf{E}(t) = \langle \alpha_\xi|\mathbf{\hat{E}}(t)|\alpha_\xi\rangle,
\end{equation}
where $\hat{\mathbf{E}}(t)$ is the electric field operator. Then we see that the system evolution follows exactly the semiclassical equation plus spontaneous emission, i.e.,
\begin{equation}
    \frac{d}{dt}\rho = -i[H-\mathbf{d}\cdot\mathbf{E}(t), \rho] + \sum_l L_l \rho L^\dagger_l - \frac{1}{2} L^\dagger_l L_l \rho - \frac{1}{2} \rho L^\dagger_l L_l.
\end{equation}

\section{Second order perturbation analysis of coherent state input}
\label{app:coherent_PT2}
Neglecting the slow spontaneous emission, the coherent state master equation (Eq. (\ref{eq:coherent_state_master_equation})) becomes
\begin{equation}
    \frac{d}{dt}\rho = -i[H-i\alpha(t)L^\dagger_{\text{inc}}+i\alpha^*(t)L_{\text{inc}}, \rho].
\end{equation}
Rotating out the Hamiltonian $H$, the interaction frame density matrix $\Tilde{\rho}(t) = e^{iHt}\rho(t)e^{-iHt}$ follows the equation
\begin{equation}
    \frac{d}{dt}\Tilde{\rho}(t) = [-\alpha(t)L^\dagger_{\text{inc}}(t) + \alpha^*(t)L_{\text{inc}}(t), \Tilde{\rho(t)}],
\end{equation}
where $L_{\text{inc}}(t)\equiv e^{iHt}L_{\text{inc}}e^{-iHt}$.
Given the initial state $\Tilde{\rho}(0)=|g\rangle\langle g|$, to second order perturbation we have,
\begin{equation}
\begin{split}
    \Tilde{\rho}(t) &= |g\rangle\langle g| + \int^t_0 dt_1 \, -\alpha(t_1)L^\dagger_{\text{inc}}(t_1)|g\rangle\langle g| - \alpha^*(t_1)|g\rangle\langle g| L_{\text{inc}}(t_1) \\
    & \quad + \int^t_0 dt_2 \int^{t_2}_0 dt_1 \, -\alpha^*(t_2)\alpha(t_1)L_{\text{inc}}(t_2)L^\dagger_{\text{inc}}(t_1)|g\rangle\langle g| - \alpha(t_2)\alpha^*(t_1)|g\rangle\langle g|L_{\text{inc}}(t_1)L^\dagger_{\text{inc}}(t_2) \\
    & \qquad\qquad\qquad\qquad\,\,\, + \alpha^*(t_2)\alpha(t_1)L^\dagger_{\text{inc}}(t_1)|g\rangle\langle g|L_{\text{inc}}(t_2) + \alpha(t_2)\alpha^*(t_1)L^\dagger_{\text{inc}}(t_2)|g\rangle\langle g|L_{\text{inc}}(t_1).
\end{split}
\end{equation}
In obtaining the equation above, terms involving $L_{\text{inc}}(t)|g\rangle\langle g|$ and $|g\rangle\langle g|L^\dagger_{\text{inc}}(t)$ were dropped. Since $L_{\text{inc}}=\sqrt{\Gamma_{\text{inc}}}|g\rangle\langle B_\text{inc}|$ (see Eq. (\ref{eq:L_def_bright_state})), these terms are identically zero. Using the fact that $e^{-iHt}|g\rangle = |g\rangle$, $L_{\text{inc}}(t)$ can be simplified as $L_{\text{inc}} e^{-iHt}$, and $L_{\text{inc}}^\dagger(t)$ can be simplified as $e^{iHt} L_{\text{inc}}^\dagger(t)$.  Switching the time index $t_1$ and $t_2$ in the first and the third terms in the double integral,
\begin{equation}
\begin{split}
    \Tilde{\rho}(t) &= |g\rangle\langle g| + \int^t_0 dt_1 \, -\alpha(t_1)e^{iHt_1}L^\dagger_{\text{inc}}|g\rangle\langle g| - \alpha^*(t_1)|g\rangle\langle g| L_{\text{inc}}e^{-iHt_1} \\
    &\quad + \int^t_0 dt_2 \int^{t_2}_0 dt_1 \,- \alpha(t_2)\alpha^*(t_1)|g\rangle\langle g|L_{\text{inc}}e^{iH(t_2-t_1)}L^\dagger_{\text{inc}} + \alpha(t_2)\alpha^*(t_1)e^{iHt_2}L^\dagger_{\text{inc}}|g\rangle\langle g|L_{\text{inc}}e^{-iHt_1} \\
    &\quad + \int^t_0 dt_1 \int^{t_1}_0 dt_2 \, -\alpha(t_2)\alpha^*(t_1)L_{\text{inc}}e^{iH(t_2-t_1)}L^\dagger_{\text{inc}}|g\rangle\langle g| + \alpha(t_2)\alpha^*(t_1)e^{iHt_2}L^\dagger_{\text{inc}}|g\rangle\langle g|L_{\text{inc}}e^{-iHt_1}.
\end{split}
\end{equation}
The first term in the first double integral is in fact equal to the first term in the second double integral. To see this, notice that
\begin{equation}
    |g\rangle\langle g|L_{\text{inc}}e^{iH(t_2-t_1)}L^\dagger_{\text{inc}} = L_{\text{inc}}e^{iH(t_2-t_1)}L^\dagger_{\text{inc}}|g\rangle\langle g| = \langle g| L_{\text{inc}}e^{iH(t_2-t_1)}L^\dagger_{\text{inc}} |g\rangle |g\rangle \langle g|.
\end{equation}
Now using the property $\int^t_0 dt_2 \int^{t_2}_0 dt_1 + \int^t_0 dt_1 \int^{t_1}_0 dt_2 = \int^t_0 dt_2 \int^t_0 dt_1$ to combine the double integrals and rotating back to the original frame, i.e.,
$\rho(t) = e^{-iHt}\Tilde{\rho}(t)e^{iHt}$, we write $\rho(t)$ in block matrix form as
\begin{equation}
    \rho(t) = 
    \begin{pmatrix}
        \big(1-\langle \beta_\alpha'(t)|\beta_\alpha'(t)\rangle\big)|g\rangle\langle g| &  |g\rangle\langle \beta_\alpha'(t)|\\
         & \\
        |\beta_\alpha'(t)\rangle\langle g| & |\beta_\alpha'(t)\rangle\langle \beta_\alpha'(t)|
    \end{pmatrix},
\end{equation}
where
\begin{equation}
    |\beta_\alpha'(t)\rangle \equiv -\int^t_0 d\tau \, \alpha(\tau) e^{-iH(t-\tau)}L^\dagger_{\text{inc}}|g\rangle.
\end{equation}
Generalization to include an initial phonon pure state follows similarly. Specifically, given an initial pure state $|g\rangle|v\rangle$ that is a product state of the chromophore ground state and a pure vibrational state $|v\rangle$, we first compute the chromophore system + vibration density matrix to the second order perturbation. Tracing out the vibrational degrees of freedom, we obtain the reduced chromophore system density matrix $\rho_{\text{phonon, pure}}(t)$ as
\begin{equation}
    \rho_{\text{phonon, pure}}(t) =
    \begin{pmatrix}
        \big(1-\langle \gamma_{\alpha,v}'(t)|\gamma_{\alpha,v}'(t)\rangle\big)|g\rangle\langle g| &  \text{Tr}_{\text{vib}}\,|g\rangle\langle \gamma_{\alpha,v}'(t)|\\
         & \\
        \text{Tr}_{\text{vib}}\,|\gamma_{\alpha,v}'(t)\rangle\langle g| & \text{Tr}_{\text{vib}}\,|\gamma_{\alpha,v}'(t)\rangle\langle \gamma_{\alpha,v}'(t)|
    \end{pmatrix},
\label{eq:coherent_PT2_pure_phonon_appendix}
\end{equation}
where 
\begin{equation}
    |\gamma_{\xi,v}'(t)\rangle \equiv -\int^t_0 d\tau \, \xi(\tau) e^{-iH_{\text{sys+vib}}(t-\tau)}L^\dagger_{\text{inc}}|g\rangle e^{-iH_{\text{vib}}\tau}|v\rangle .
\end{equation}
If the initial phonon state is a thermal mixture of pure states $\sum_v P_v |v\rangle\langle v|$, where each pure state $|v\rangle$ has the Boltzmann weight $P_v$, then the reduced chromophore system state in the second order perturbation is simply a thermal mixture of the states Eq. (\ref{eq:coherent_PT2_pure_phonon_appendix}).
\begin{equation}
    \rho_{\text{phonon, mixed}}(t) = \sum_v P_v
    \begin{pmatrix}
        \big(1-\langle \gamma_{\alpha,v}'(t)|\gamma_{\alpha,v}'(t)\rangle\big)|g\rangle\langle g| &  \text{Tr}_{\text{vib}}\,|g\rangle\langle \gamma_{\alpha,v}'(t)|\\
         & \\
        \text{Tr}_{\text{vib}}\,|\gamma_{\alpha,v}'(t)\rangle\langle g| & \text{Tr}_{\text{vib}}\,|\gamma_{\alpha,v}'(t)\rangle\langle \gamma_{\alpha,v}'(t)|
    \end{pmatrix}.
\end{equation}

\section{Detailed proof of the long time emission behavior (eq.\ref{eq:observation2_1} and \ref{eq:observation2_2})}
\label{app:proof_obs2}
We first note that on restriction to the ground and singly excited subspaces of the system Hilbert space, the HEOM has two steady state solutions. One is the ground state $\rho_\text{g} = |g\rangle\langle g|$, and the other lies in the singly excited subspace. We denote the normalized steady state in the singly excited subspace as $\rho_{\text{st}}$. We express the HEOM in a vectorized form as $d\mathbf{v}/dt = \mathbf{A}\mathbf{v}$, where the vector $\mathbf{v}$ contains the vectorized physical system density matrix and all vectorized HEOM auxiliary density matrices, and $\mathbf{A}$ is a matrix such that $\mathbf{A}\mathbf{v}$ gives the HEOM time derivative. The two steady states $\mathbf{v}_1$ and $\mathbf{v}_2$ satisfy $\mathbf{A}\mathbf{v}_1=\mathbf{A}\mathbf{v}_2=0$, and are degenerate eigenvectors of $\mathbf{A}$ with eigenvalue $0$. $\mathbf{v}_1$ consists of $\rho_\text{g}$ in the physical density matrix and $0$ in all other auxiliary density matrices. $\mathbf{v}_2$ consists of $\rho_{\text{st}}$ in the physical density matrix and takes some nonzero values in other auxiliary density matrices. 
\par
After the single photon pulse has passed (i.e.. when $\xi(t)$ becomes negligibly small), the Fock state indices in Eq. (\ref{eq:Fock+HEOM_final}) decouple from each other, and the physical $(1,1)$ component of the Fock state hierarchy evolves with the HEOM plus the Lindblad dissipators that account for spontaneous emission. We therefore write the system dynamics as $d\mathbf{v}/dt = (\mathbf{A}+\mathbf{D})\mathbf{v}$, where $\mathbf{D}$ is the Lindblad dissipator $\sum_l \mathcal{D}[L_l]$. Since the energy scale of $\mathbf{D}$ is much smaller than the energy scale of $\mathbf{A}$, we can think of $\mathbf{D}$ as a perturbation on $\mathbf{A}$ that breaks the degeneracy of $\mathbf{v}_1$ and $\mathbf{v}_2$. $\mathbf{v}_1$ remains the eigenvector with zero eigenvalue, since one can check directly that $\mathbf{D}\mathbf{v}_1=0$. Following a degenerate perturbation theory approach, to the lowest order the other perturbed eigenvector $\mathbf{v}_2'$ can be written as a linear combination of the unperturbed eigenvectors plus a correction of order $\epsilon\sim|\mathbf{D}|/|\mathbf{A}|\sim \tau_{\text{sys+vib}}/\tau_{\text{emission}}$:
\begin{equation}
    \mathbf{v}_2' = c_1\mathbf{v}_1 + c_2\mathbf{v}_2 + \mathcal{O}(\epsilon).
\label{eq:perturbed_eigvec}
\end{equation}
The corresponding eigenvalue (denoted as $-\Gamma_\text{long time}$ for reasons to be clear in a moment) is small in magnitude, on the energy scale of spontaneous emission. Note that we do not assume the eigenvectors are orthogonal, since $\mathbf{A}$ and $\mathbf{A}+\mathbf{D}$ are in general not normal operators. 
\par
At long times, when the transients of HEOM decay away, the vectorized system plus auxiliary density matrices take the form
\begin{equation}
    \mathbf{v}(t) = d_1 \mathbf{v}_1 + d_2 \mathbf{v}_2' e^{-\Gamma_{\text{long time}}t}.
\end{equation}
Using Eq. (\ref{eq:perturbed_eigvec}), we write the physical density matrix at long times as
\begin{equation}
    \rho(t) = d_1 \rho_{\text{g}} + d_2 e^{-\Gamma_{\text{long time}}t} (c_1\rho_{\text{g}} + c_2 \rho_{\text{st}} + \mathcal{O}(\epsilon)).
\label{eq:long_time_rho_phenom}
\end{equation}
Since the excited state will eventually decay to the ground state, $\Gamma_{\text{long time}}$ has a positive real part. In fact, we will see below that to the lowest order $\Gamma_{\text{long time}}$ is purely real. Using the fact that $\rho(\infty)=\rho_\text{g}$ and $\text{Tr}\rho = 1$, we obtain $d_1 = 1$ and $c_1=-c_2$, and hence
\begin{equation}
    \rho(t) = \rho_g + b e^{-\Gamma_{\text{long time}}t} (\rho_{\text{st}} - \rho_{\text{g}} + \mathcal{O}(\epsilon)),
\label{eq:long_time_rho_1}
\end{equation}
where $b=d_2 c_2$. Thus the excited portion of the system density matrix follows a single exponential decay into the ground state.
\par To determine the value of $\Gamma_{\text{long time}}$, we first take the time derivative of Eq. (\ref{eq:long_time_rho_1})
\begin{equation}
    \frac{d}{dt}\rho(t) = -b\Gamma_{\text{long time}} e^{-\Gamma_{\text{long time}}t}(\rho_{\text{st}}-\rho_{\text{g}} + \mathcal{O}(\epsilon)).
\label{eq:long_time_rho_deriv_phenom}
\end{equation}
On the other hand, substituting Eq. (\ref{eq:long_time_rho_1}) into the Fock state + HEOM master equation (Eq. (\ref{eq:Fock+HEOM_final})), we have
\begin{equation}
    \frac{d}{dt}\rho(t)= \big(\text{HEOM}+\sum_l \mathcal{D}[L_l]\big)\big(\rho_{\text{g}} + b e^{-\Gamma_{\text{long time}}t} (\rho_{\text{st}} - \rho_{\text{g}}+ \mathcal{O}(\epsilon))\big),
\label{eq:long_time_rho_deriv_micro}
\end{equation}
where ``HEOM" is used as shorthand for the Hamiltonian evolution term together with the HEOM part of the Fock state + HEOM master equation, Eq. (\ref{eq:Fock+HEOM_final}).

The rate of change of the total excited subspace probability is
\begin{equation}
    \frac{dP}{dt} = \text{Tr}(\Pi_{\text{ex}}\frac{d\rho}{dt}),
\label{eq:dP/dt}
\end{equation}
where $\Pi_{\text{ex}}\equiv \sum_j |j\rangle\langle j|$ is the excited subspace projector. From Eq. (\ref{eq:long_time_rho_deriv_phenom}), we have
\begin{equation}
    \frac{dP}{dt} = b\Gamma_{\text{long time}} e^{-\Gamma_{\text{long time}}t}(1 + \mathcal{O}(\epsilon)),
\label{eq:dP/dt_1}
\end{equation}
and from Eq. (\ref{eq:long_time_rho_deriv_micro}), we have
\begin{equation}
    \frac{dP}{dt} = b e^{-\Gamma_{\text{long time}}t}\sum_l\text{Tr}(L_l^\dagger L_l \rho_{\text{st}} + \mathcal{O}(\epsilon)).
\label{eq:dP/dt_2}
\end{equation}
The HEOM term does not contribute to $dP/dt$, since HEOM does not change the system excitation number.
Comparing Eq. (\ref{eq:dP/dt_1}) to Eq. (\ref{eq:dP/dt_2}), we then arrive at the identification
\begin{equation}
    \Gamma_{\text{long time}} = (1+\mathcal{O}(\epsilon))\sum_l \text{Tr} (L^\dagger_l L_l \rho_{\text{st}}).
\label{eq:long_time_main_result_1}
\end{equation}

\section{Absorption probability with phonon in the long pulse regime}
\label{app:long_pulse_phonon}
In our typical HEOM parameter regimes, the time scale for the system to reach the steady state, $\tau_{\text{steady}}$, is on the order of $100\,\text{fs}$. Then if $\tau_{\text{steady}} \ll \tau_{\text{pulse}}$, the system essentially remains in a steady state with respect to the HEOM during $\tau_{\text{pulse}}$. Using the notation of appendix \ref{app:proof_obs2}, we write the physical density matrix as
\begin{equation}
    \rho^{\Vec{0}}_{1,1}(t) = P(t)\rho_{\text{st}} + (1-P(t))\rho_{\text{g}},
\label{eq:long_pulse_system_ansatz}
\end{equation}
where $P(t)$ is the excitation probability. Substituting this ansatz into the Fock state + HEOM master equations, Eq. (\ref{eq:Fock+HEOM_final}), and applying Eq. (\ref{eq:dP/dt}), we can write the time dependence of the excitation probability as
\begin{equation}
    \frac{d}{dt}P(t) = -\xi(t) \text{Tr}(L^\dagger_{\text{inc}} \rho^{\Vec{0}}_{0,1}) + \text{c.c.}
\label{eq:long_pulse_excitation_prob_deriv}
\end{equation}
The time derivative of $P(t)$ depends on $\rho^{\Vec{0}}_{0,1}$. To solve for $\rho^{\Vec{0}}_{0,1}$, we first write the equations for the auxiliary density matrices with Fock state index $(0,1)$ in the vectorized form 
\begin{equation}
    \frac{d}{dt}\mathbf{v}_{0,1}(t) = -\xi^*(t)\mathbf{v}_{0,0} L_{\text{inc}}  + \mathbf{A} \mathbf{v}_{0,1},
\end{equation}
where $\mathbf{v}_{m,n}$ represents the vectorized form of all $\rho^{\Vec{n}}_{m,n}$.
$\mathbf{v}_{0,0}$ is independent of time and consists of $|g\rangle\langle g|$ in the $\Vec{n}=\Vec{0}$ component and $0$ in all other components. The notation $\mathbf{v}_{0,0}L_{\text{inc}}$ means right multiplying every auxiliary density matrix in $\mathbf{v}_{0,0}$ by $L_{\text{inc}}$.
Solving formally for $\mathbf{v}_{0,1}$,
\begin{equation}
    \mathbf{v}_{0,1}(t) = -\int^t_0 d\tau \, e^{\mathbf{A}(t-\tau)} \xi^*(\tau) \big(\mathbf{v}_{0,0} L_{\text{inc}}\big).
\end{equation}
Since the HEOM does not connect different excitation subspaces, and the only two HEOM steady states are in the $|\text{ground}\rangle\langle \text{ground}|$ and $|\text{excited}\rangle\langle \text{excited}|$ blocks, the fact that $|g\rangle \langle g|L$ is in the $|\text{ground}\rangle\langle \text{excited}|$ block implies that $\mathbf{v}_{0,0}L$ only contains the transients of HEOM that decay to zero on the time scale of $\tau_{\text{steady}}$. Therefore the factor $e^{\mathbf{A}(t-\tau)}$ only contributes significantly when $\tau\in [t-\mathcal{O}(\tau_{\text{steady}}), t]$. Within this time interval, the pulse temporal profile $\xi(\tau)$ is essentially constant. Using Eq. (\ref{eq:L_def_bright_state}), we can then approximate $\mathbf{v}_{0,1}$ as
\begin{equation}
    \mathbf{v}_{0,1}(t) = -\sqrt{\Gamma_{\text{inc}}}\xi^*(t) \int^{\mathcal{O}(\tau_{\text{steady}})}_0 d\tau \, e^{\mathbf{A}\tau}\big(\mathbf{v}_{0,0} |g\rangle\langle B_{\text{inc}}|\big) .
\label{eq:long_time_abs_v01}
\end{equation}
The physical HEOM index $\Vec{n}=\Vec{0}$ component of the integrand $e^{\mathbf{A}\tau}\big(\mathbf{v}_{0,0} |g\rangle\langle B_{\text{inc}}|\big)$ is strictly in the $|\text{ground}\rangle\langle \text{excited}|$ block for all $\tau$, and we write the integrand as $|g\rangle \langle \zeta(\tau)|$, where $|\zeta(\tau)\rangle$ is some unnormalized state in the singly excited subspace. This is because the action of the HEOM, $e^{\mathbf{A}\tau}$, does not change the excitation number, and the $\Vec{n}=\Vec{0}$ component of the initial state $\mathbf{v}_{0,0}|g\rangle\langle B_{\text{inc}}|$ (i.e., $|g\rangle\langle B_{\text{inc}}|$) lies in the $|\text{ground}\rangle\langle \text{excited}|$ block. We now write the $\Vec{n}=\Vec{0}$ component of the integral in Eq. (\ref{eq:long_time_abs_v01}) in the form
\begin{equation}
    \int^{\mathcal{O}(\tau_{\text{steady}})}_0 d\tau\,|g\rangle\langle\zeta(\tau)|,
\end{equation}
where $|\zeta(0)\rangle = |B_\text{inc}\rangle$, the normalized bright state, and $|\zeta(\tau)\rangle$ decays to $0$ on the order of $\tau=\tau_{\text{steady}}$. Therefore the value of this integral is in the $|\text{ground}\rangle\langle \text{excited}|$ block and has magnitude on the order of $\tau_{\text{steady}}$. We write this order of magnitude estimate as $\tau_{\text{steady}}|g\rangle\langle \phi|$, where $|\phi\rangle$ is some normalized excited state induced by the HEOM.
Now we can express the order of magnitude estimate of the $\Vec{n}=\Vec{0}$ component of $\mathbf{v}_{0,1}$ in Eq. (\ref{eq:long_time_abs_v01}) as
\begin{equation}
    \rho^{\Vec{0}}_{0,1}(t) \sim -\sqrt{\Gamma_{\text{inc}}}\xi^*(t) \tau_{\text{steady}} |g\rangle\langle \phi|.
\end{equation}
Substituting this into Eq. (\ref{eq:long_pulse_excitation_prob_deriv}), we find
\begin{equation}
    \frac{d}{dt} P(t) \sim \Gamma_{\text{inc}}|\xi(t)|^2 \tau_{\text{steady}} \big| \langle \phi |B_{\text{inc}}\rangle \big|^2.
\end{equation}
Because $|\phi\rangle$ arises from the HEOM 
and $|B_{\text{inc}}\rangle$ arises from the dipole orientations, they are quite independent of each other, and we expect $| \langle \phi |B_{\text{inc}}\rangle |^2 \sim 1/N$, where $N$ is the number of chromophores. (To obtain this scaling we used the fact that the average square overlap of two independent, uniformly distributed normalized vectors in an N-dimensional Hilbert space is $1/N$.) 
Integrating $P(t)$ over the pulse duration, we have 
\begin{equation}
    \text{long pulse abs. prob. with phonon} \sim \frac{\Gamma_{\text{inc}} \tau_{\text{steady}}}{N}.
\label{eq:long_pulse_main}
\end{equation}
We have now arrived at the important result that in the long pulse regime, i.e., $ \tau_{\text{pulse}} \gg \tau_{\text{steady}}$, the absorption probability in the presence of phonon coupling is independent of the pulse duration.

\bibliographystyle{unsrt}
\typeout{}
\bibliography{references.bib}

\end{document}